\documentclass[fleqn,10pt]{wlscirep}
\usepackage{tabularx}
\usepackage{multirow}
\usepackage[FIGTOPCAP,nooneline]{subfigure}
\usepackage{url}
\usepackage{graphicx}
\usepackage{amsmath}
\usepackage[cmintegrals]{newtxmath}
\usepackage{bm}
\usepackage{authblk}

\newcommand{\figsizea}{4cm}
\newcommand{\figsizeb}{4cm}
\newcommand{\figsizec}{4cm}

\newcolumntype{X}{>{\centering\arraybackslash}p{3cm}}
\newcolumntype{Y}{>{\centering\arraybackslash}p{2cm}}
\newcolumntype{Z}{>{\centering\arraybackslash}p{3.5cm}}
\newcolumntype{A}{>{\centering\arraybackslash}p{3cm}}
\newcolumntype{B}{>{\centering\arraybackslash}p{2cm}}
\newcolumntype{C}{>{\centering\arraybackslash}p{1.8cm}}
\newcolumntype{D}{>{\centering\arraybackslash}p{1.75cm}}
\newcolumntype{E}{>{\raggedleft\arraybackslash}p{2.1cm}}
\newcolumntype{F}{>{\centering\arraybackslash}p{0.9cm}}
\newcolumntype{G}{>{\centering\arraybackslash}p{1.25cm}}

\title{
Randomizing hypergraphs preserving degree correlation and local clustering
}

\author[1]{Kazuki Nakajima}
\author[1]{Kazuyuki Shudo}
\author[2, 3, 4, *]{Naoki Masuda}

\affil[1]{Department of Mathematical and Computing Science, Tokyo Institute of Technology}
\affil[2]{Department of Mathematics, State University of New York at Buffalo}
\affil[3]{Computational and Data-Enabled Science and Engineering Program, State University of New York at Buffalo}
\affil[4]{Faculty of Science and Engineering, Waseda University}
\affil[*]{Corresponding author. E-mail address: naokimas@buffalo.edu}

\begin{abstract}
Many complex systems involve direct interactions among more than two entities and can be represented by hypergraphs, in which hyperedges encode higher-order interactions among an arbitrary number of nodes.
To analyze structures and dynamics of given hypergraphs, a solid practice is to compare them with those for randomized hypergraphs that preserve some specific properties of the original hypergraphs. 
In the present study, we propose a family of such reference models for hypergraphs, called the hyper $dK$-series, by extending the so-called $dK$-series for dyadic networks to the case of hypergraphs.
The hyper $dK$-series preserves up to the individual node's degree, node's degree correlation, node's redundancy coefficient, and/or the hyperedge's size depending on the parameter values.
Furthermore, we numerically find that higher-order hyper $dK$-series more accurately preserves the shortest path length and degree distribution of the one-mode projection of the original hypergraph, which the method does not intend to preserve.
We also apply the hyper $dK$-series to numerical simulations of epidemic spreading and evolutionary game dynamics on empirical social hypergraphs.
We find that the hyperedge's size affects these dynamics more than any of the node's properties and that the node's degree correlation and redundancy in the empirical hypergraphs promote cooperation.
\end{abstract}


\begin{document}

\flushbottom
\maketitle
%
%
\thispagestyle{empty}


\section{Introduction}
Networks are a representation of complex systems that consist of nodes and pairwise interactions among the nodes \cite{boccaletti, barrat2008, latora2017, newman_networks}.
Various mathematical and computational methods have enabled us to study the structure and dynamics of network data across disciplines. 
Many networks share structural patterns, such as heterogeneous distributions of the node's degree (i.e., number of other nodes to which a node is directly connected), an abundance of triangles, correlation in terms of the degree of adjacent pairs of nodes, community structure, and many more.
These and other structural properties affect dynamic processes on networks such as epidemic spreading, evolution of cooperation, and synchronization.

Real-world complex systems often involve unit interactions among more than two nodes.
Examples include group conversations in social contact networks \cite{stehle, mastrandrea}, multiple recipients of single emails \cite{klimt}, co-authoring in collaboration networks \cite{newman2001_2, patania, vasilyeva2021}, joint interactions among proteins in biological systems \cite{klamt, estrada}, and many more \cite{benson, battiston}.
These complex systems can be expressed as hypergraphs composed of nodes and hyperedges, where a hyperedge represents interaction among two or more nodes.
A major method for analyzing hypergraphs is to project them to dyadic networks (i.e., conventional networks, in which each edge connects a pair of nodes) and then analyze them \cite{newman2001_2, newman2001_3, barabasi2002}. 
However, a growing body of evidence suggests the limitations of describing the structure and dynamics of networks including higher-order interactions only using pairwise interactions \cite{patania, benson, ramasco, gomez, benson2016, giusti, grilli, lambiotte2019, chodrow, schaub, yoon, lee2021}.
In line with this, various measurements, dynamical process models, and theories have been developed for hypergraphs, particularly in recent years \cite{battiston}.

In general, a reference model for networks produces synthetic networks that preserve some specific properties of the given network and randomize other properties of the given network \cite{cimini2019}.
Regardless of the type of networks (e.g., dyadic networks or hypergraphs), it is a recommended practice that one compares the structure and dynamics of a network at hand with those for randomized networks produced by reference models.
Such an analysis helps us to reveal whether or not the given network has a certain structure relative to the random case and how the structural properties not preserved by the reference network model impacts dynamics on networks.
For dyadic networks, a family of standard reference model is the configuration models that preserve the degree of each node or its expectation \cite{molloy, newman2001, fosdick}.
The configuration models have been used for finding higher-order structural properties of various networks that the node's degree or its distribution does not imply \cite{watts, milo, newman2002, newman2006, boccaletti}.
Furthermore, such findings have led to the development of reference models that preserve some higher-order properties of the input network, e.g., the degree correlation and the clustering coefficient of the node \cite{maslov, serrano, mahadevan, newman2009, karrer, stanton, gjoka_2_5_k, bassler, orsini}.

For hypergraphs, the properties of hyperedges as well as those of nodes are considered to affect their structure and dynamics. 
The existing reference models for hypergraphs preserve only up to the degree of each node and the size of each hyperedge (i.e., number of nodes that belong to each hyperedge) of a given hypergraph \cite{newman2001, saracco2015, boroojeni, saracco2017, chodrow}.
In the present study, we propose a family of reference models for hypergraphs, called the hyper $dK$-series.
The original $dK$-series is a nested family of reference models that preserve local properties of nodes of the given dyadic network \cite{mahadevan, gjoka_2_5_k, orsini}.
The hyper $dK$-series extends the $dK$-series to the case of hypergraphs.
The hyper $dK$-series preserves local properties of nodes and hyperedges of the given hypergraph to tunable extents.
Then, we showcase its use in investigating epidemic spreading \cite{de} and evolutionary game dynamics \cite{alvarez} models on hypergraphs.
Our code for the hyper $dK$-series is available at \url{https://github.com/kazuibasou/hyper-dk-series}.

\section{Preliminaries}

\subsection{Hypergraph and Bipartite Graph}
We represent a network including higher-order interactions among two or more entities as an unweighted hypergraph that consists of a set of nodes $V = \{v_1, \ldots, v_N\}$ and a set of hyperedges $E = \{e_1, \ldots, e_M\}$, where $N$ is the number of nodes, and $M$ is the number of hyperedges.
We assume that the original hypergraph, for which we generate sample hypergraphs using reference models, contains no multiple edges.
Each hyperedge $e_j \in E$ is a subset of $V$ with arbitrary cardinality $|e_j|$.

\begin{figure}[t]
  \begin{center}
	\includegraphics[scale=0.4]{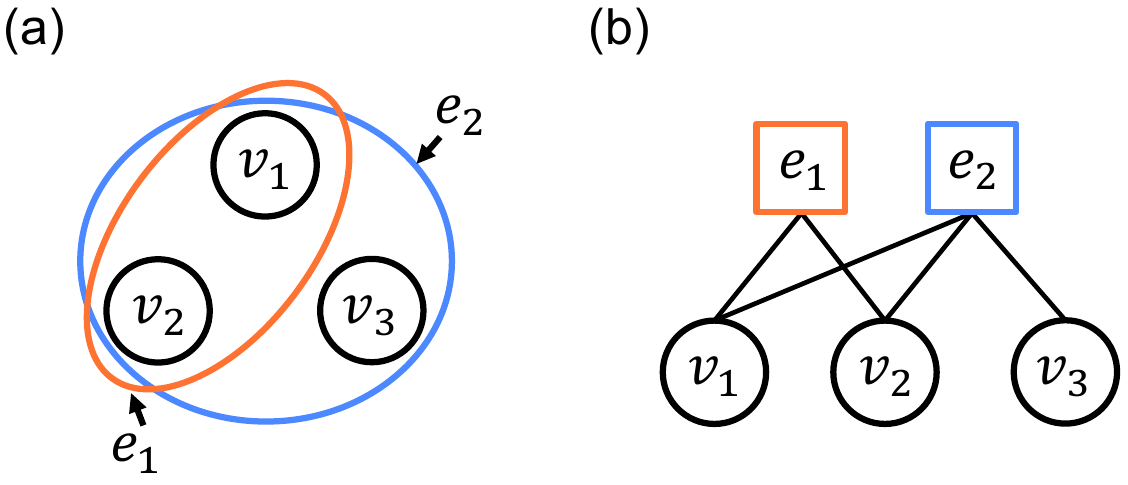}
  \end{center}
  \caption{Hypergraph and the corresponding bipartite graph. (a) A hypergraph that consists of $V = \{v_1, v_2, v_2\}$ and $E = \{e_1, e_2\}$, where $e_1 = \{v_1, v_2\}$ and $e_2 = \{v_1, v_2, v_3\}$. (b) The corresponding bipartite graph, which consists of $V$, $E$, and $\mathcal{E} = \{(v_1, e_1), (v_1, e_2), (v_2, e_1), (v_2, e_2), (v_3, e_2)\}$.}
  \label{hypergraph_example}
\end{figure}

We denote by $G = (V, E, \mathcal{E})$ the bipartite graph that corresponds to the given hypergraph, where $\mathcal{E}$ is a set of edges in the bipartite graph.
An edge $(v_i, e_j)$ exists between each node $v_i$ and each hyperedge $e_j$ if and only if $v_i$ belongs to the hyperedge $e_j$ in the hypergraph.
We denote by $\mathcal{M} = |\mathcal{E}|$ the number of edges in $G$.
We show in Fig. \ref{hypergraph_example} a hypergraph and its bipartite-graph representation.

\subsection{Local Properties of Nodes and Hyperedges}
In this section, we describe local properties of bipartite graph $G$ some of which our reference models preserve.
We denote the incidence matrix of $G$ by $B = (B_{ij})$, where $i=1, \ldots, N,\ j=1, \ldots, M$, $B_{ij} = 1$ if $(v_i, e_j) \in \mathcal{E}$, and $B_{ij} = 0$ otherwise.
Let $k_i = \sum_{j=1}^M B_{ij}$ be the degree of node $v_i$.
We denote the size of hyperedge $e_j$, i.e., the number of nodes that belong to hyperedge $e_j$, by $s_j = \sum_{i=1}^N B_{ij}$.

We define the joint degree distribution of two nodes that share at least one hyperedge, which extends the joint degree distribution for dyadic networks defined in Refs. \cite{orsini, mahadevan}.
Let $m(k, k')$ denote the number of hyperedges that nodes with degree $k = 1, \ldots, M$ and nodes with degree $k' = k, \ldots, M$ share.
For example, in a bipartite graph shown in Fig. \ref{hypergraph_example}(b), one obtains $m(1, 2) = 2$ because node $v_1$ with degree $k_1=2$ and node $v_3$ with degree $k_3=1$ share a hyperedge $e_2$, and node $v_2$ with degree $k_2=2$ and node $v_3$ share a hyperedge $e_2$.
Similarly, one obtains $m(1, 1) = 0$ and $m(2, 2) = 2$.
We define the pairwise joint degree distribution of the node, denoted by $P(k, k')$, as
\begin{align}
P(k, k') = \frac{2 m(k, k')}{\sum_{j=1}^M s_j(s_j-1)}.
\label{eq:1}
\end{align}
Note that $P(k, k')$ is normalized, i.e., $\sum_{k = 1}^M \sum_{k'=k}^M P(k, k') = 1$.
We also define the average degree of the nearest neighbors of nodes with degree $k$, which extends the definition for dyadic networks defined in Refs. \cite{pastor, boccaletti}, by
\begin{align}
k_{\text{nn}}(k) = \frac{\sum_{k'=1}^M k' P(k, k')}{\sum_{k'=1}^M P(k, k')}.
\label{eq:2}
\end{align}
Equations \eqref{eq:1} and \eqref{eq:2} are consistent with the corresponding definitions for dyadic networks when $s_j = 2$ for each hyperedge $e_j \in E$.

We also examine quadruple relationships around a node in a bipartite graph, which is similar to local clustering (i.e., abundance of triangles) in dyadic networks.
The redundancy coefficient of node $v_i$, denoted by $r_i$, quantifies the amount of quadratic relationships around the node in a bipartite graph \cite{latapy}.
It is the fraction of pairs of hyperedges to which $v_i$ belongs such that at least one different node also belongs to both hyperedges.
Formally, if $k_i > 1$, we define
\begin{align}
r_i = \frac{2}{k_i(k_i-1)} \sum_{j=1}^M \sum_{j'=1}^{j-1} B_{i, j} B_{i, j'} 1_{\{|\Gamma|\ >\ 0\}}
\label{eq:3}
\end{align}
where we define $\Gamma = \{v_{i'} \in V \backslash \{v_i\}\ |\ B_{i', j} B_{i', j'} > 0\}$ and $1_{\{cond\}}$ denotes an indicator function that returns 1 if a condition $cond$ holds, and it returns 0 otherwise.
We define $r_i = 0$ if $k_i \in \{0, 1\}$.
The degree-dependent redundancy coefficient of the node is the average of the redundancy coefficient over the nodes with degree $k$, i.e.,
\begin{align}
r(k) = \frac{1}{n(k)} \sum_{i=1,\ k_i=k}^N r_i,
\label{eq:4}
\end{align}
where $n(k)$ is the number of nodes with degree $k$.

One can also define the pairwise joint size distribution of the hyperedge and the redundancy coefficient of the hyperedge in the same way as for the node.
However, we do not introduce them because we construct reference models that preserve up to the size distribution of the hyperedge.
This choice stands on our observation that it is practically difficult to generate randomized bipartite graphs preserving up to pairwise correlation and quadratic relationships for both nodes' degrees and hyperedges' sizes.
If one is interested in preserving the size correlation and redundancy for hyperedges instead of the corresponding quantities for nodes, one can apply our algorithm described in the following text after interchanging the nodes and hyperedges in the bipartite-graph representation of the hypergraph.

\section{Reference models for hypergraphs --- hyper $dK$-series}

\begin{table}[t]
\caption{Properties of nodes and hyperedges corresponding to each $d_v$ and $d_e$ value. The hyper $dK$-series with $(d_v, d_e) = (2, 1)$, for example, preserves the quantities for $d_v = 0,1,2$, and $d_e = 0,1$ shown in this table.}
  \begin{center}
    \begin{tabular}{|G|c|} \hline
      Parameter value & \raisebox{-1.75mm}{Properties to be preserved} \\ \hline
      $d_v = 0$ & Average degree of the node \\ \hline
      $d_v = 1$ & Degree of each node \\ \hline
      $d_v = 2$ & \begin{tabular}{c}Pairwise joint degree distribution of the node \end{tabular} \\ \hline
      $d_v = 2.5$ & \begin{tabular}{c}Degree-dependent redundancy coefficient of the node\end{tabular} \\ \hline
      $d_e = 0$ & Average size of the hyperedge \\ \hline
      $d_e = 1$ & Size of each hyperedge \\ \hline
    \end{tabular}
  \end{center}
  \label{table:1}
\end{table}

In this section, we propose a family of reference models for hypergraphs that preserve local properties of nodes and hyperedges in the given hypergraph to different extents.
We extend a class of reference models for dyadic networks called the $dK$-series \cite{mahadevan, gjoka_2_5_k, orsini} to the case of hypergraphs.
The $dK$-series preserves some local properties of nodes (i.e., degree distribution, joint degree distribution, or degree-dependent clustering coefficient) of a given dyadic network.

The proposed model, which we refer to as hyper $dK$-series, produces a bipartite graph that preserves the joint degree distributions of the node in the subgraphs of size $d_v \in \{0, 1, 2, 2.5\}$ or less and the size distributions of the hyperedge in the subgraphs of size $d_e \in \{0, 1\}$ or less in the given bipartite graph $G$. 
We list the quantity corresponding to each $d_v$ and $d_e$ value in Table \ref{table:1}. 
By definition, the hyper $dK$-series with $d_v =  0$ preserves the numbers of edges in $G$, or equivalently, the average degree of the node.
The hyper $dK$-series with $d_v = 1$ preserves the degree of each node.
With $d_e = 0$ and $d_e = 1$, the hyper $dK$-series similarly preserves the average size of hyperedges and the size of each hyperedge, respectively.
With $d_v = 2$, it preserves the degree of each node and aims to preserve the pairwise joint degree distribution of the node.
With $d_v = 2.5$, it intends to preserve the joint degree distributions of nodes in the subgraphs of size between $d_v = 2$ and $d_v = 3$.
By definition, this means that the hyper $dK$-series preserves the degree of each node, approximately preserves the pairwise joint degree distribution of the node, and approximately preserves the degree-dependent redundancy coefficient of the node.
Like the $dK$-series for dyadic networks \cite{mahadevan, gjoka_2_5_k, orsini}, the hyper $dK$-series have an inclusiveness property. 
In other words, the hyper $dK$-series with given values of $d_v$ and $d_e$ preserve quantities that any hyper $dK$-series with $(d_v', d_e')$, where $d'_v \le d_v$ and $d'_e \le d_e$, preserve.

\subsection{$d_v \in \{0, 1\}$} \label{section:3.1}
In this section, we describe generation of bipartite graphs using the hyper $dK$-series with $d_v \in \{0, 1\}$ and $d_e \in \{0, 1\}$.
We distinguish between the original bipartite graph, denoted by $G = (V, E, \mathcal{E})$, and the bipartite graph produced by the hyper $dK$-series, denoted by $\tilde{G} = (V, E, \tilde{\mathcal{E}})$.
We allow $\tilde{G}$ to have multiple edges between nodes and hyperedges and to have multiple connected components, which are allowed in previous studies as well \cite{latapy, newman2001}.
We define a component of a bipartite graph as any of its maximal subgraphs in which any two nodes are connected to each other by a path within the subgraph.
Our algorithm of the hyper $dK$-series starts with a bipartite graph with $N$ nodes, $M$ hyperedges, and no edge.

\begin{figure*}[t]
  \begin{center}
	\includegraphics[scale=0.4]{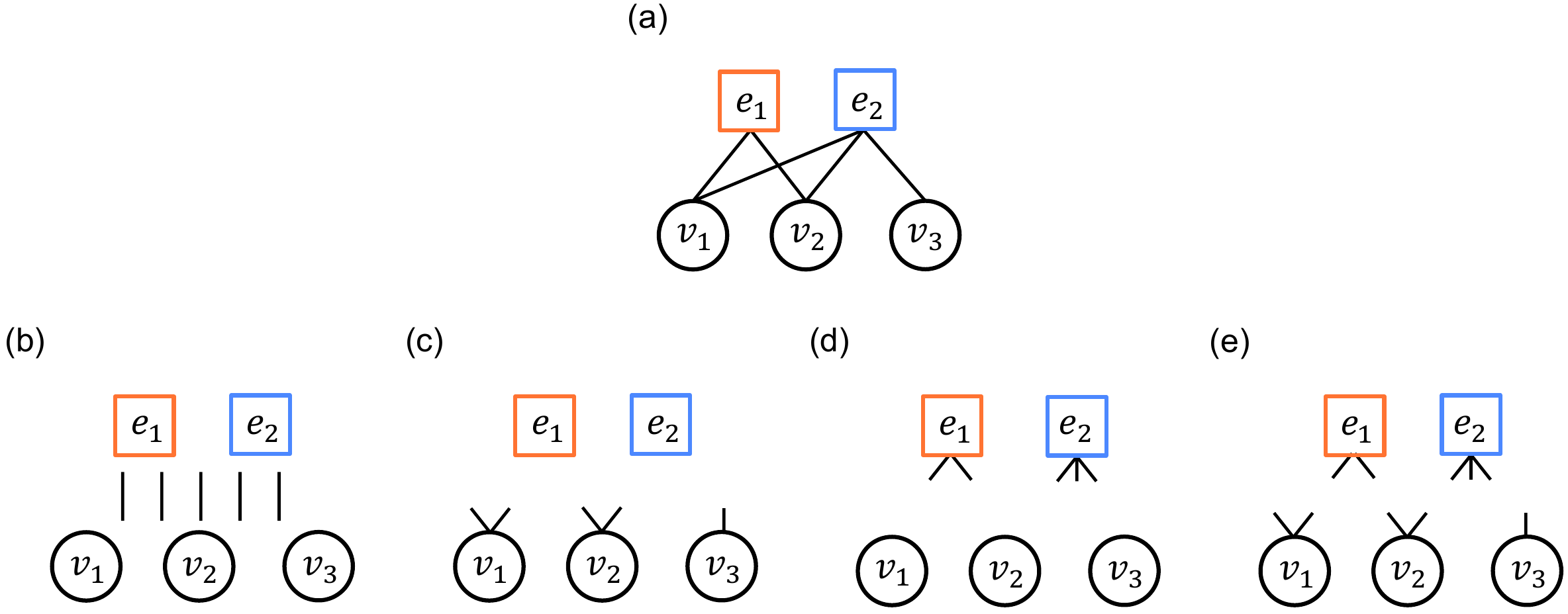}
  \end{center}
  \caption{An example schematically showing the algorithm of the hyper $dK$-series with $d_v \in \{0, 1\}$ and $d_e \in \{0, 1\}$. (a) A bipartite graph. (b) $(d_v, d_e) = (0, 0)$. (c) $(d_v, d_e) = (1, 0)$. (d) $(d_v, d_e) = (0, 1)$. (e) $(d_v, d_e) = (1, 1)$.}
  \label{model_stub}
\end{figure*}

When $(d_v, d_e) = (0, 0)$, we sequentially add edges to construct $\tilde{G}$ as follows. 
We select a node uniformly randomly, i.e., with probability $1/N$ and a hyperedge uniformly at random, i.e., with probability $1/M$, and connect them (Fig. \ref{model_stub}(b)).
We repeat this procedure $\mathcal{M}$ times.
The generated bipartite graph has $N$ nodes, $M$ hyperedges, and $\mathcal{M}$ edges, and hence preserves the average nodal degree and the average size of the hyperedge.
When $(d_v, d_e) = (1,0)$, we first attach $k_i$ half-edges to each node $v_i$  (Fig. \ref{model_stub}(c)).
Then, we connect each of the $\mathcal{M}$ half-edges to a hyperedge chosen uniformly at random, i.e., with probability $1/M$.
The case of $(d_v, d_e) = (0, 1)$ is parallel to that of $(d_v, d_e) = (1,0)$.
Specifically, we first attach $s_j$ half-edges to each hyperedge $e_j$ (Fig. \ref{model_stub}(d)) and then connect each of the $\mathcal{M}$ half-edges to a node chosen uniformly at random, i.e., with probability $1/N$.
When $(d_v, d_e) = (1,1)$, we first attach $k_i$ half-edges to each node $v_i$ and $s_j$ half-edges to each hyperedge $e_j$ (Fig. \ref{model_stub}(e)).
Then, we select a free (i.e., yet available) half-edge attached to a node and a free half-edge attached to a hyperedge uniformly at random and connect them to create a hyperedge. 
We repeat these steps until we exhaust all the free half-edges.

The hyper $dK$-series with $d_v \in \{0, 1\}$ and $d_e \in \{0, 1\}$ are the same as the existing reference models for bipartite graphs.
Specifically, the hyper $dK$-series with $(d_v, d_e) = (1, 1)$ is a standard configuration model for bipartite graphs \cite{newman2001, fosdick}, which one often uses as a reference model for bipartite graphs \cite{saracco2015, latapy, guillaume, pena, tarissan, payrato2019} and hypergraphs \cite{chodrow}.
The hyper $dK$-series with $(d_v, d_e) = (0, 0)$ is the bipartite version of the Erd{\H{o}}s-R{\'e}nyi random graph \cite{er}.
The hyper $dK$-series with $(d_v, d_e) = (0, 1)$ and $(1, 0)$ has also been used as a reference model for bipartite graphs \cite{saracco2017} and hypergraphs \cite{zhou2011}.

\subsection{$d_v \in \{2, 2.5\}$}
The hyper $dK$-series with $d_v \leq 1$ and $d_e \leq 1$ exactly preserves up to the degree of each node and the size of each hyperedge.
However, it is practically difficult to construct a bipartite graph that exactly preserves the pairwise joint degree distribution of the node by starting from the empty network and adding edges.
The intuitive explanation for this difficulty is as follows.
Consider an edge, of which one end has already been attached to a node $v$ with degree $k$. 
Suppose that we connect the other end of this edge to hyperedge $e$ of size $s$.
If $s \ge 3$, then $m(k, k')$, i.e., the number of hyperedges that a node with degree $k$ and a node with degree $k'$ share simultaneously changes for multiple values of $k'$ in general.
This fact makes it difficult to connect edges between nodes and hyperedges one by one while exactly preserving the node's pairwise joint degree distribution, i.e., $P(k, k')$, for all $k$ and $k'$.

This problem is similar to the one for dyadic networks; it is difficult to construct dyadic networks that exactly preserve higher-order structures than the pairwise joint degree distribution of the node \cite{mahadevan, orsini, gjoka_2_5_k}.
To mitigate this problem, the algorithm of the $dK$-series for dyadic networks uses the so-called targeting-rewiring process with the aim of preserving the pairwise joint degree distribution and the triadic relationships, i.e., the degree-dependent clustering coefficient of the node.
In the targeting-rewiring process, one repeatedly rewires edges in the generated network such that the final network exactly preserves the pairwise joint degree distribution and approximately preserves the degree-dependent clustering coefficient of the input network.

\begin{figure*}[t]
  \begin{center}
	\includegraphics[scale=0.275]{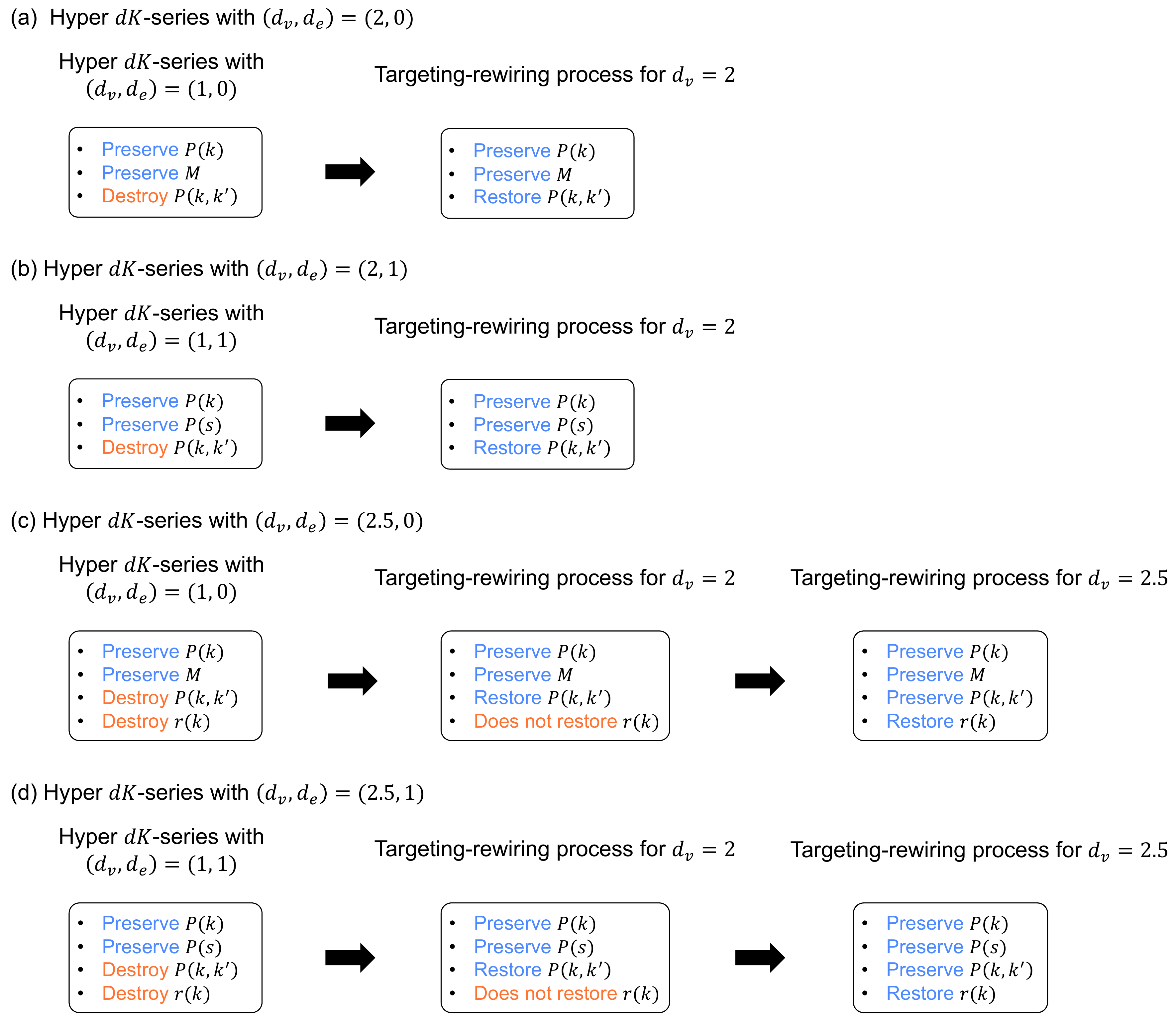}
  \end{center}
  \caption{Workflow of the hyper $dK$-series with $d_v \in \{2, 2.5\}$ and $d_e \in \{0, 1\}$. $M$ represents the number of hyperedges; $P(k)$ represents the degree distribution of the node; $P(s)$ represents the size distribution of the hyperedge; $P(k, k')$ represents the joint degree distribution of the node; $r(k)$ represents the degree-dependent redundancy coefficient of the node.}
  \label{workflow_hyper_dk_series_d_v_2_2_5}
\end{figure*}

We extend the targeting-rewiring process for $dK$-series to the case of bipartite graphs to realize the algorithm of hyper $dK$-series with $d_v \in \{2, 2.5\}$.
We show the composition of the hyper $dK$-series with $d_v \in \{2, 2.5\}$, which involves the targeting-rewiring process, in Fig. \ref{workflow_hyper_dk_series_d_v_2_2_5}.
Specifically, the hyper $dK$-series with $d_v = 2$ starts by generating a bipartite graph using the hyper $dK$-series with $d'_v = 1$ and the given $d_e \in \{0, 1\}$ (see Fig. \ref{workflow_hyper_dk_series_d_v_2_2_5}(a) and \ref{workflow_hyper_dk_series_d_v_2_2_5}(b)).
The generated network preserves the degree of each node and either the average size of hyperedges or the size of each hyperedge depending on whether $d_e = 0$ or $d_e = 1$, respectively.
Then, we run the targeting-rewiring process for $d_v = 2$, which amounts to repeatedly rewiring edges such that the randomized hypergraph approximately restores the joint degree distribution of the original hypergraph while exactly preserving the degree of each node.

\begin{figure}[t]
  \begin{center}
	\includegraphics[scale=0.3]{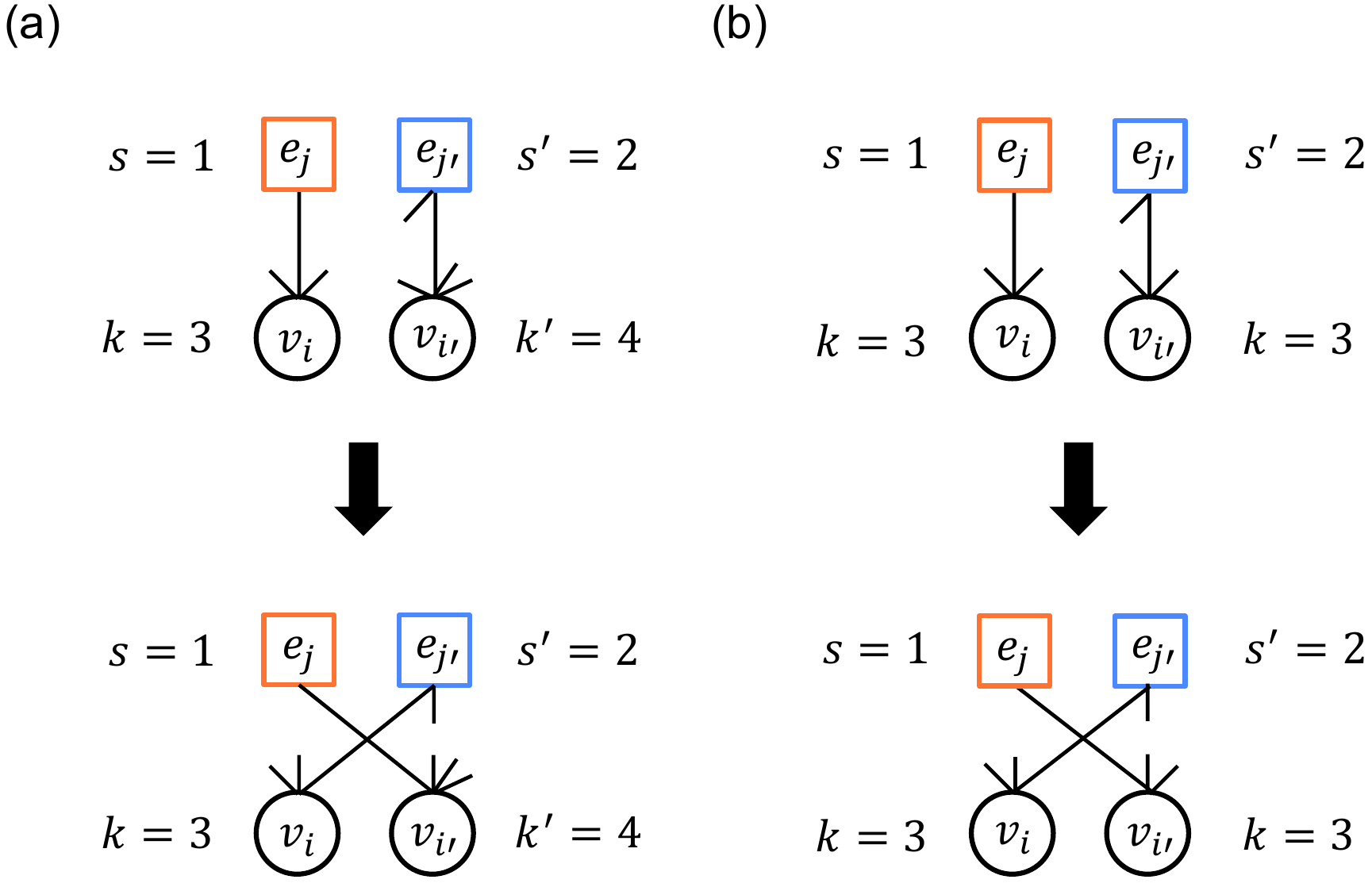}
  \end{center}
  \caption{Rewiring of two edges in the targeting-rewiring process. (a) $d_v = 2$. (b) $d_v = 2.5$. In (a), we allow $k \neq k'$. In (b), we require $k = k'$. Note that the algorithm for $d_v=2.5$ undergoes the rewiring process for $d_v = 2$ shown in (a) before one runs the rewiring process shown in (b).}
  \label{rewiring}
\end{figure}

The targeting-rewiring process for $d_v = 2$ runs as follows.
We first select a pair of edges $(v_i, e_j)$ and $(v_{i'}, e_{j'})$ such that $i \neq i'$ and $j \neq j'$ uniformly at random (see Fig. \ref{rewiring}(a)).
Then, we replace $(v_i, e_j)$ and $(v_{i'}, e_{j'})$ by $(v_i, e_{j'})$ and $(v_{i'}, e_j)$ if and only if a distance between the original and present pairwise joint degree distribution, denoted by $D_{2}$, decreases if we rewire the edges.
Using the normalized $L^1$ distance, we define $D_{2}$ by
\begin{align}
D_2 
&= \frac{\sum_{k=1}^M \sum_{k'=k}^{M} |P'(k, k') - P(k, k')|}{\sum_{k=1}^M \sum_{k'=k}^{M} P(k, k')} \notag \\
&= \sum_{k=1}^M \sum_{k'=k}^{M} \left|\frac{2 m'(k, k')}{\sum_{j=1}^M s'_j(s'_j-1)} - \frac{2m(k, k')}{\sum_{j=1}^M s_j(s_j-1)} \right|, \label{eq:5}
\end{align}
where $P'(k, k')$, $m'(k, k')$, and $s_j^{\prime}$ represent the pairwise joint degree distribution of the node, the number of hyperedges that nodes with degree $k$ and nodes with degree $k'$ share, and the size of hyperedge $e_j$, respectively, for the rewired hypergraph.
To derive the second line in Eq. \eqref{eq:5}, we have used $\sum_{k=1}^M \sum_{k'=k}^{M} P(k, k') = 1$.
We repeat the rewiring attempts $R$ times until $D_2$ becomes sufficiently small and hardly decreases by further rewiring.
We set $R = 500 \mathcal{M}$. 

The rewiring preserves the normalization factor, $\sum_{j=1}^M s'_j (s'_j-1)$, because the rewiring does not alter $s'_j$ for any $j=1, \ldots, M$.
This property makes it easy to calculate $D_2$.
In other words, for each edge $(v, e)$ to be added or removed by the rewiring, it is sufficient to calculate how the number of hyperedges, $m'(k, k')$, where $k$ and $k'$ are the degrees of two nodes belonging to hyperedge $e$, changes (see Eq. \eqref{eq:5}).

It is also difficult to construct bipartite graphs that exactly preserve the degree-dependent redundancy coefficient of the node, $r(k)$, over the values of $k$.
This is because the redundancy coefficients of multiple nodes simultaneously change if one adds or removes an edge in general.
Therefore, for $d_v = 2.5$, we further repeatedly rewire edges of the hypergraph generated by the hyper $dK$-series with $d_v = 2$ as follows. 
(We call this procedure targeting-rewriting for $d_v = 2.5$. See also Figs. \ref{workflow_hyper_dk_series_d_v_2_2_5}(c) and \ref{workflow_hyper_dk_series_d_v_2_2_5}(d).)
We first select a pair of edges $(v_i, e_j)$ and $(v_{i'}, e_{j'})$ such that $i \neq i'$, $j \neq j'$, and $k_i = k_{i'}$ uniformly at random (see Fig. \ref{rewiring}(b)).
Then, we replace $(v_i, e_j)$ and $(v_{i'}, e_{j'})$ by $(v_i, e_{j'})$ and $(v_{i'}, e_j)$ if and only if the distance defined by
\begin{align}
D_{2.5} = \frac{\sum_{k=1}^M |r'(k) - r(k)|}{\sum_{k=1}^M r(k)},
\label{eq:6}
\end{align}
where $r'(k)$ represents the degree-dependent redundancy coefficient of the node for the rewired hypergraph, decreases after the rewiring. 
We repeat the rewiring attempts $R = 500 \mathcal{M}$ times. 

It is easy to calculate $D_{2.5}$ upon a rewiring attempt.
To explain this, we rewrite Eq. \eqref{eq:6} as
\begin{align}
D_{2.5} 
&= \frac{\sum_{k=1}^M \frac{1}{n(k)} | \sum_{i=1,\ k_i=k}^N (r'_i - r_i)|}{\sum_{k=1}^M \frac{1}{n(k)} \sum_{i=1,\ k_i=k}^N r_i}, \label{eq:7}
\end{align}
where $r'_i$ represents the redundancy coefficient of node $v_i$ for the rewired hypergraph. 
To derive Eq. \eqref{eq:7}, we have used the fact that the rewiring exactly preserves the degree of each node.
Equation \eqref{eq:7} implies that it is sufficient to only calculate the change in $r'_i$ for the nodes that are involved in the rewiring (i.e., $v_i$ and $v_{i'}$) and those that share at least one hyperedge with either $v_i$ or $v_{i'}$.

The first subprocess comprising the hyper $dK$-series with $d_v \in \{2, 2.5\}$ is to generate a randomized hypergraph using the hyper $dK$-series with $d_v = 1$ (see Fig. \ref{workflow_hyper_dk_series_d_v_2_2_5}). 
This process preserves the node's degree distribution and destroys the degree correlation and redundancy of the node. 
The second subprocess comprising the hyper $dK$-series with $d_v \in \{ 2, 2.5 \}$ is the targeting-rewiring process. 
This process also preserves the node's degree distribution.
Therefore, the entire procedure of the hyper $dK$-series with $d_v \in \{2, 2.5\}$ preserves the node's degree. 
Furthermore, the targeting-rewriting with $d_v = 2$ and $d_v = 2.5$ makes the degree correlation and redundancy, respectively, approach those of the original hypergraph, which has been lost in the course of the first subprocess. 
Therefore, the entire hyper $dK$-series with $d_v = 2$ and $d_v = 2.5$ approximately preserves the degree correlation and the redundancy, respectively.

The targeting-rewiring process for $d_v=2.5$ also preserves the degree correlation, i.e., $P'(k, k')$ for each $k$ and $k'$, for the following two reasons.
First, owing to the constraint that $k_i = k_{i'}$, the rewiring preserves $m'(k, k')$, i.e., the number of hyperedges that nodes with degree $k$ and nodes with degree $k'$ share, for any $k$ and $k'$.
Second, the rewiring preserves the normalization factor $\sum_{j=1}^M s'_j (s'_j - 1)$ as in the case of $d_v = 2$.

The targeting-rewiring process for $d_v =$ 2 or 2.5 preserves the size of each hyperedge of the randomized hypergraph.
However, with $(d_v, d_e) = (2, 0)$ or $(2.5, 0)$, the hyper $dK$-series does not preserve the size of each hyperedge of the input hypergraph.
This is because we first generate a bipartite graph with $(d_v, d_e) = (1, 0)$, which destroys the size distribution of hyperedges, prior to the targeting-rewiring (see Figs. \ref{workflow_hyper_dk_series_d_v_2_2_5}(a) and \ref{workflow_hyper_dk_series_d_v_2_2_5}(c)).

\subsection{An alternative algorithm for $(d_v, d_e) = (2,1)$: Randomizing rewiring}
For $(d_v, d_e) = (2, 1)$, we have an alternative to the targeting-rewiring process, which is an extension of the so-called randomizing-rewiring process in $dK$-series for dyadic networks \cite{mahadevan, orsini} to the case of bipartite graphs. 
The randomizing rewiring produces bipartite graphs that exactly preserve both nodal degree distribution and $P(k, k')$.
In randomizing rewiring, the initial bipartite graph is a replica of the original bipartite graph $G$.
Then, we select a pair of edges, $(v_i, e_j)$ and $(v_{i'}, e_{j'})$, such that $i \neq i'$, $j \neq j'$, and $k_i = k_{i'}$ uniformly at random, and then replace $(v_i, e_j)$ and $(v_{i'}, e_{j'})$ by $(v_i, e_{j'})$ and $(v_{i'}, e_j)$.
The rewiring preserves the degree of each node, $P(k, k')$, and the size of each hyperedge.
We repeat this rewiring procedure $R'$ times, where $R'$ is sufficiently large, and use the final result as $\tilde{G}$.
We set $R' = 100 \mathcal{M}$ because we have found up to our numerical efforts that the overlap of edges of $G$ and those of the rewired hypergraph converges sufficiently before $R'=100\mathcal{M}$.

The randomizing rewiring has an advantage over the targeting rewiring in that it exactly preserves both the degree of each node and $P(k, k')$ of the input bipartite graph; the targeting rewiring only approximately preserves $P(k, k')$.
However, in contrast to the case of dyadic networks for which the randomizing rewiring is efficient \cite{mahadevan, orsini}, the randomizing rewiring for the hyper $dK$-series has two drawbacks.
First, it is only practical with $(d_v, d_e) = (2, 1)$.
On one hand, although we can easily extend the randomizing rewiring to the case of $d_v\leq 1$ and $d_e\leq 1$, efficient algorithms for generating bipartite graphs exactly preserving the quantities with $d_v\leq 1$ and $d_e\leq 1$ already exist, as we described in Section \ref{section:3.1}.
On the other hand, it is practically difficult to apply the randomizing rewiring in the case of $(d_v, d_e) = (2, 0), (2.5, 0)$, and $(2.5, 1)$ because a proposed random rewiring that respects the constraints imposed by the given $(d_v, d_e)$ rarely preserves $P(k, k')$.
Second, the overlap of the edges in $G$ and those in the rewired hypergraph converges to a nonnegligibly large value with the randomizing rewiring with $(d_v, d_e) = (2, 1)$.
In other words, the randomizing rewiring does not sufficiently randomly shuffle the edges of the original bipartite graph even if one carries out the rewiring many times. 
We show numerical evidence of this phenomenon in Appendix A.
Therefore, we use the targeting rewiring in the following analyses when $(d_v, d_e) = (2, 1)$. 

\section{Results}
\subsection{Data}
In this section, we apply the hyper $dK$-series to four empirical hypergraphs.
The NDC-classes hypergraph, which we refer to as the drug hypergraph in the following text, is a drug network constructed from the National Drug Code Directory \cite{benson}.
Its nodes are class labels, such as serotonin reuptake inhibitor, and a hyperedge is a set of class labels associated with a single drug.
The Enron hypergraph is an email communication network \cite{benson, klimt}, in which a node is an email address, and a hyperedge is a set of all recipient addresses of an email.
The primary-school hypergraph is a social contact network, where nodes are individuals (i.e., students or teachers), and a hyperedge represents an event in which a set of individuals are in face-to-face contact event with each other \cite{benson, stehle}.
The high-school hypergraph is also a social contact network, where nodes are students, and a hyperedge is a face-to-face contact event among a set of students \cite{benson, mastrandrea}.
We preprocessed each data set by first removing multiple hyperedges in the original hypergraph, and then by extracting the largest connected component.
Table \ref{table:2} shows properties of the largest connected component, which we use in the following analysis, for the four data sets.

\begin{table*}[t]
\caption{Properties of the empirical data sets. $N$: number of nodes, $M$: number of hyperedges, $\mathcal{M}$: number of edges in the corresponding bipartite graph, $\bar{k}$: average degree of the node, $\bar{s}$: average size of the hyperedge, $\bar{r}$: average redundancy coefficient of the node, $\bar{l}$: average shortest path length between nodes.}
\label{table:2}
\begin{center}
	\begin{tabular}{| l | c c c c c c c | c |}\hline
	Data & $N$ & $M$ & $\mathcal{M}$ & $\bar{k}$ & $\bar{s}$ & $\bar{r}$ & $\bar{l}$ & References \rule[-2.5pt]{0pt}{12.5pt} \\ \hline
	drug & 628 & 816 & 5,688 & 9.06 & 6.97 & 0.70 & 3.53 & \cite{benson} \\
	Enron & 143 & 1,512 & 4,550 & 31.82 & 3.01 & 0.35 & 2.08 & \cite{benson, klimt} \\ 
	primary-school & 242 & 12,704 & 30,729 & 126.98 & 2.42 & 0.06 & 1.73 & \cite{benson, stehle} \\
	high-school & 327 & 7,818 & 18,192 & 55.63 & 2.33 & 0.07 & 2.16 & \cite{benson, mastrandrea} \\ \hline
  	\end{tabular}
\end{center}
\end{table*}

\subsection{Structural Properties}
For each empirical hypergraph, we compare six structural properties among the given hypergraph and hypergraphs generated by the hyper $dK$-series with $d_v \in \{0, 1, 2, 2.5\}$ and $d_e \in \{0, 1\}$.
We also analyze an existing reference model for bipartite graphs, the B2K \cite{boroojeni}, as a benchmark.
In terms of the terminology of hypergraphs, the B2K model preserves the degree of each node, the size of each hyperedge, and the number of hyperedges with size $s$ to which nodes with degree $k$ belong for each $k$ and $s$.

\begin{figure*}
      \begin{minipage}{0.31\hsize}
        \begin{center}
        \subfigure[]{
          \includegraphics[height=\figsizea]{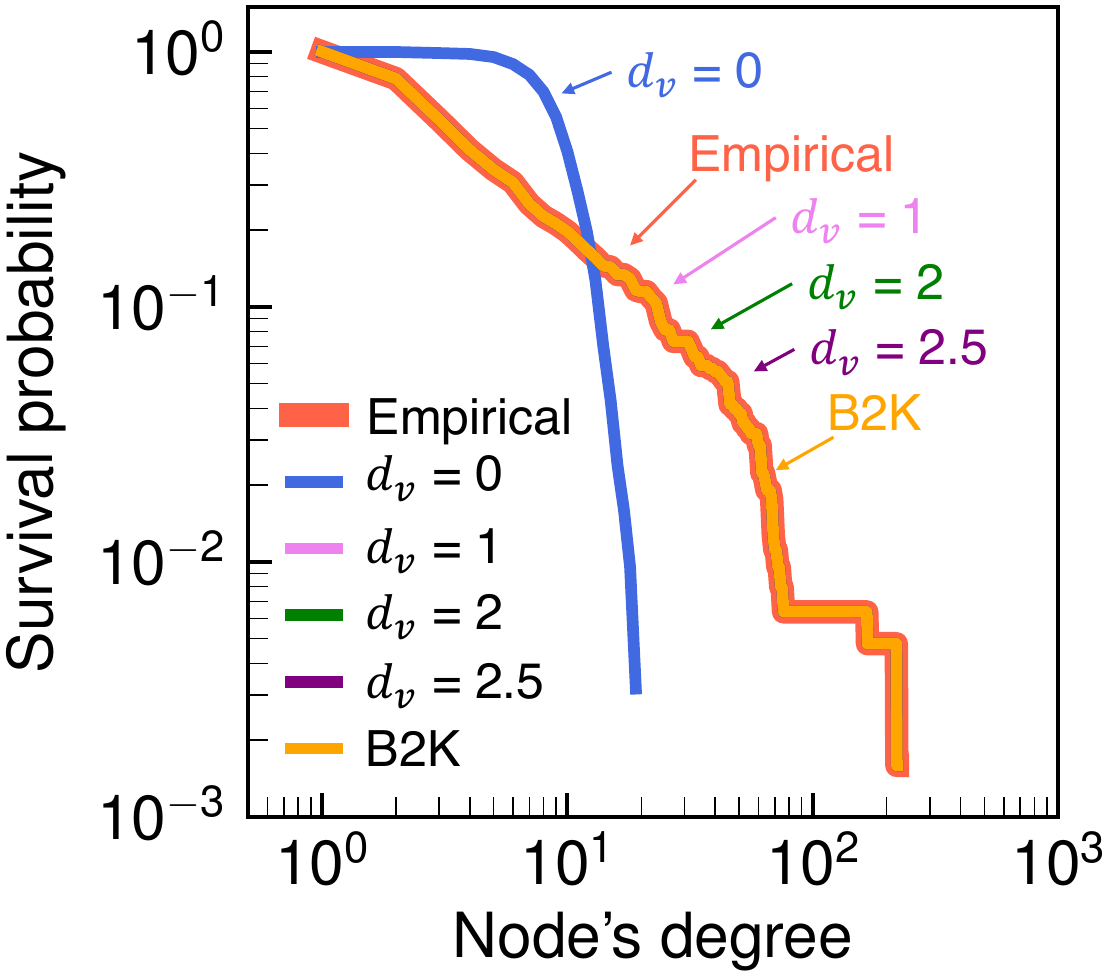}
          }
        \end{center}
      \end{minipage}
      \hspace{1mm}
      \begin{minipage}{0.31\hsize}
        \begin{center}
        \subfigure[]{
        \includegraphics[height=\figsizea]{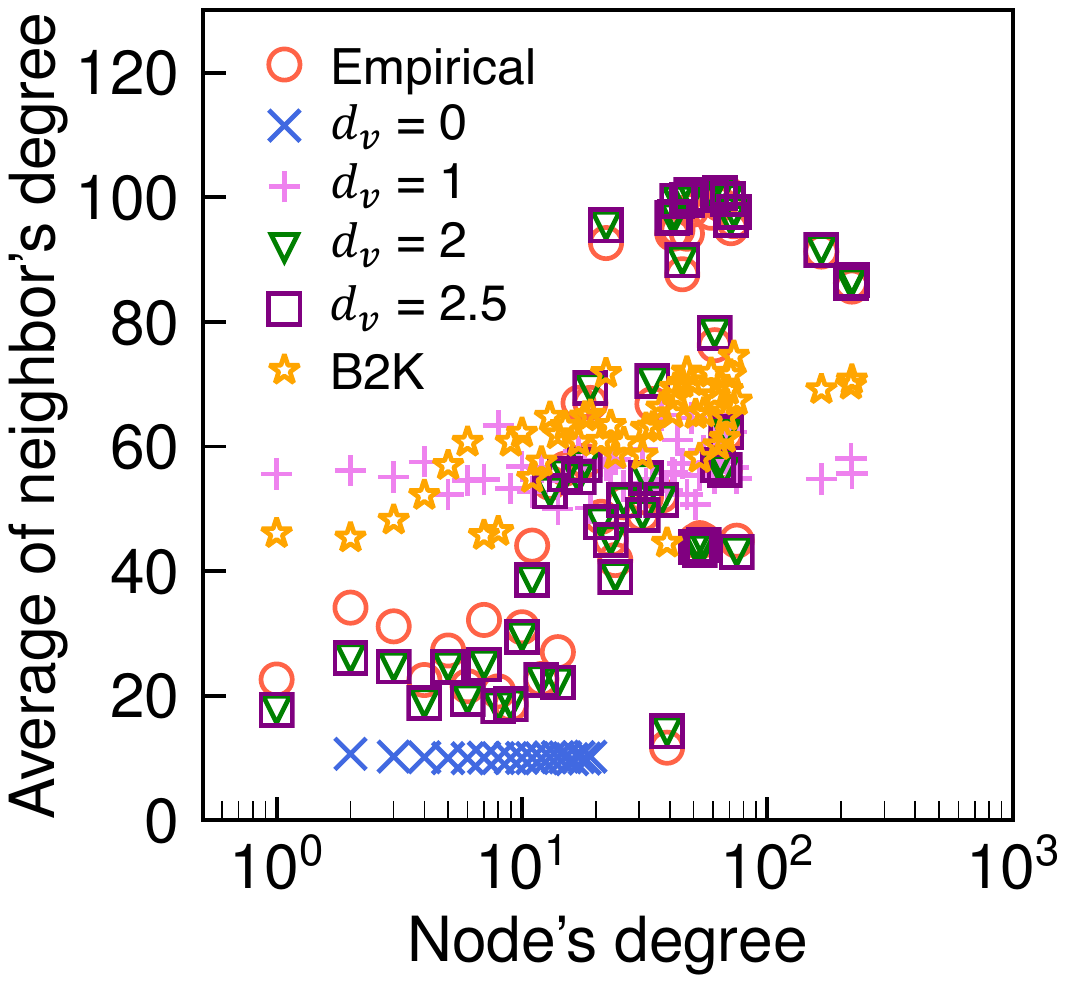}
        }
        \end{center}
      \end{minipage}
      \hspace{1mm}
      \begin{minipage}{0.31\hsize}
        \begin{center}
        \subfigure[]{
          \includegraphics[height=\figsizea]{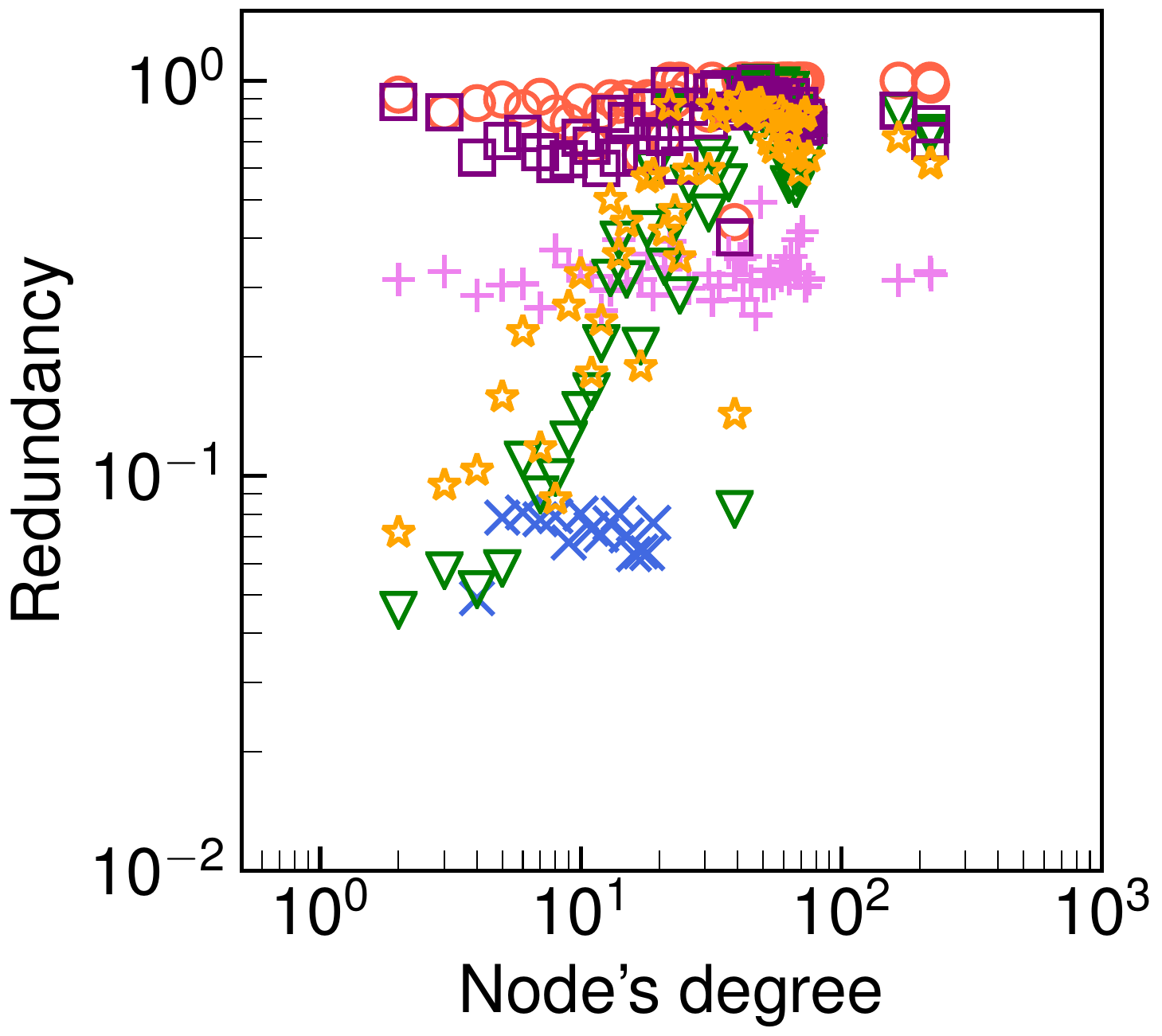}}
        \end{center}
      \end{minipage}
      \vspace{2mm}
      \\
      \begin{minipage}{0.31\hsize}
        \begin{center}
        \subfigure[]{
          \includegraphics[height=\figsizea]{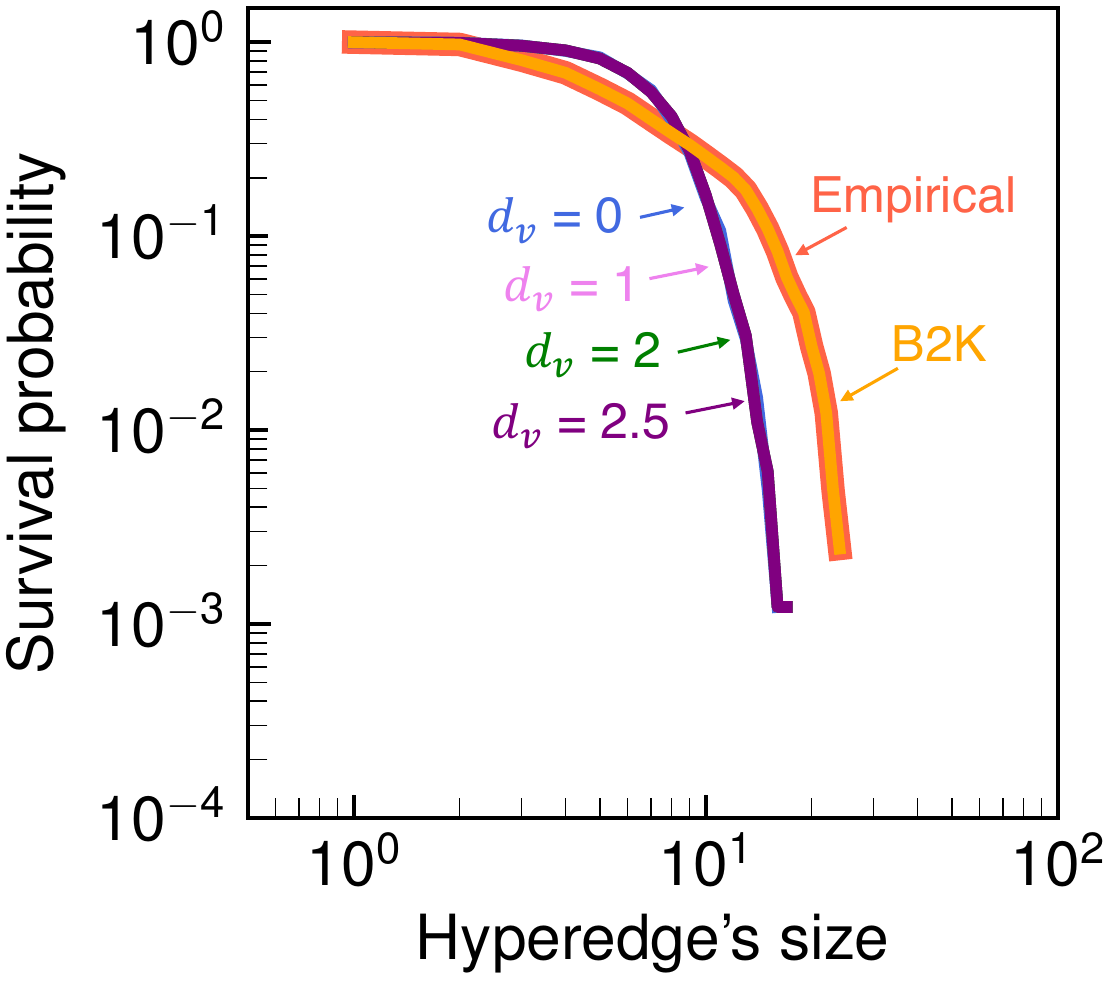}}
        \end{center}
      \end{minipage}
      \hspace{0.5mm}
      \begin{minipage}{0.31\hsize}
        \begin{center}
        \subfigure[]{
          \includegraphics[height=\figsizea]{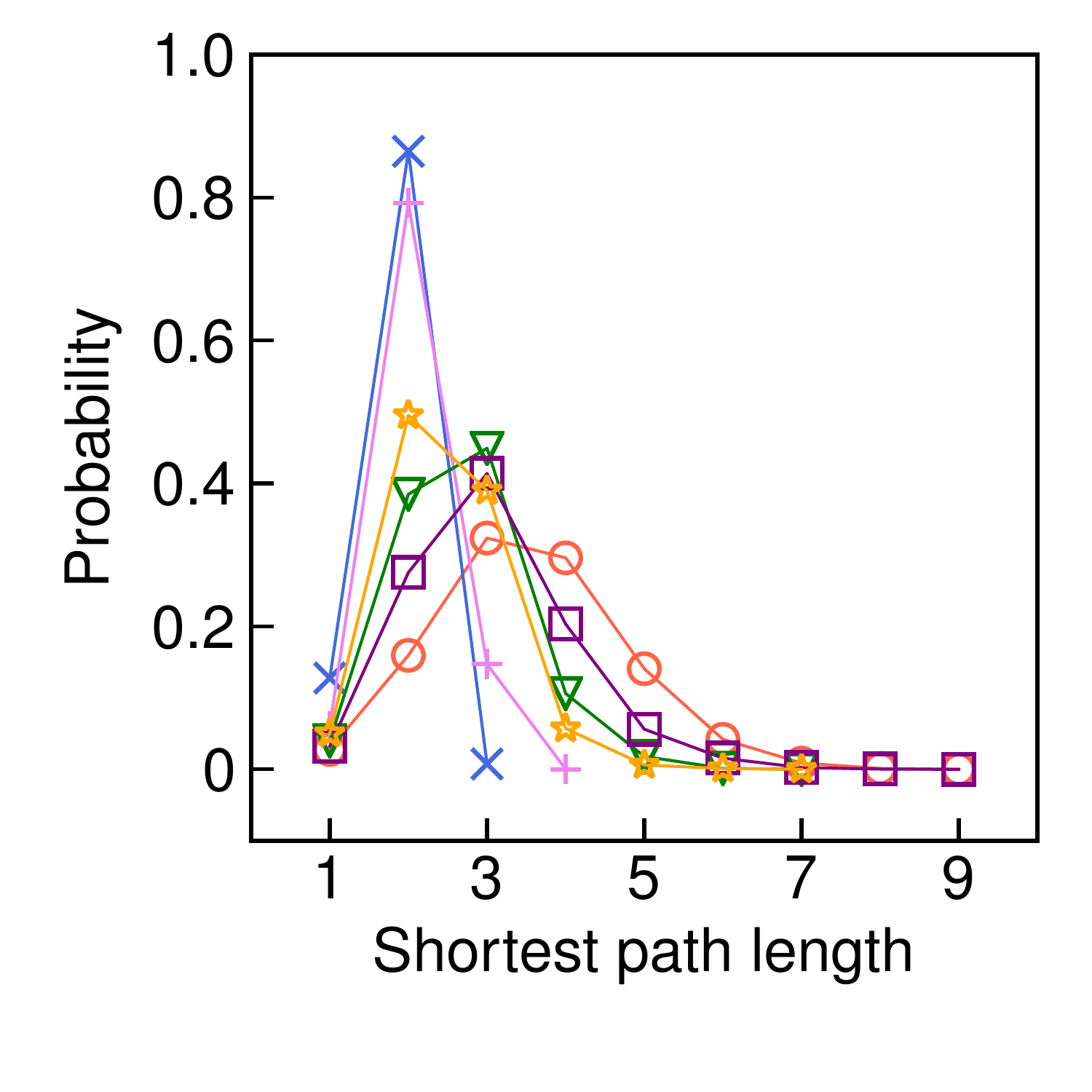}}
        \end{center}
      \end{minipage}
      \hspace{1mm}
      \begin{minipage}{0.31\hsize}
        \begin{center}
        \subfigure[]{
        \includegraphics[height=\figsizea]{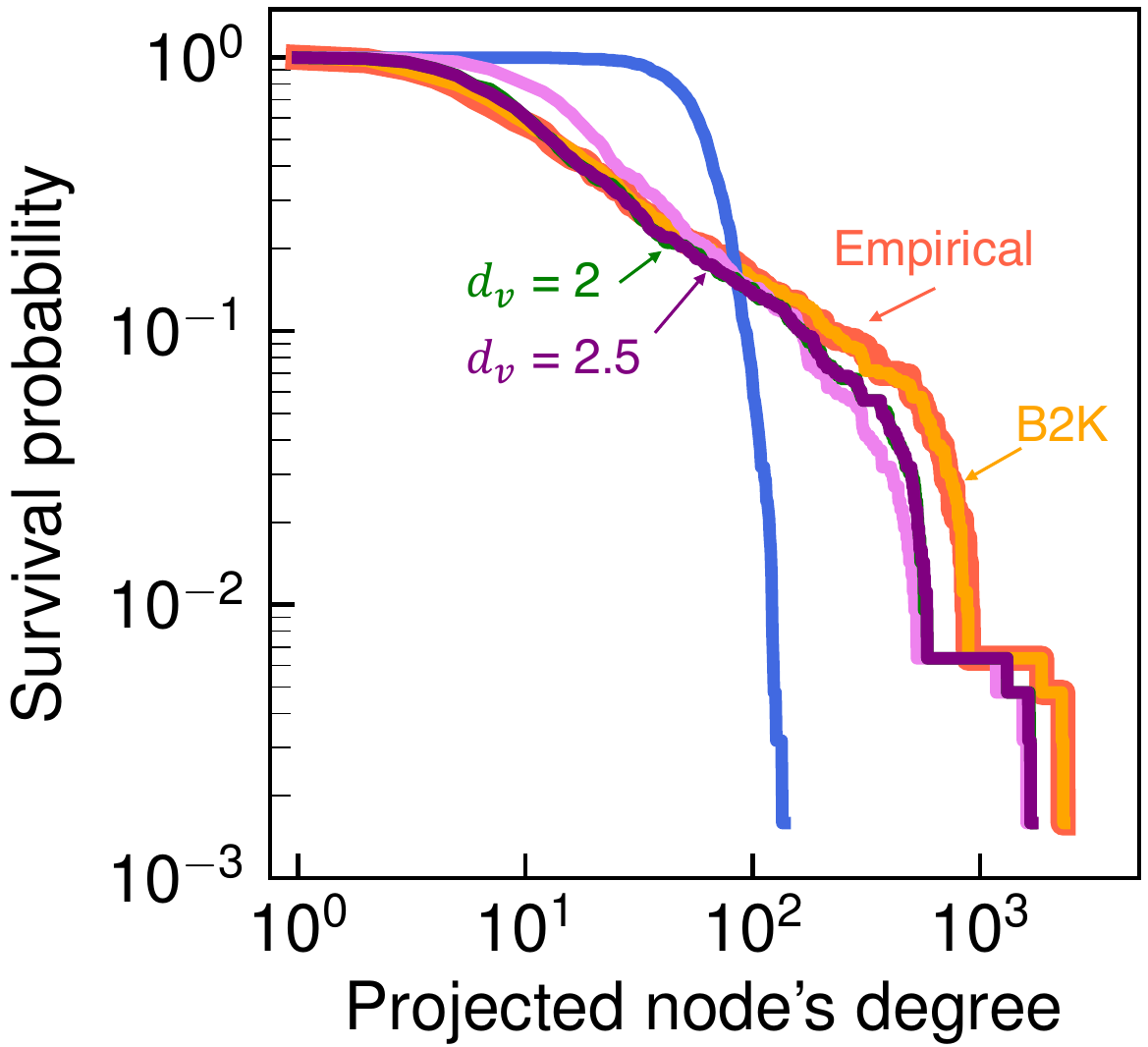}}
        \end{center}
      \end{minipage}
      \vspace{2mm}
      \\
      \begin{minipage}{0.31\hsize}
        \begin{center}
        \subfigure[]{
          \includegraphics[height=\figsizea]{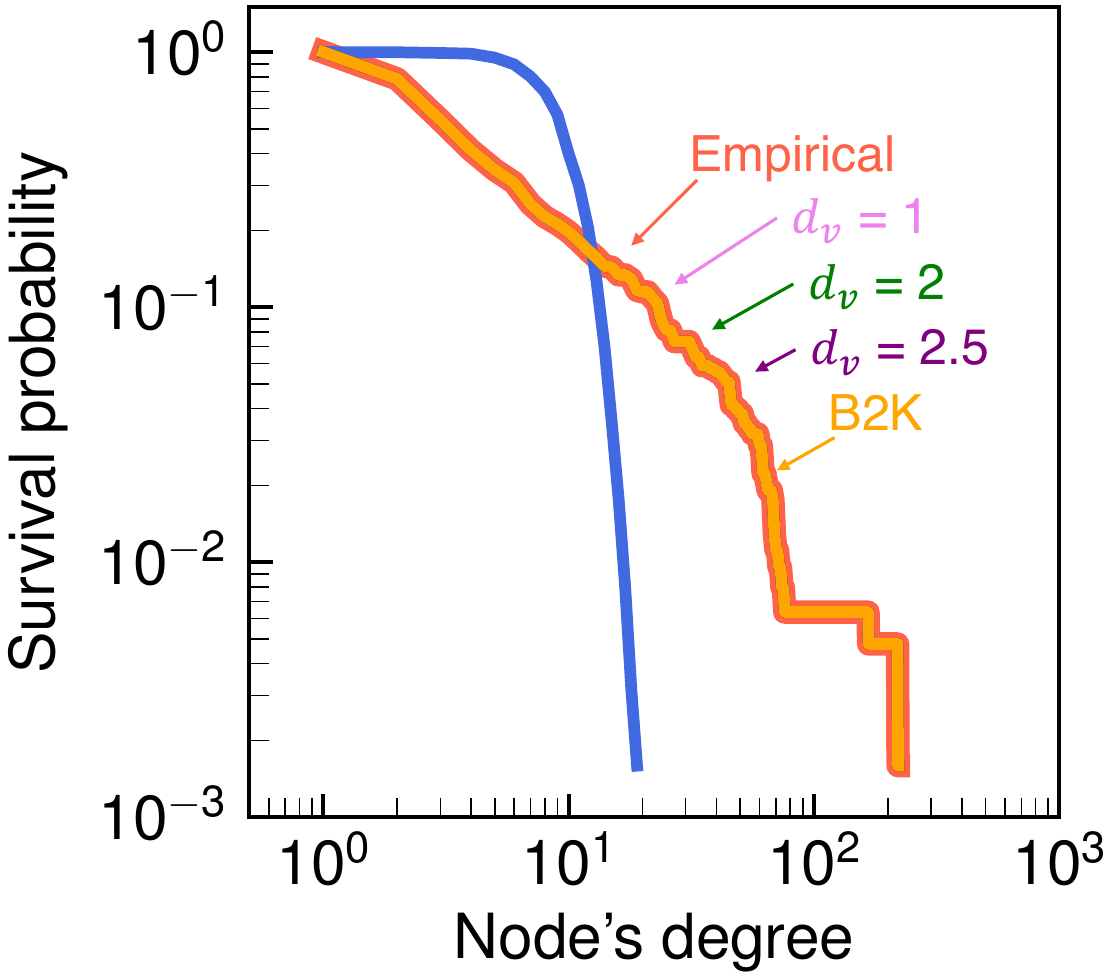}}
        \end{center}
      \end{minipage}
      \hspace{1mm}
      \begin{minipage}{0.31\hsize}
        \begin{center}
        \subfigure[]{
          \includegraphics[height=\figsizea]{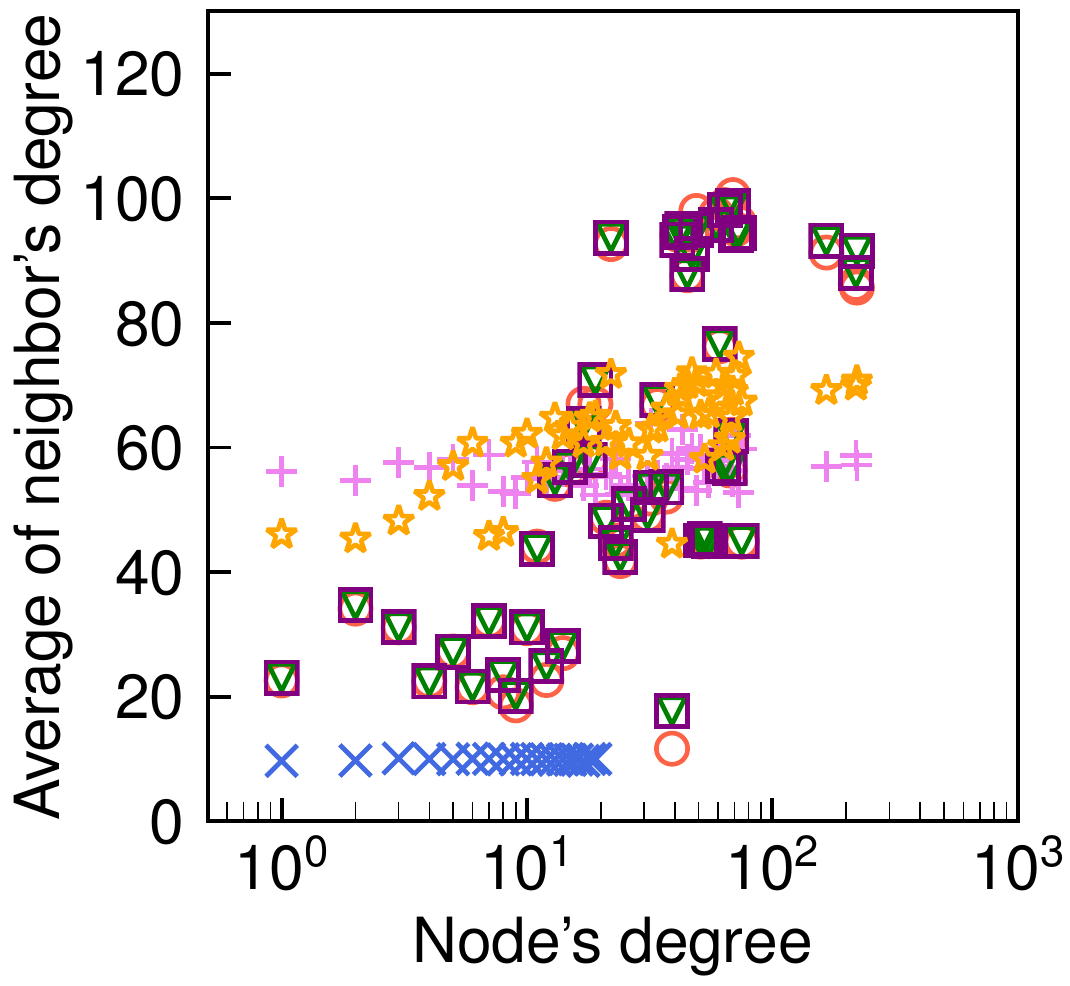}}
        \end{center}
      \end{minipage}
      \hspace{1mm}
      \begin{minipage}{0.31\hsize}
        \begin{center}
        \subfigure[]{
          \includegraphics[height=\figsizea]{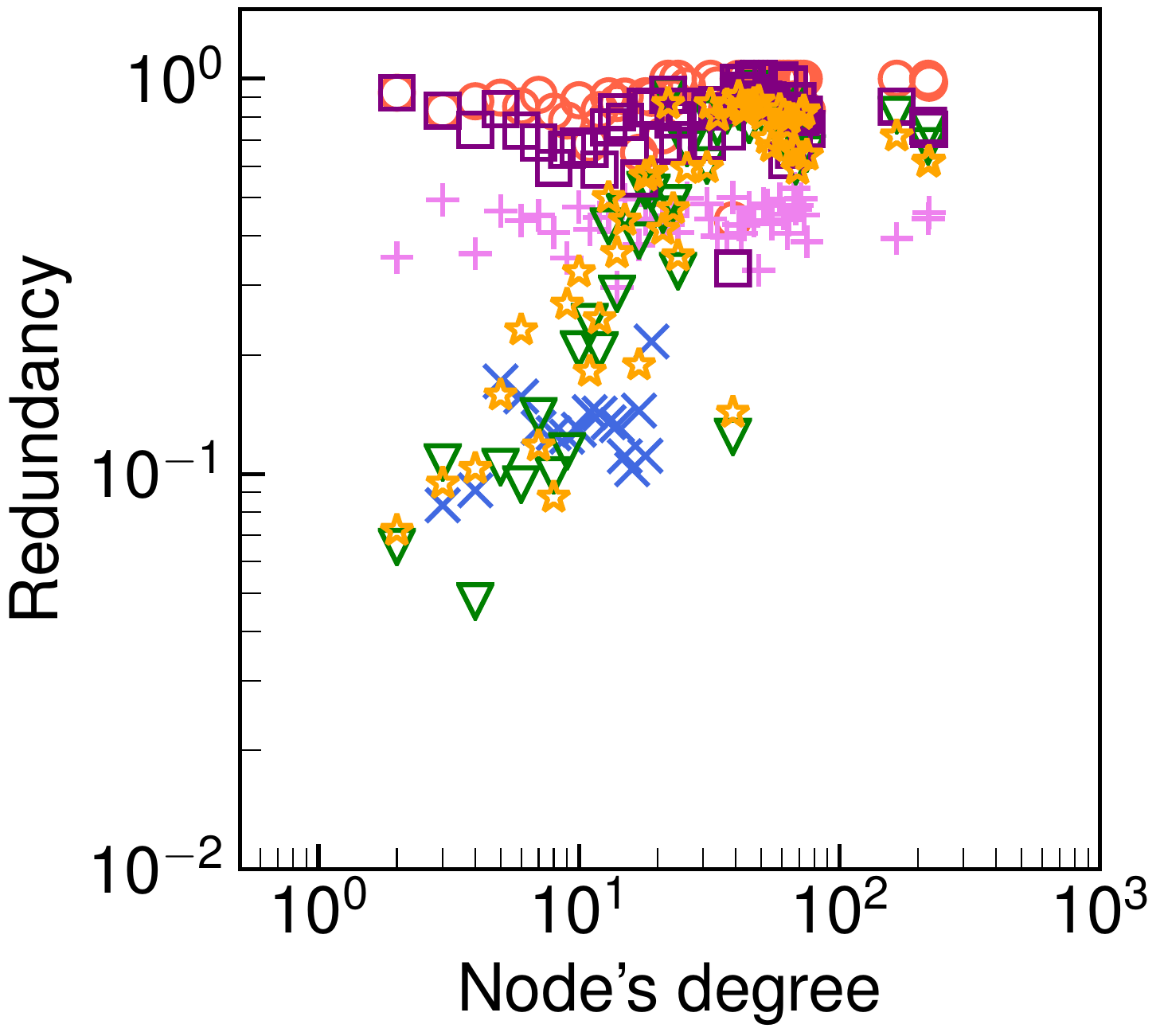}}
        \end{center}
      \end{minipage}
      \vspace{2mm}
      \\
      \begin{minipage}{0.31\hsize}
        \begin{center}
        \subfigure[]{
        \includegraphics[height=\figsizea]{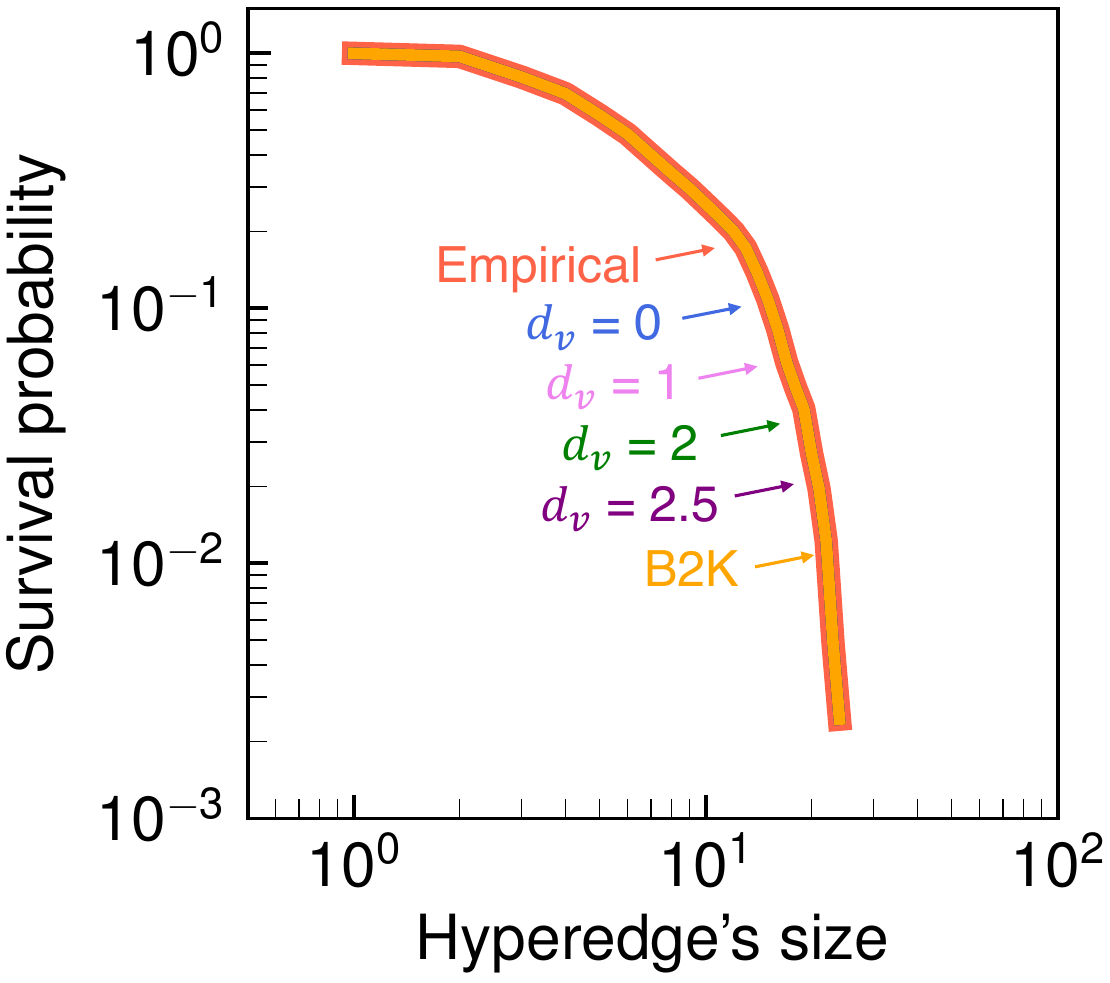}}
        \end{center}
      \end{minipage}
      \hspace{1mm}
      \begin{minipage}{0.31\hsize}
        \begin{center}
        \subfigure[]{
          \includegraphics[height=\figsizea]{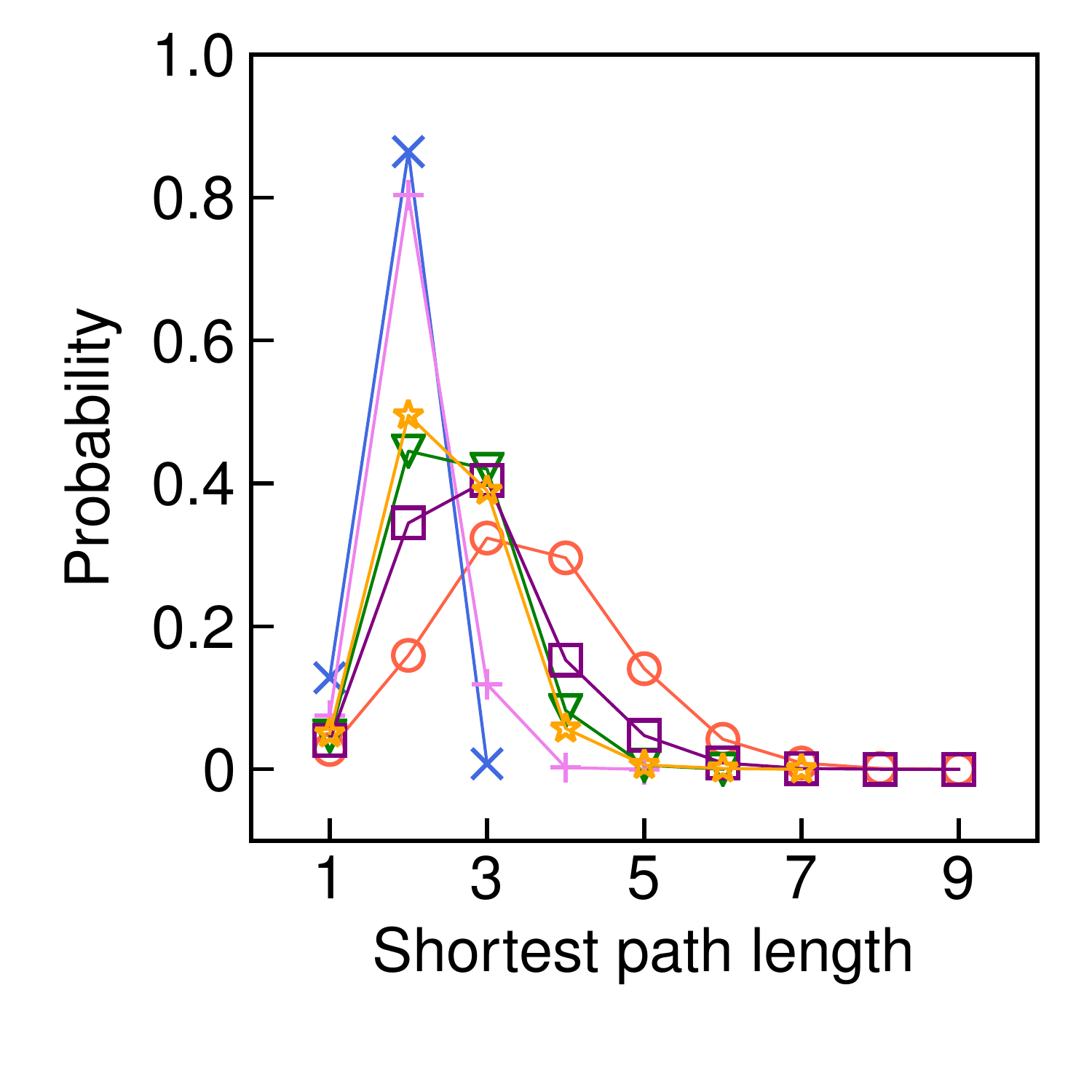}}
        \end{center}
      \end{minipage}
      \hspace{1mm}
      \begin{minipage}{0.31\hsize}
        \begin{center}
        \subfigure[]{
          \includegraphics[height=\figsizea]{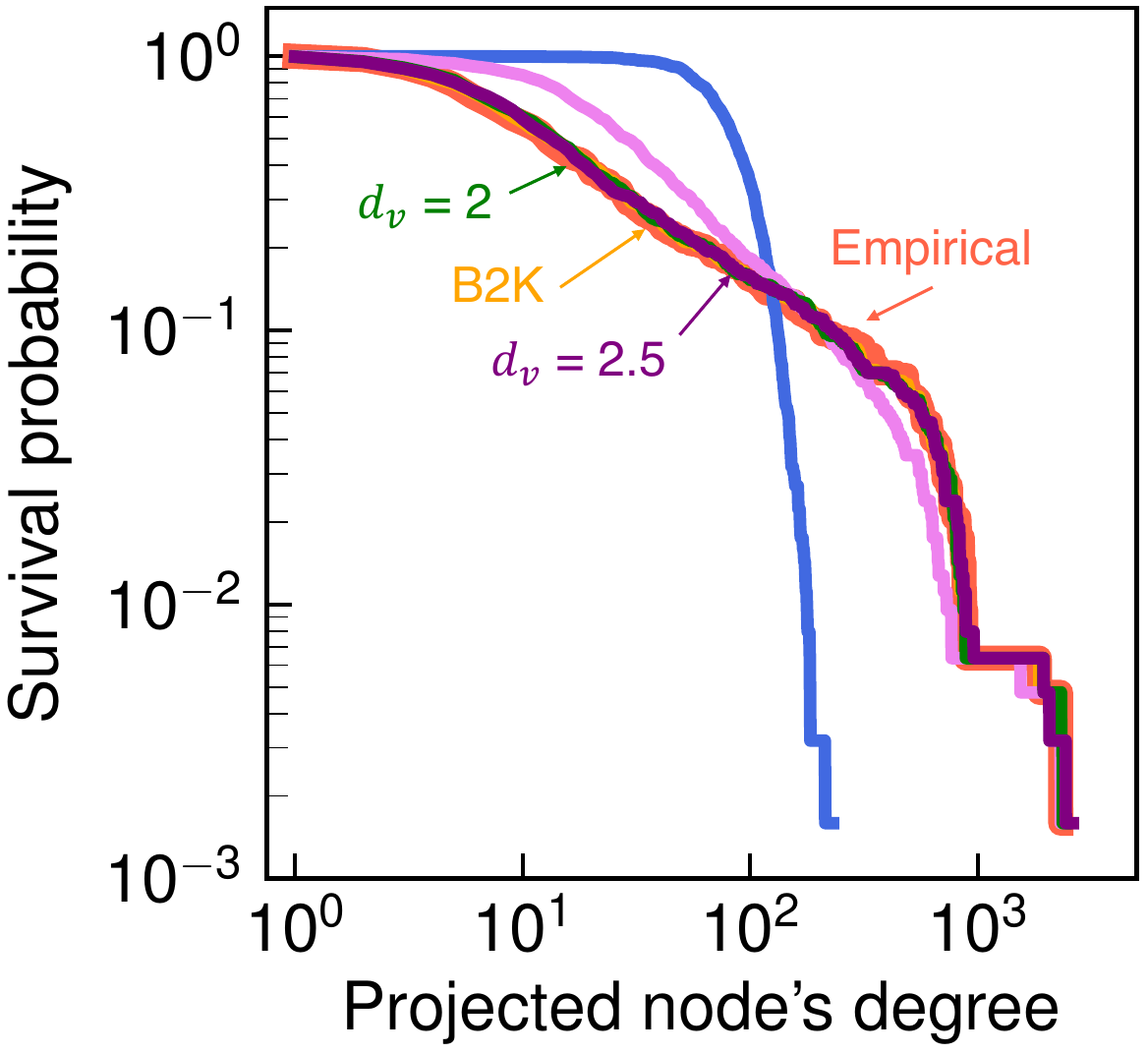}}
        \end{center}
      \end{minipage}
      \caption{Structural properties of the drug hypergraph, the networks generated by the hyper $dK$-series, and the B2K model.
We use the hyper $dK$-series with $d_e = 0$ in (a)--(f) and $d_e = 1$ in (g)--(l). Panels (a) and (g): cumulative degree distribution of the node, (b) and (h): average degree of nearest neighbors of nodes with degree $k$, (c) and (i): degree-dependent redundancy coefficient of the node, (d) and (j): cumulative size distribution of the hyperedge, (e) and (k): distribution of shortest path length between nodes, and (f) and (l): cumulative degree distribution of the one-mode projection. We define the shortest path length between two nodes as the smallest number of hyperedges on the path between the two nodes among all the paths.
The average shortest path length is the average of the shortest average path between a pair of nodes over all pairs of nodes in the largest connected component. 
The largest connected component of randomized hypergraphs contains almost all nodes for all the four empirical hypergraphs (see section S1 in the supplementary material for details). 
We indicate the curves by the arrow and label wherever multiple curves completely or heavily overlap each other. }
      \label{fig:4}
\end{figure*}

Figure \ref{fig:4} compares the six structural properties between the drug hypergraph, the hyper $dK$-series, and the B2K model.
The results for the hyper $dK$-series with $d_e = 0$ and various values of $d_v$ together with those for the original drug hypergraph and the B2K model are shown in Fig. \ref{fig:4}(a)--\ref{fig:4}(f).
We make the following observations. 
First, Fig. \ref{fig:4}(a) indicates that the hyper $dK$-series with $d_v \in \{1, 2, 2.5\}$ but not $d_v=0$ exactly preserves the degree of each node (and therefore the degree distribution) of the drug hypergraph, which is expected.
Second, Fig. \ref{fig:4}(b) indicates that the hyper $dK$-series with $d_v \in \{2, 2.5\}$ but not $d_v \in \{0, 1\}$ approximately preserves the average degree of the nearest neighbors of nodes with degree $k$, denoted by $k_{\text{nn}}(k)$, in the input hypergraph.
Because $k_{\text{nn}}(k)$ is a derivative of the pairwise joint distribution of the node's degree, $P(k, k')$, which the hyper $dK$-series with $d_v \ge 2$ intends to preserve, this result is also expected. 
The hyper $dK$-series with $d_v \in \{0, 1\}$ produces networks without noticeable degree correlation of the node (see Fig. \ref{fig:4}(b)).
Third, as expected, the hyper $dK$-series with $d_v = 2.5$ but not with smaller $d_v$ values approximately preserves the degree-dependent redundancy coefficient of the node, $r(k)$, of the empirical hypergraph (see Fig. \ref{fig:4}(c)).
Fourth, as expected, the hyper $dK$-series with any $d_v \in \{0, 1, 2, 2.5\}$ and $d_e = 0$ does not preserve the distribution of the size of the hyperedge of the original hypergraph; it only preserves the average size of the hyperedge (see Fig. \ref{fig:4}(d)).
Fifth, the hyper $dK$-series with a larger value of $d_v$ better approximates the distribution of the shortest path length between node pairs although the hyper $dK$-series is not designed to preserve this quantity (see Fig. \ref{fig:4}(e)).
Finally, we show in Fig. \ref{fig:4}(f) the cumulative degree distribution of the one-mode projection, where each pair of nodes in the projected network are adjacent if they belong to at least one common hyperedge, and the multiplicity of the edge is equal to the number of hyperedges that the two nodes share \cite{ramasco2006, li2007}.
The hyper $dK$-series progressively better approximates the cumulative degree distribution of the one-mode projection when $d_v$ is larger, whereas the results are similar between $d_v = 2$ and $d_v = 2.5$. 
Note that the hyper $dK$-series is not designed to preserve the degree distribution of the one-mode projection.

We show in Fig. \ref{fig:4}(g)--\ref{fig:4}(l)  the results for the hyper $dK$-series with $d_e = 1$ and various values of $d_v$ together with those for the B2K model.
The results for the empirical hypergraph and the B2K model shown in these figures are the same as those shown in Fig. \ref{fig:4}(a)--\ref{fig:4}(f).
We make the following observations.
First, as expected, the results shown in Fig. \ref{fig:4}(g)--\ref{fig:4}(i) are similar to those shown in Fig. \ref{fig:4}(a)--\ref{fig:4}(c). 
In other words, the hyper $dK$-series with $d_v \geq 1$ preserves the degree distribution of the node, that with $d_v \geq 2$ additionally preserves $k_{\text{nn}}(k)$, and that with $d_v = 2.5$ additionally preserves $r(k)$. 
Second, Fig. \ref{fig:4}(j) indicates that the hyper $dK$-series preserves the distribution of the size of hyperedge, which is because we set $d_e=1$.
Third, similar to the case of $d_e = 0$, the hyper $dK$-series with a larger $d_v$ value better approximates the distribution of the shortest path length between nodes (see Fig. \ref{fig:4}(k)).
A comparison between Figs. \ref{fig:4}(e) and \ref{fig:4}(k) suggests that the approximation accuracy is not notably different between $d_e=0$ and $d_e = 1$.
Finally, a comparison between Figs. \ref{fig:4}(f) and \ref{fig:4}(l) suggests that the hyper $dK$-series with $d_v \geq 2$ and $d_e = 1$ more accurately approximates the cumulative degree distribution of the one-mode projection than the hyper $dK$-series with the same $d_v$ value and $d_e = 0$ and than that with $d_v \le 1$ and $d_e=1$.
This is presumably because the node's degree in the one-mode projection depends not only on the degree of the node in the original hypergraph but also on the size of each hyperedge to which the node belongs.

The B2K model exactly preserves the distributions of node's degree and hyperedge's size, as expected (see Figs. \ref{fig:4}(a) and \ref{fig:4}(d)).
However, it little preserves the node's degree correlation and the redundancy coefficient of the empirical network (see Figs. \ref{fig:4}(b) and \ref{fig:4}(c)).
Therefore, roughly speaking, the complexity of the B2K model is somewhere between that of the hyper $dK$-series with $(d_v,d_e)=(1,1)$ and that with $(d_v,d_e)=(2,1)$.
We also find that the B2K model accurately preserves the degree distribution of the one-mode projection (see Fig. \ref{fig:4}(f)) although the B2K model does not intend to preserve it.

\begin{table*}
\caption{Distance between the empirical hypergraphs and those generated by the reference models (i.e., hyper $dK$-series and B2K model). In the table, $P(k)$ represents the cumulative degree distribution of the node; $k_{\text{nn}}(k)$ represents the average degree of the nearest neighbors of nodes with degree $k$; $r(k)$ represents the degree-dependent redundancy coefficient of the node; $P(s)$ represents the cumulative size distribution of the hyperedge; $P(l)$ represents the distribution of the shortest path length between nodes; $P(\breve{k})$ represents the cumulative degree distribution of the one-mode projection.}
\label{table:3}
\begin{center}
	\begin{tabular}{D || E | F F F F F F}\hline
	Data & \centering Model & $P(k)$ & $k_{\text{nn}}(k)$ & $r(k)$ & $P(s)$ & $P(l)$ & $P(\breve{k})$ \rule[0mm]{0mm}{3.5mm} \\ \hline
	\multirow{9}{*}{drug} & $(d_v, d_e) = (0, 0)$ & 0.605 & 0.948 & 0.977 & 0.250 & 1.582 & 0.669 \\ 
	& $(1, 0)$ & 0.000 & 0.396 & 0.625 & 0.252 & 1.335 & 0.234 \\ 
	& $(2, 0)$ & 0.000 & 0.041 & 0.427 & 0.252 & 0.765 & 0.093 \\ 
	& $(2.5, 0)$ & 0.000 & 0.041 & 0.139 & 0.252 & 0.440 & 0.088 \\ 
	& $(0, 1)$ & 0.598 & 0.945 & 0.957 & 0.000 & 1.610 & 0.701 \\ 
	& $(1, 1)$ & 0.000 & 0.397 & 0.502 & 0.000 & 1.409 & 0.311 \\ 
	& $(2, 1)$ & 0.000 & 0.022 & 0.393 & 0.000 & 0.783 & 0.049 \\ 
	& $(2.5, 1)$ & 0.000 & 0.022 & 0.137 & 0.000 & 0.582 & 0.043 \\ 
	& \centering B2K & 0.000 & 0.326 & 0.394 & 0.000 & 0.850 & 0.027 \\ \hline
	\multirow{9}{*}{Enron} & $(d_v, d_e) = (0, 0)$ & 0.427 & 0.821 & 0.955 & 0.163 & 0.623 & 0.385 \\ 
	& $(1, 0)$ & 0.000 & 0.195 & 0.767 & 0.163 & 0.487 & 0.075 \\ 
	& $(2, 0)$ & 0.000 & 0.012 & 0.400 & 0.163 & 0.432 & 0.090 \\ 
	& $(2.5, 0)$ & 0.000 & 0.012 & 0.058 & 0.163 & 0.331 & 0.083 \\ 
	& $(0, 1)$ & 0.426 & 0.808 & 0.948 & 0.000 & 0.671 & 0.393 \\ 
	& $(1, 1)$ & 0.000 & 0.195 & 0.747 & 0.000 & 0.483 & 0.080 \\ 
	& $(2, 1)$ & 0.000 & 0.030 & 0.498 & 0.000 & 0.434 & 0.057 \\ 
	& $(2.5, 1)$ & 0.000 & 0.030 & 0.175 & 0.000 & 0.352 & 0.047 \\ 
	& \centering B2K & 0.000 & 0.192 & 0.729 & 0.000 & 0.474 & 0.052 \\\hline
	\multirow{9}{*}{primary-school} & $(d_v, d_e) = (0, 0)$ & 0.374 & 0.832 & 0.924 & 0.304 & 0.860 & 0.704 \\ 
	& $(1, 0)$ & 0.000 & 0.088 & 0.547 & 0.305 & 0.705 & 0.372 \\ 
	& $(2, 0)$ & 0.000 & 0.007 & 0.370 & 0.305 & 0.371 & 0.358 \\ 
	& $(2.5, 0)$ & 0.000 & 0.007 & 0.206 & 0.305 & 0.346 & 0.357 \\ 
	& $(0, 1)$ & 0.377 & 0.834 & 0.970 & 0.000 & 0.537 & 0.390 \\ 
	& $(1, 1)$ & 0.000 & 0.089 & 0.807 & 0.000 & 0.434 & 0.041 \\ 
	& $(2, 1)$ & 0.000 & 0.014 & 0.563 & 0.000 & 0.244 & 0.031 \\ 
	& $(2.5, 1)$ & 0.000 & 0.014 & 0.112 & 0.000 & 0.035 & 0.032 \\ 
	& \centering B2K & 0.000 & 0.088 & 0.811 & 0.000 & 0.421 & 0.019 \\ \hline
	\multirow{9}{*}{high-school} & $(d_v, d_e) = (0, 0)$ & 0.308 & 0.698 & 0.908 & 0.326 & 0.534 & 0.622 \\ 
	& $(1, 0)$ & 0.000 & 0.111 & 0.724 & 0.326 & 0.528 & 0.364 \\ 
	& $(2, 0)$ & 0.000 & 0.009 & 0.434 & 0.326 & 0.511 & 0.321 \\ 
	& $(2.5, 0)$ & 0.000 & 0.009 & 0.030 & 0.326 & 0.505 & 0.322 \\ 
	& $(0, 1)$ & 0.312 & 0.692 & 0.963 & 0.000 & 0.534 & 0.345 \\ 
	& $(1, 1)$ & 0.000 & 0.112 & 0.894 & 0.000 & 0.515 & 0.073 \\ 
	& $(2, 1)$ & 0.000 & 0.025 & 0.792 & 0.000 & 0.497 & 0.050 \\ 
	& $(2.5, 1)$ & 0.000 & 0.025 & 0.092 & 0.000 & 0.440 & 0.051 \\ 
	& \centering B2K & 0.000 & 0.102 & 0.884 & 0.000 & 0.499 & 0.019 \\ \hline
	\end{tabular}
\end{center}
\end{table*}

To be quantitative, we measure the distance in the distribution of each of the six quantities between the empirical hypergraph and each type of synthetic hypergraph for each data set.
For the degree distribution of the node, the size distribution of the hyperedge, and the degree distribution of one-mode projection, we calculate the Kolmogorov-Smirnov distance between the cumulative distribution for the original bipartite graph and that for the generated bipartite graph.
The Kolmogorov-Smirnov distance between two cumulative distributions, denoted by $F_1(x)$ and $F_2(x)$, is given by $\sup_{x} |F_1(x) - F_2(x)|$.
For $k_{\text{nn}}(k)$, $r(k)$, and the distribution of the shortest path length between nodes (which we denote by $P(\ell)$ for the shortest path length $\ell$), we calculate the normalized $L^1$ distance between the vector corresponding to the original bipartite graph, denoted by $\bm{x} = (x_1, x_2, \ldots, x_L)$, and that corresponding to the synthetic bipartite graph, denoted by $\tilde{\bm{x}} = (\tilde{x}_1, \tilde{x}_2, \ldots, \tilde{x}_L)$.
Specifically, we set $x_k = k_{\text{nn}}(k)$ with $k=1, \ldots, M$, $x_k = r(k)$ with $k=1, \ldots, M$, or $x_k = P(\ell)$ with $k=1, \ldots, N-1$, and similar for $\tilde{\bm{x}}$.
The distance between $\bm{x}$ and $\tilde{\bm{x}}$ is defined by $\sum_{i=1}^L |\tilde{x}_i - x_i| / \sum_{i=1}^L |x_i|$.
We calculate the distance average of each property over the independent 100 runs for each model.
In each model, we generate an independent bipartite graph for each run.

We show the distance measurement results in Table \ref{table:3}.
The following observations apply to all the data sets unless we state otherwise.
First, we verify that the degree distribution of the node is the same between the empirical data and the hyper $dK$-series with $d_v \geq 1$ and the B2K model.
Second, the hyper $dK$-series with $d_v = 2$ realize a considerably small distance to the empirical data in terms of $k_{\text{nn}}(k)$.
Third, the hyper $dK$-series with $d_v = 2.5$ yields a small distance to the empirical data in terms of $r(k)$.
Fourth, the distribution of hyperedge's size is the same between the empirical data, any hyper $dK$-series with $d_e = 1$, and the B2K model.
Fifth, for both $d_e=0$ and $d_e=1$, the hyper $dK$-series is more similar to the empirical data in terms of the distribution of shortest path length between nodes (i.e., $P(\ell)$) when $d_v$ is larger.
However, with the exception of primary-school hypergraph, the relative error between the hyper $dK$-series and the empirical hypergraph in terms of $P(\ell)$ is large (i.e., $> 30\%$) even with $(d_v, d_e) = (2.5,1)$.
Finally, the hyper $dK$-series with $(d_v,d_e)=(2,1)$, $(2.5,1)$ and the B2K model are close to the empirical data in terms of the degree distribution of the one-mode projection. 
All these results are consistent with those shown in Fig. \ref{fig:4}.
We also statistically tested how significantly the hyper $dK$-series changes each structural property of a given hypergraph (see section S2 in the supplementary material).

\begin{figure}
\begin{minipage}{0.48\hsize}
        \begin{center}
        \subfigure[]{
          \includegraphics[height=\figsizeb]{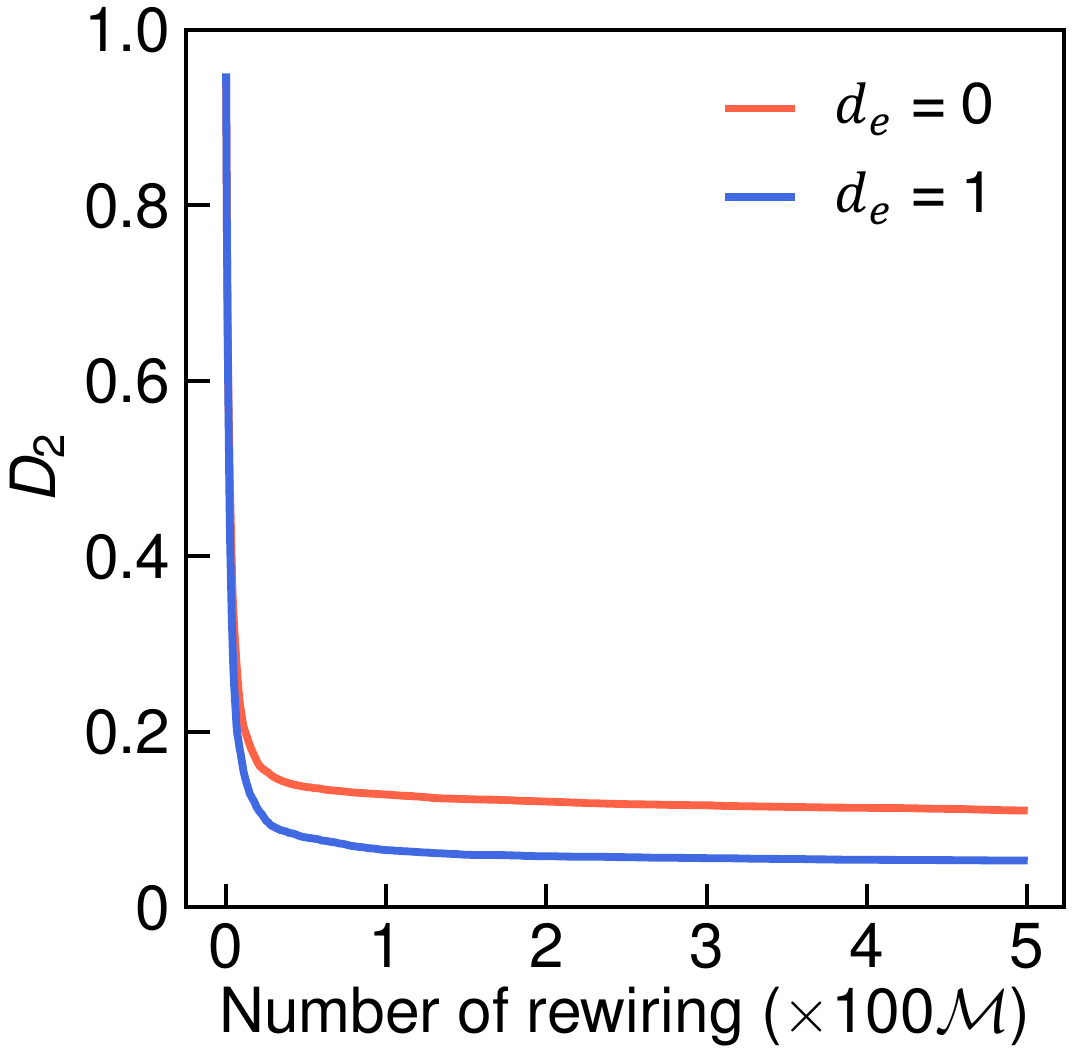}}
        \end{center}
\end{minipage}
        \hspace{1mm}
\begin{minipage}{0.48\hsize}
        \begin{center}
        \subfigure[]{
          \includegraphics[height=\figsizeb]{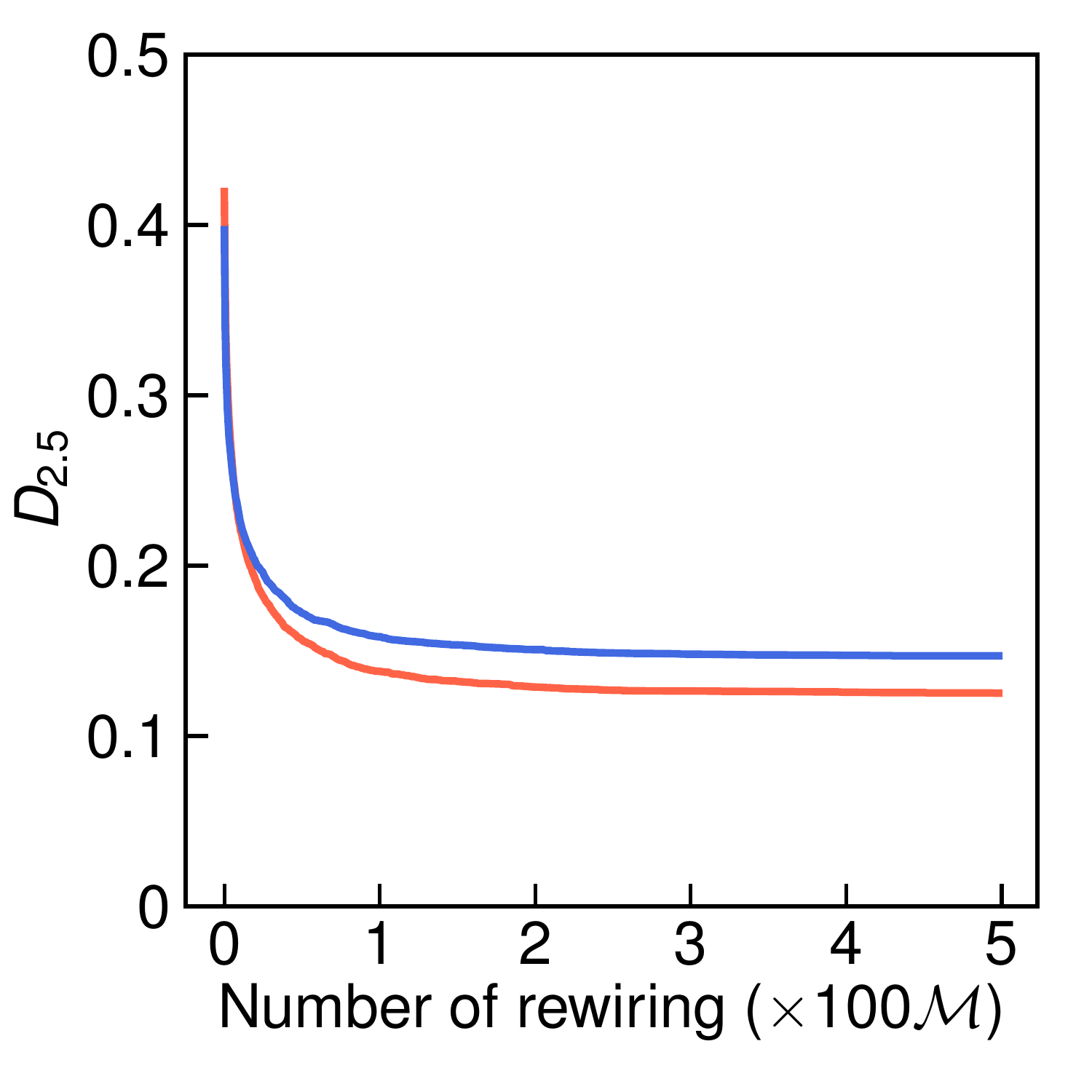}}
        \end{center}
\end{minipage}
      \caption{Distance between the original and synthetic hypergraphs in the targeting-rewiring process for the drug data set. (a) $d_v = 2$. (b) $d_v = 2.5$.}
      \label{fig:5}
\end{figure}

\begin{figure*}[t]
      \begin{minipage}{0.24\hsize}
        \begin{center}
        \subfigure[]{
          \includegraphics[height=\figsizec]{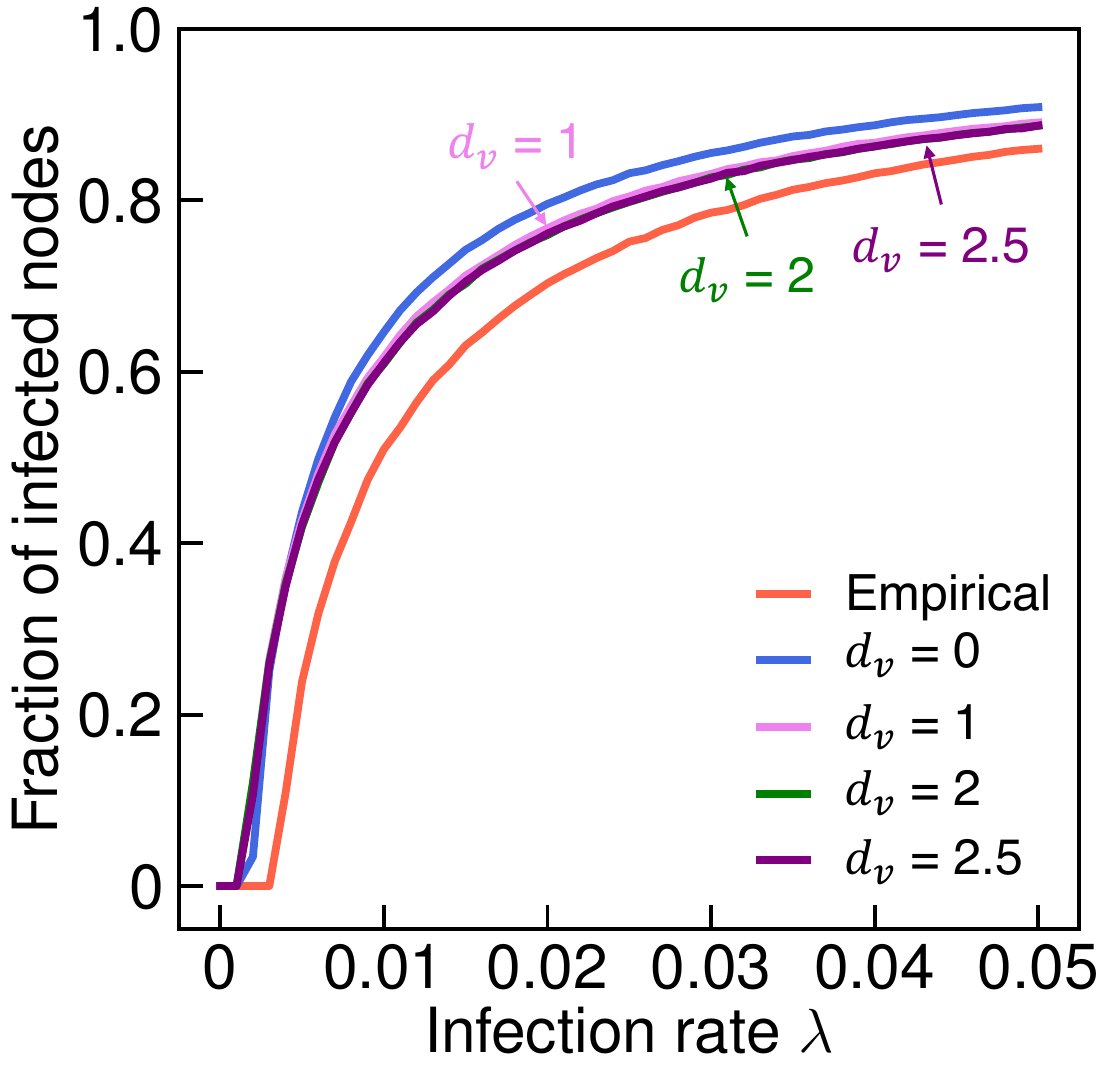}}
        \end{center}
      \end{minipage}
      \begin{minipage}{0.24\hsize}
        \begin{center}
        \subfigure[]{
        \includegraphics[height=\figsizec]{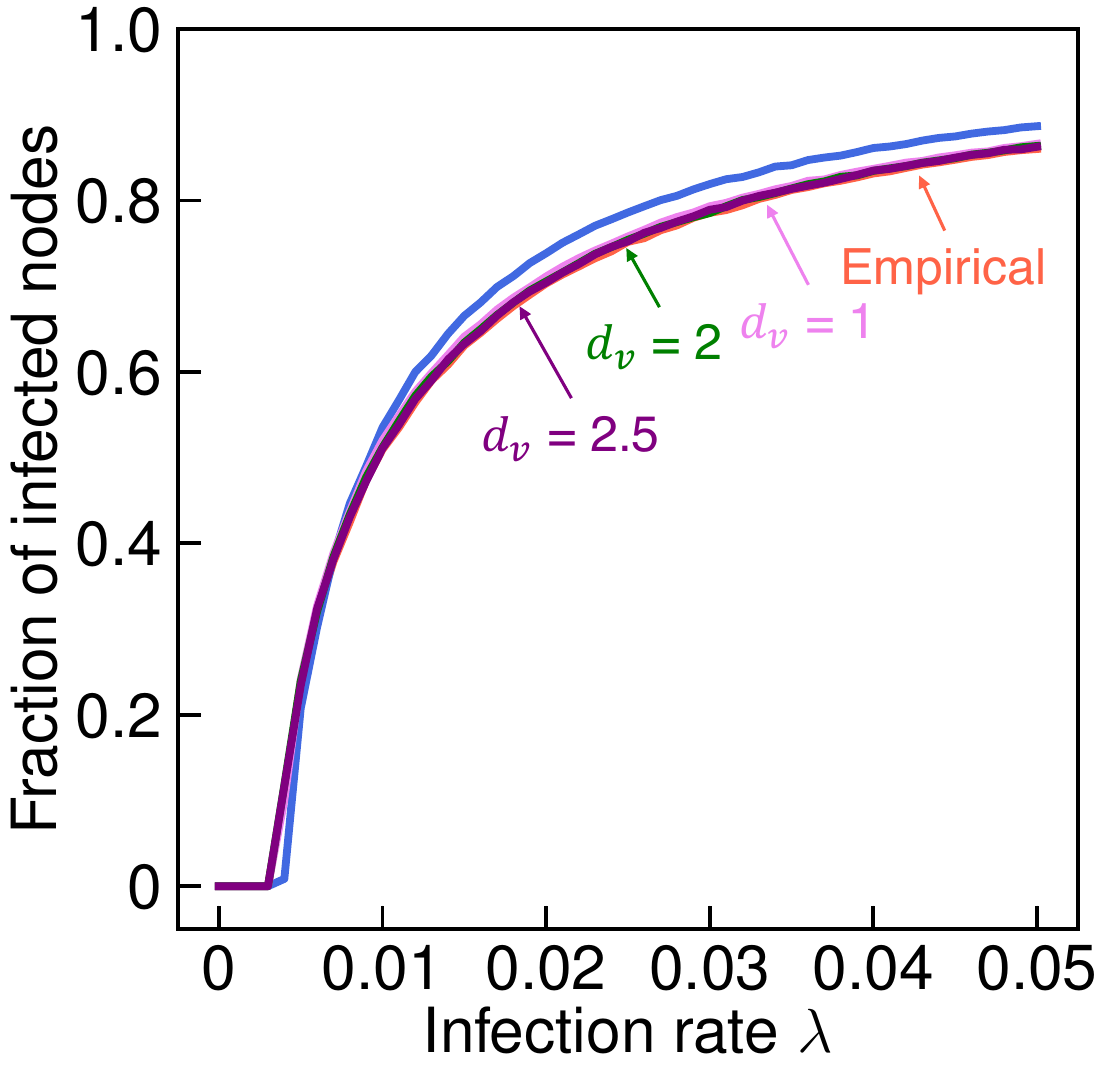}}
        \end{center}
      \end{minipage}
      \begin{minipage}{0.24\hsize}
        \begin{center}
        \subfigure[]{
          \includegraphics[height=\figsizec]{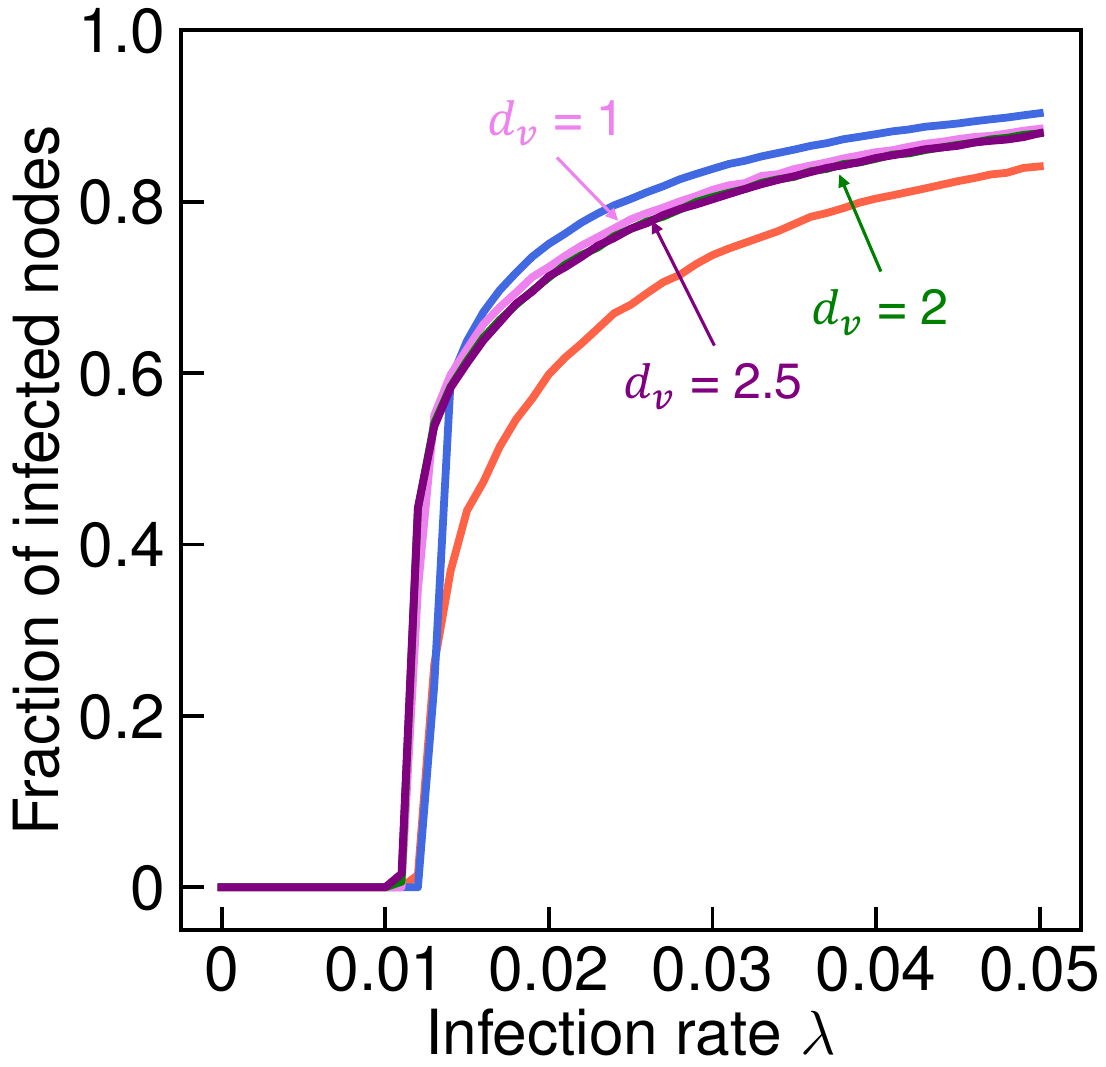}}
        \end{center}
      \end{minipage}
      \begin{minipage}{0.24\hsize}
        \begin{center}
        \subfigure[]{
          \includegraphics[height=\figsizec]{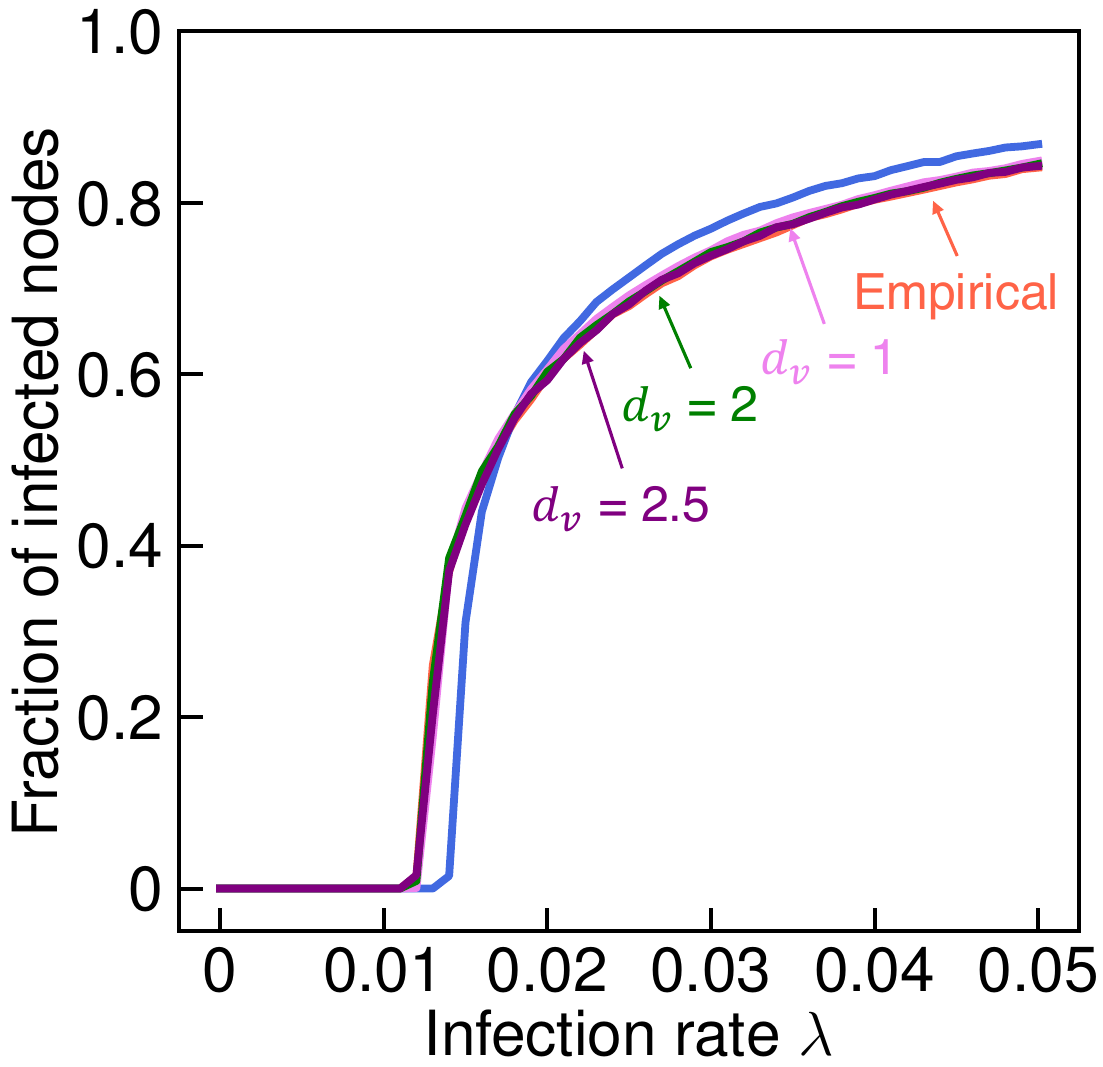}}
        \end{center}
      \end{minipage}
      
      \begin{minipage}{0.24\hsize}
        \begin{center}
        \subfigure[]{
          \includegraphics[height=\figsizec]{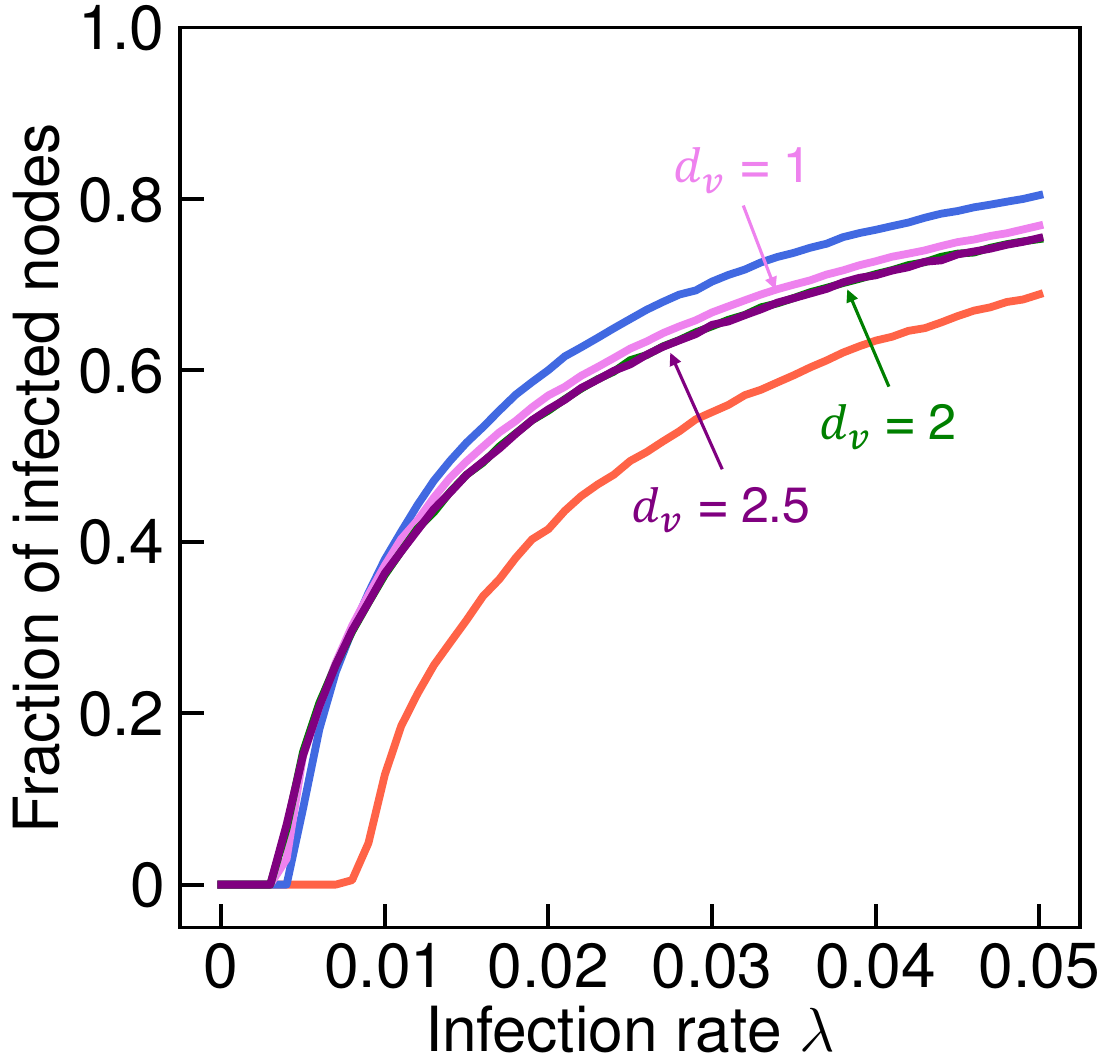}}
        \end{center}
      \end{minipage}
      \begin{minipage}{0.24\hsize}
        \begin{center}
        \subfigure[]{
        \includegraphics[height=\figsizec]{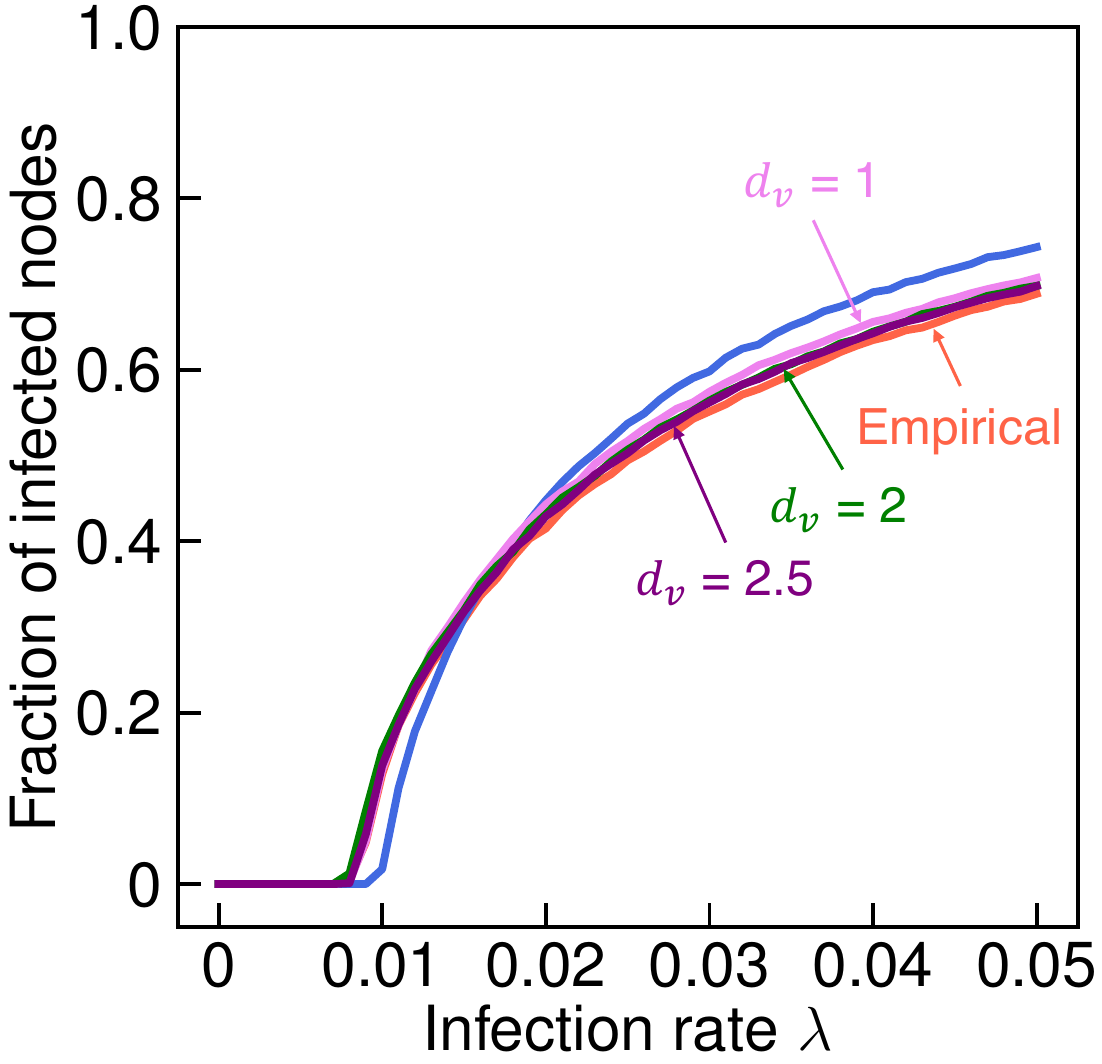}}
        \end{center}
      \end{minipage}
      \begin{minipage}{0.24\hsize}
        \begin{center}
        \subfigure[]{
          \includegraphics[height=\figsizec]{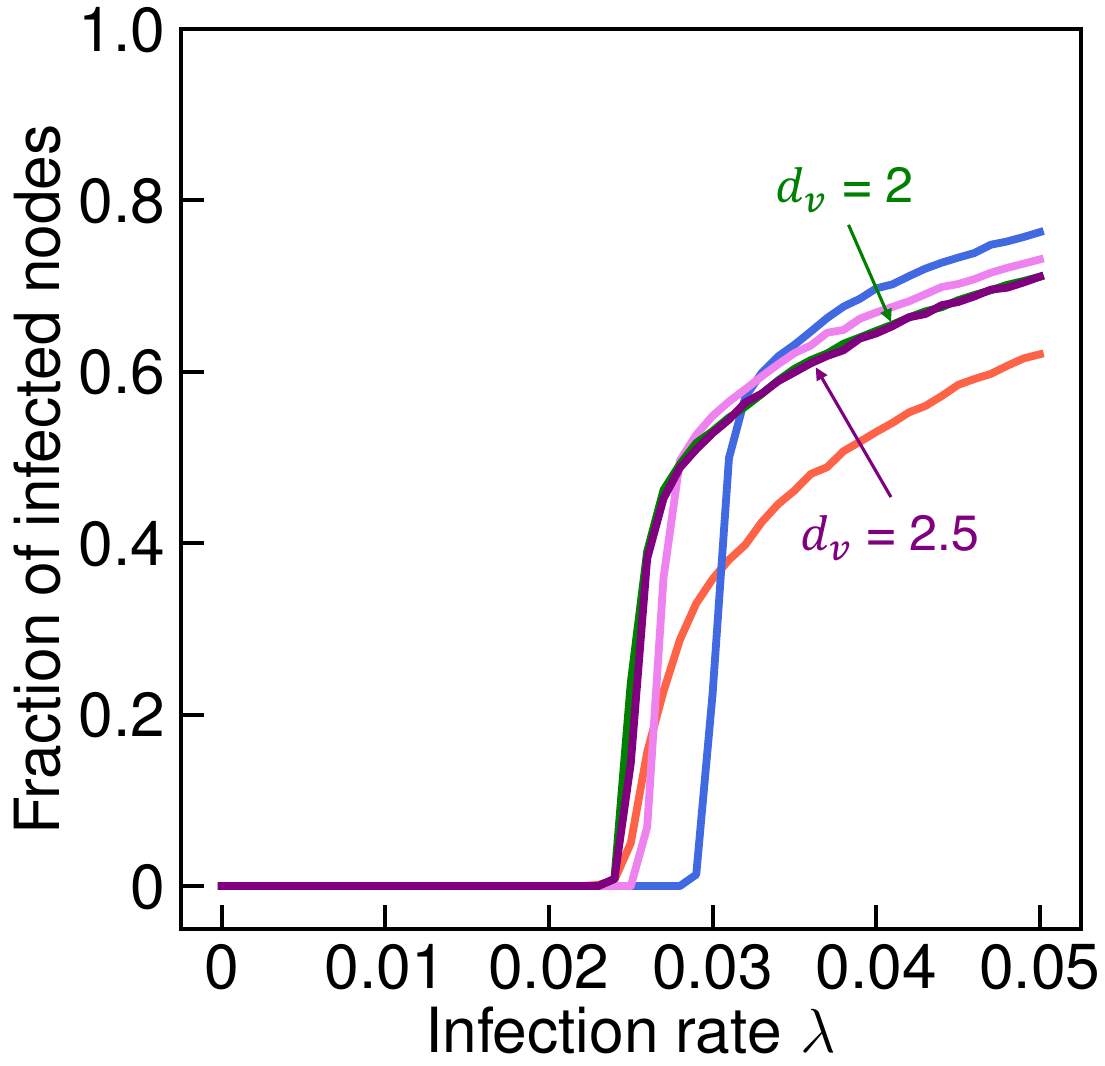}}
        \end{center}
      \end{minipage}
      \begin{minipage}{0.24\hsize}
        \begin{center}
        \subfigure[]{
          \includegraphics[height=\figsizec]{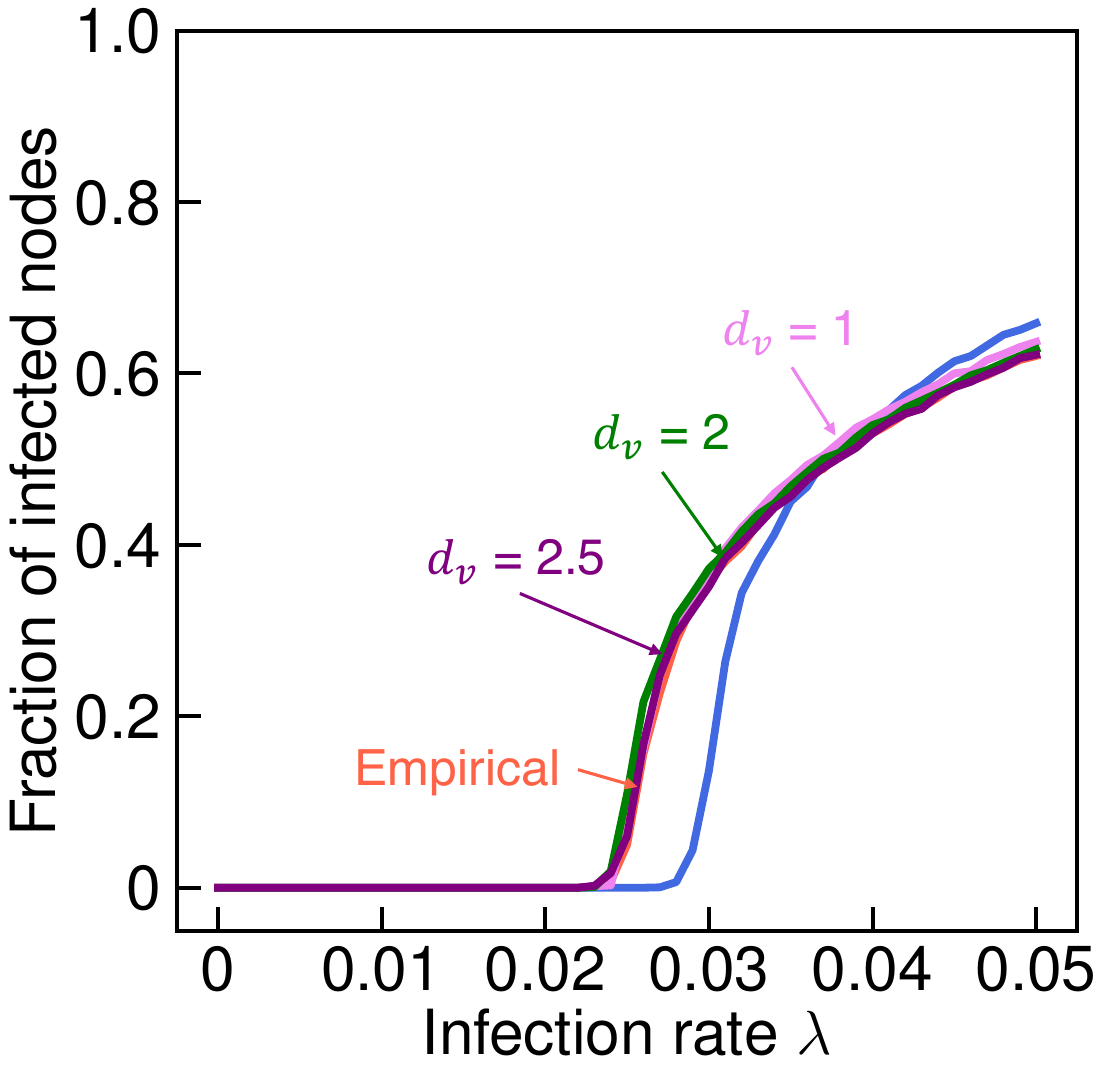}}
        \end{center}
      \end{minipage}
      \caption{Fraction of infected nodes in the SIS model on hypergraphs. The results for primary-school data set are shown in (a)--(d), and those for the high-school data set are shown in (e)--(h). We set $(d_e,\theta) = (0, 0.1)$ in (a) and (e); $(d_e,\theta) = (1, 0.1)$ in (b) and (f); $(d_e,\theta) = (0, 0.5)$ in (c) and (g); $(d_e,\theta) = (1, 0.5)$ in (d) and (h). We indicate the curves by the arrow and label wherever multiple curves heavily overlap each other. }
      \label{fig:6}
\end{figure*}

To examine if the targeting rewiring introduces sufficient randomization, we measure the distance measures $D_2$ and $D_{2.5}$, which are defined in Eqs. \eqref{eq:5} and \eqref{eq:6}, as a function of the number of rewiring attempts, $R$, for the hyper $dK$-series with $d_v \in \{2, 2.5\}$ and $d_e \in \{0,1\}$.
The results for the drug data set are shown in Fig. \ref{fig:5}.  
For both $d_e=0$ and $d_e=1$, $D_2$ rapidly decreased to values that are $\approx$ 15\% larger than the final value in the first $\approx$ $100 \mathcal{M}$ targeting rewiring attempts.
Then, $D_2$ continued to decrease slowly towards the final value.
Similarly, $D_{2.5}$ in the case of both $d_e = 0$ and $d_e = 1$ rapidly decreased to values that are $\approx$ 10\% larger than the final values in the first $100 \mathcal{M}$ targeting rewiring attempts and then slowly decayed towards the final values.
We confirmed that the trajectories of $D_2$ and $D_{2.5}$ were similar for the other three data sets.

\subsection{Epidemic Spreading}
A primary application of the hyper $dK$-series is to simulations of dynamical or other processes on hypergraphs. 
Specifically, comparisons between the results on the original and synthetic hypergraphs generally help us to understand particular structural properties of the hypergraph that impact the processes on hypergraphs. 
For example, comparisons between a dynamical process on networks generated by the hyper $dK$-series with $d_v=0$ and with $d_v=1$ will reveal the effect of the node's degree distribution. 
This is because the hyper $dK$-series with $d_v=0$ destroys the degree distribution of the original hypergraph, whereas that with $d_v=1$ preserves it. 
Likewise, comparisons between $d_v=1$ and $d_v=2$ will reveal the effects of degree correlation; comparisons between $d_v=2$ and $d_v=2.5$ will reveal the effects of redundancy; comparisons between $d_e = 0$ and 1 will reveal the effects of the hyperedge's size distribution.
We showcase the application of the hyper $dK$-series with epidemic spreading and evolutionary game dynamics models. 

In this section, we examine a susceptible-infected-susceptible (SIS) model on hypergraphs in continuous time \cite{de}.
Each node is in either the susceptible state or the infectious state at any time $t$.
Each infectious node recovers and becomes susceptible according to a Poisson process with rate $\delta$.
A fundamental assumption underlying the present model, which distinguishes it from other SIS models on hypergraphs \cite{bodo, suo2018, jhun2019}, is that the contagion process is critical-mass dynamics, which generalizes a previous model \cite{iacopini}.
Let $\rho_j$ denote the fraction of infectious nodes in hyperedge $e_j \in E$.
For each hyperedge $e_j$, each susceptible node in $e_j$ becomes infected at rate $\lambda_j$ if and only if $\rho_j \ge \theta$, where $\theta$ is a parameter. 
We set $\delta = 1$ and $\lambda_j = \lambda \log_{2} |e_j|$, where $\lambda$ is a parameter \cite{de}. 

We assume that all the nodes are initially infectious and run the SIS model on the primary-school and high-school hypergraphs until $t = 100$.
We confirmed that the fraction of infected nodes converges to an approximate stationary value before $t = 100$.
For the given $\theta$ and $\lambda$ values, we average the fraction of infected nodes over $95 \le t \le 100$ and over 100 runs. 
In the case of the hyper $dK$-series, we generate an independent bipartite graph for each run.

\begin{figure*}[t]
      \begin{minipage}{0.24\hsize}
        \begin{center}
        \subfigure[]{
          \includegraphics[height=\figsizec]{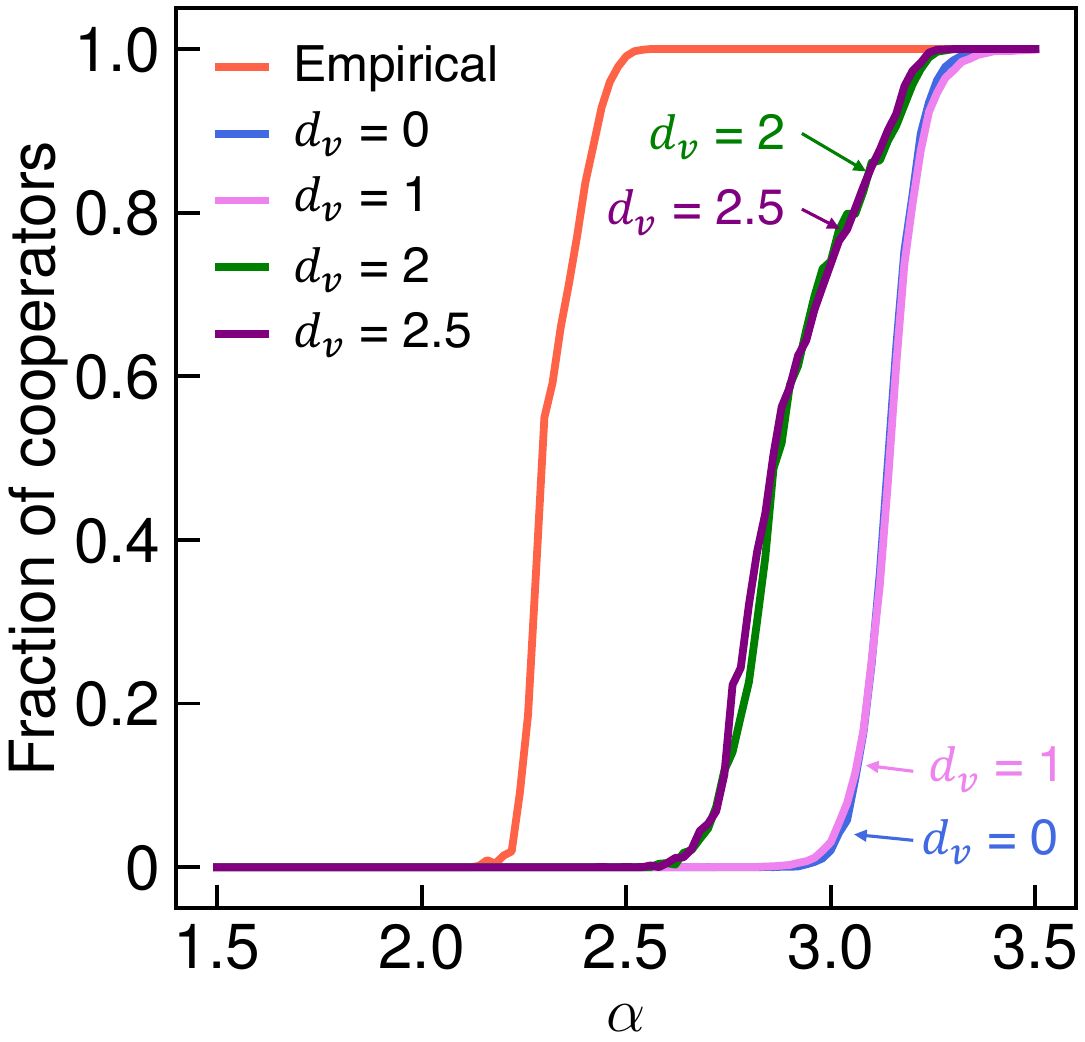}}
        \end{center}
      \end{minipage}
      \begin{minipage}{0.24\hsize}
        \begin{center}
        \subfigure[]{
        \includegraphics[height=\figsizec]{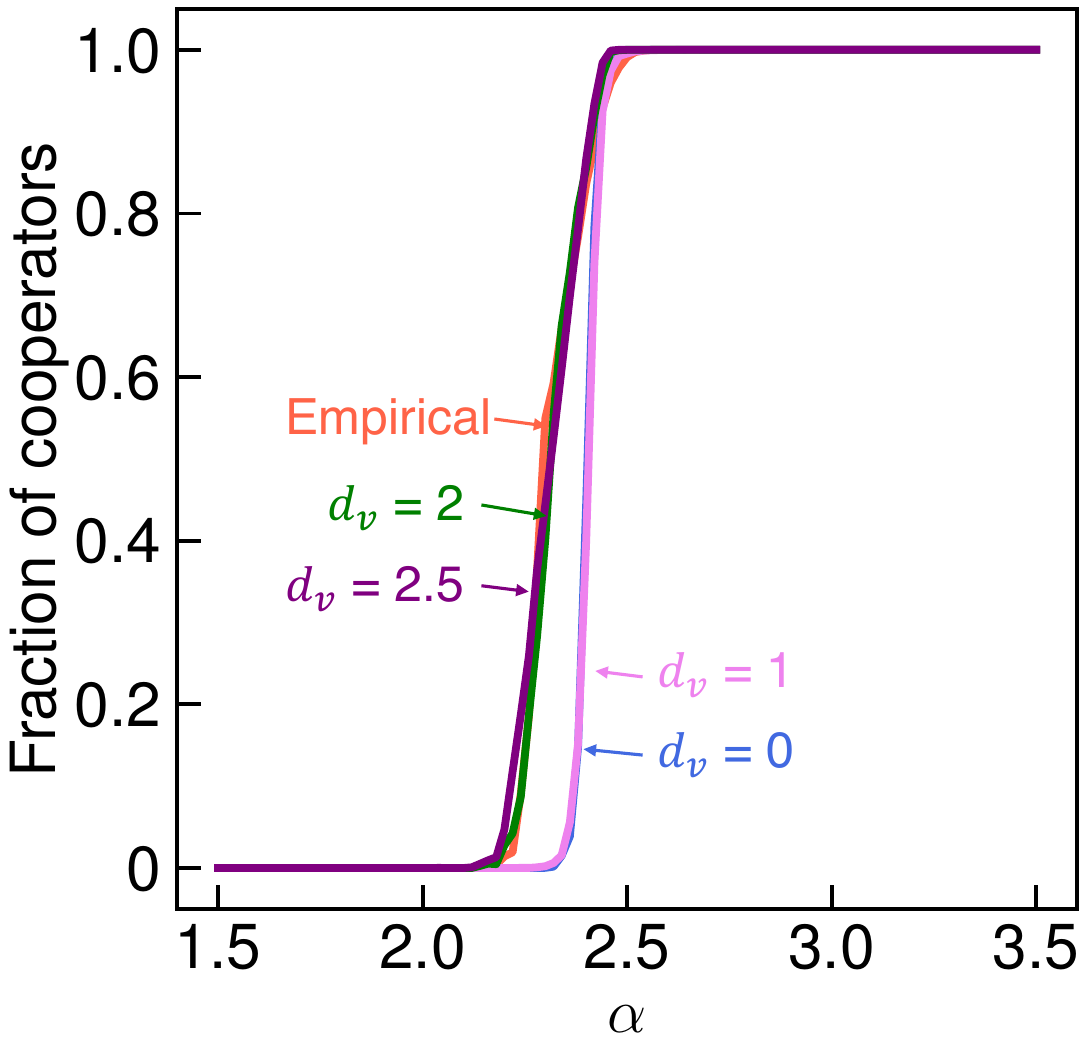}}
        \end{center}
      \end{minipage}
      \begin{minipage}{0.24\hsize}
        \begin{center}
        \subfigure[]{
          \includegraphics[height=\figsizec]{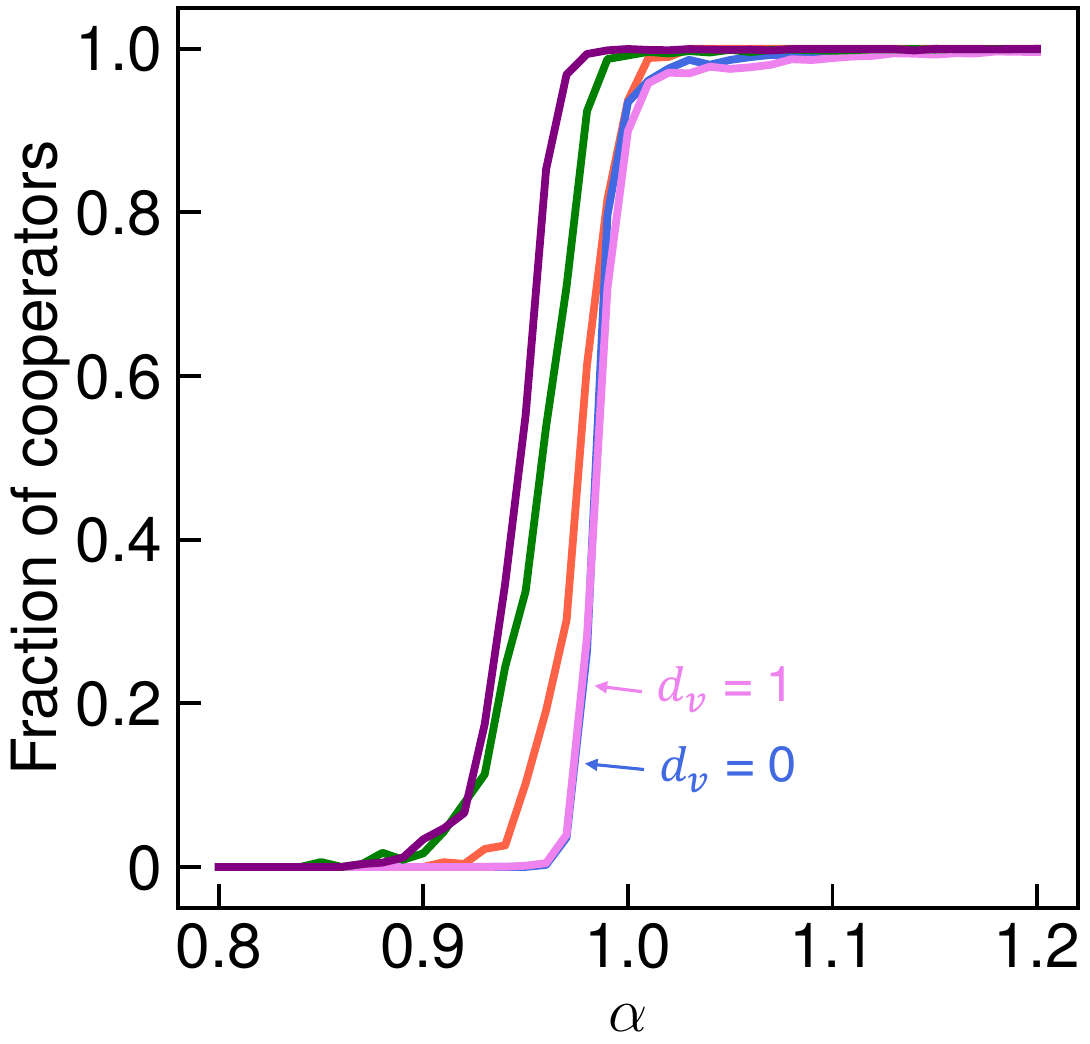}}
        \end{center}
      \end{minipage}
      \begin{minipage}{0.24\hsize}
        \begin{center}
        \subfigure[]{
          \includegraphics[height=\figsizec]{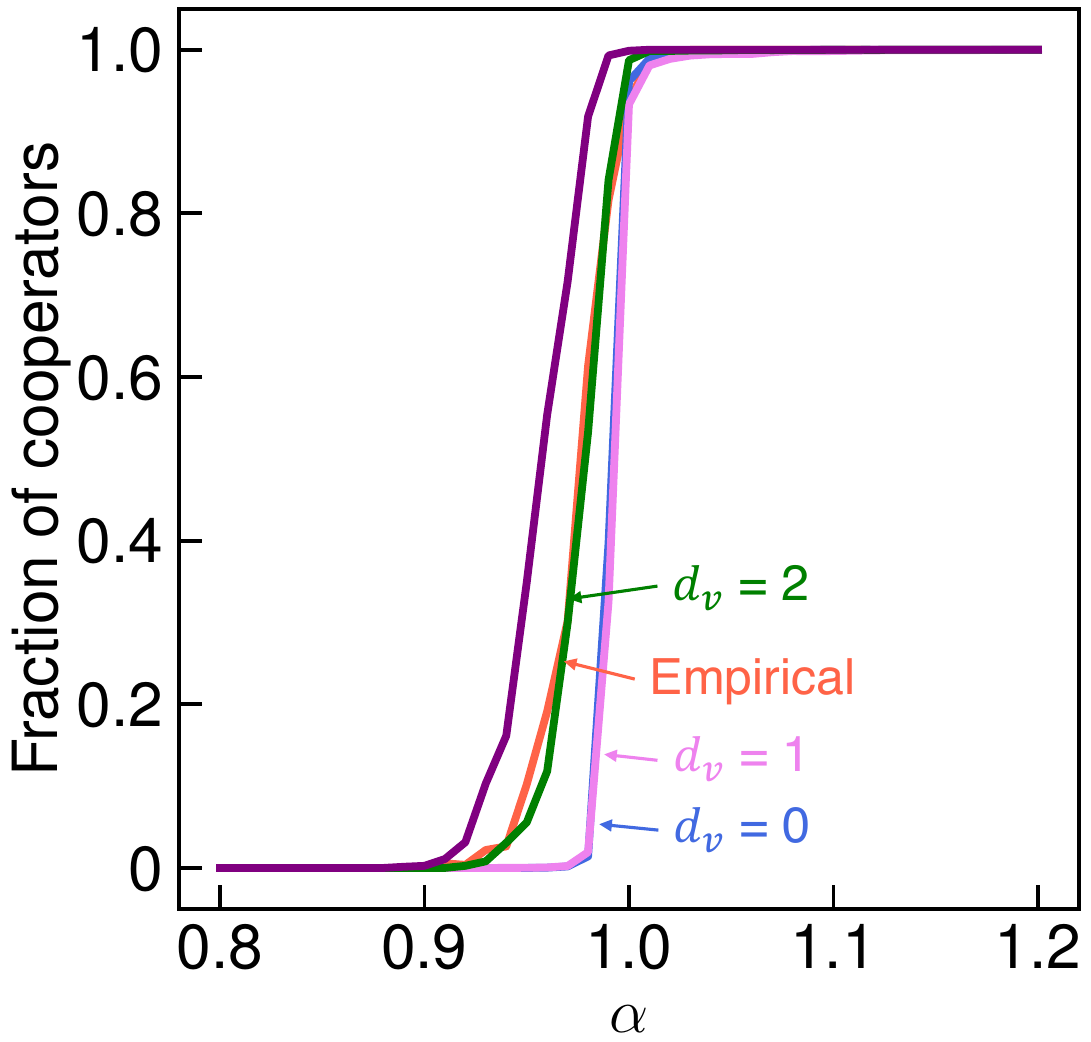}}
        \end{center}
      \end{minipage}
      
      \begin{minipage}{0.24\hsize}
        \begin{center}
        \subfigure[]{
          \includegraphics[height=\figsizec]{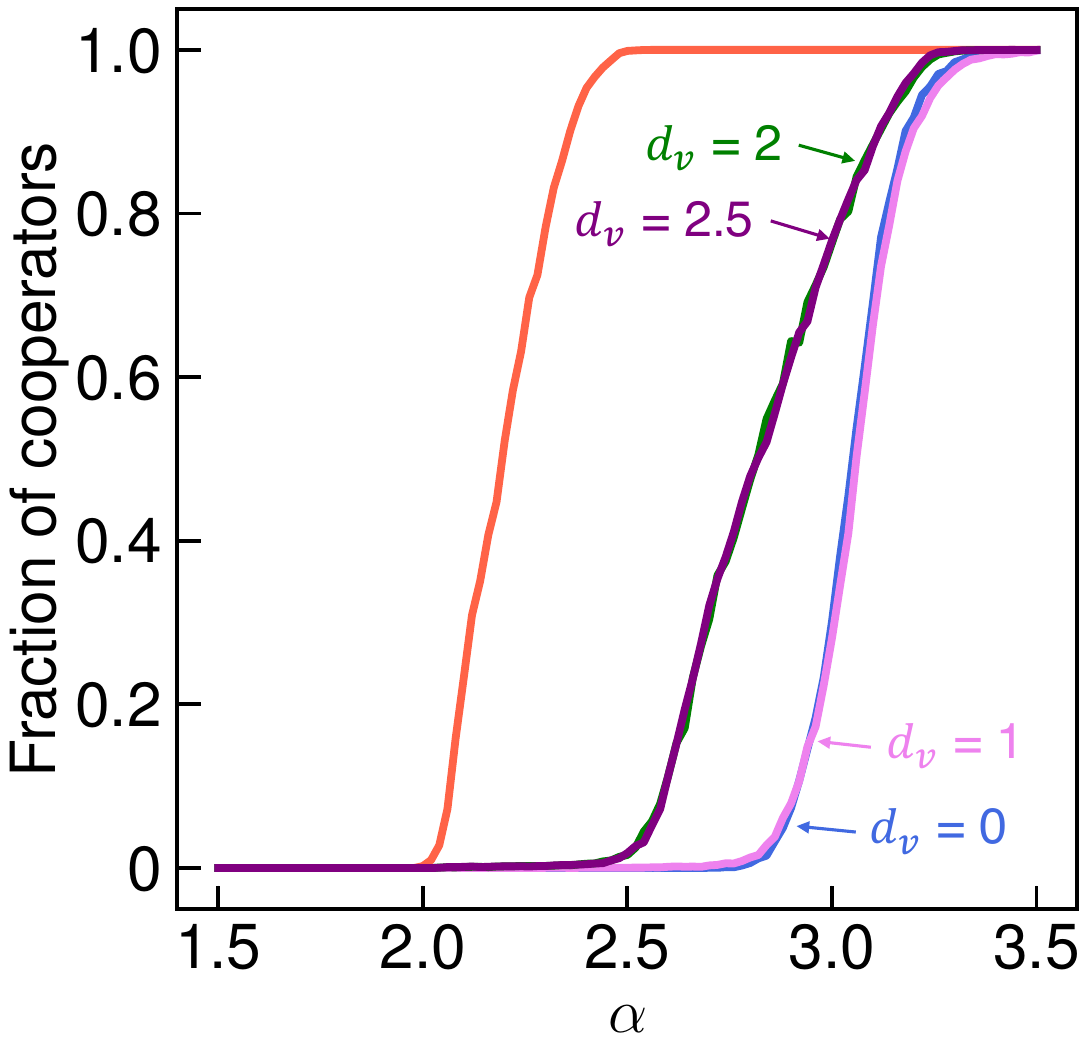}}
        \end{center}
      \end{minipage}
      \begin{minipage}{0.24\hsize}
        \begin{center}
        \subfigure[]{
        \includegraphics[height=\figsizec]{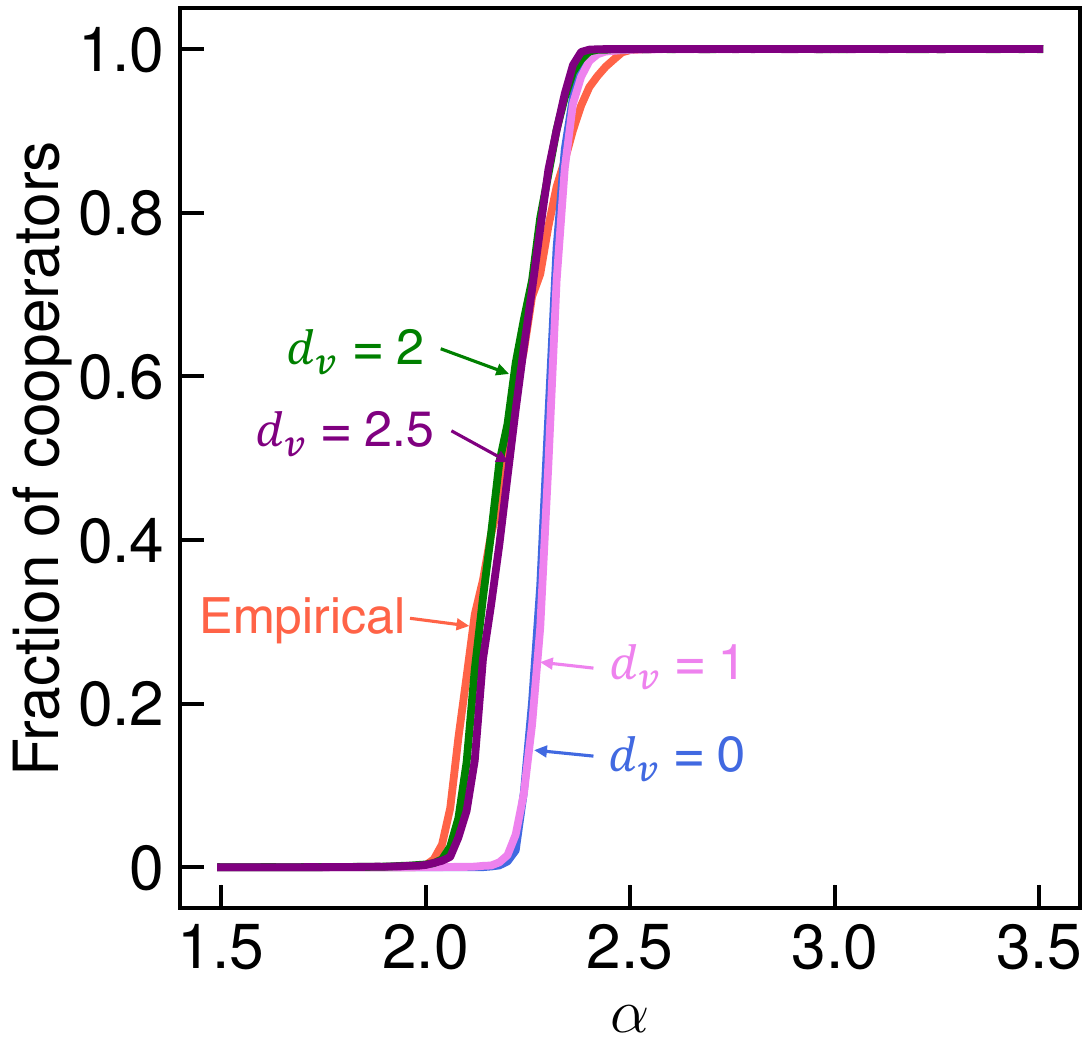}}
        \end{center}
      \end{minipage}
      \begin{minipage}{0.24\hsize}
        \begin{center}
        \subfigure[]{
          \includegraphics[height=\figsizec]{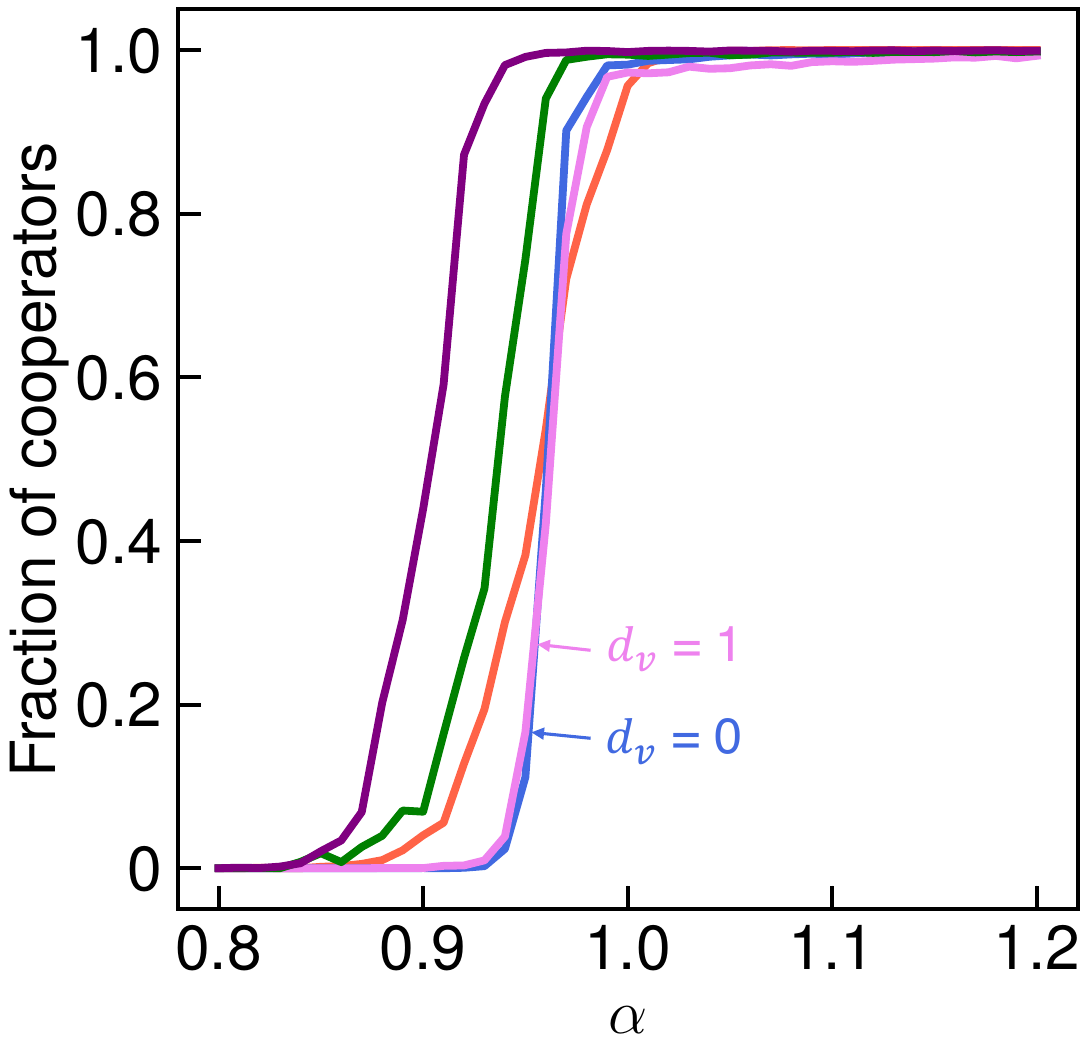}}
        \end{center}
      \end{minipage}
      \begin{minipage}{0.24\hsize}
        \begin{center}
        \subfigure[]{
          \includegraphics[height=\figsizec]{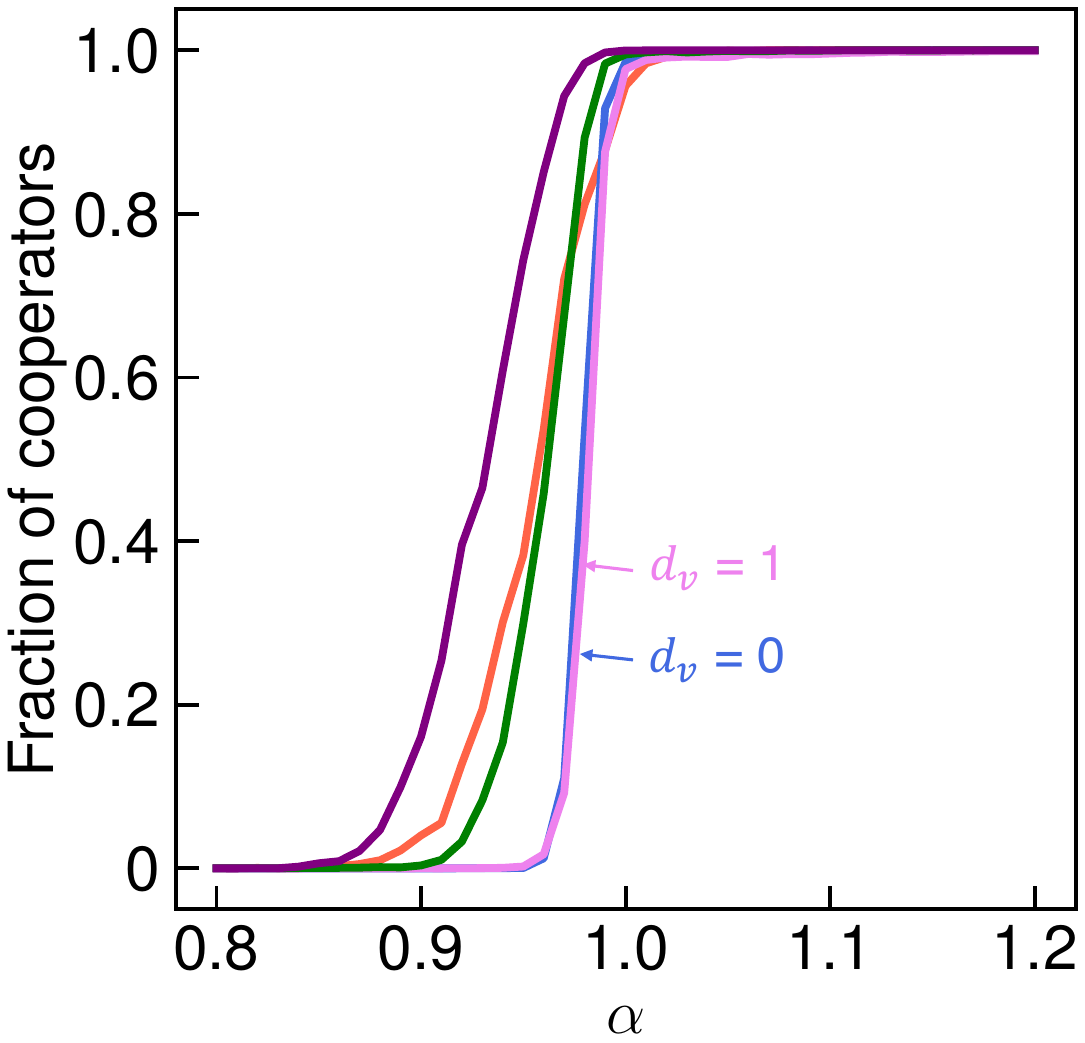}}
        \end{center}
      \end{minipage}
      \caption{Evolution of cooperation in the public goods game on hypergraphs. Panels (a)--(d) show the fraction of cooperators for the primary-school data set, and panels (e)--(h) are for the high-school data set. We set $(d_e,\beta) = (0, 0)$ in (a) and (e); $(d_e,\beta) = (1, 0)$ in (b) and (f); $(d_e,\beta) = (0, 1)$ in (c) and (g); and $(d_e,\beta) = (1, 1)$ in (d) and (h). We indicate the curves by the arrow and label wherever multiple curves heavily overlap each other.}
      \label{fig:7}
\end{figure*}

In Figs. \ref{fig:6}(a) and \ref{fig:6}(b), we set $\theta=0.1$ and compare the fraction of infected nodes among the primary-school hypergraph and hypergraphs generated by the corresponding hyper $dK$-series.
We set $d_e = 0$ in Fig. \ref{fig:6}(a) and $d_e = 1$ in Fig. \ref{fig:6}(b).
The results for the empirical hypergraph shown in Figs. \ref{fig:6}(a) and \ref{fig:6}(b) are the same.
We make the following observations.
First, the hyper $dK$-series with $d_e = 0$ considerably overestimates the fraction of infected nodes and underestimates the epidemic threshold for the empirical hypergraph for any $d_v$.
Second, the fraction of infected nodes in the hyper $dK$-series with $d_e=1$ is closer to that in the empirical hypergraph than with $d_e = 0$.
Third, the hyper $dK$-series with $(d_v,d_e)=(1,1)$, $(2,1)$, and $(2.5,1)$ accurately estimate the fraction of infected nodes and the epidemic threshold in the empirical hypergraph and almost to the same extent. 
In other words, the hyper $dK$-series with $(d_v, d_e) = (1,1)$ is necessary and sufficient for reproducing the fraction of infected nodes as a function of the infection rate. 
These results indicate that the size of each hyperedge, or equivalently, its distribution, is a main determinant of the epidemic spreading  more than are the node's local properties with $d_v>1$, such as the degree correlation and redundancy coefficient, and mesoscopic or macroscopic structure of the hypergraph.
These results qualitatively remain the same for a different threshold value, i.e., $\theta = 0.5$ (see Figs. \ref{fig:6}(c) and \ref{fig:6}(d)) and for the high-school hypergraph (see Figs. \ref{fig:6}(e)--\ref{fig:6}(h)).

\subsection{Evolutionary Dynamics}

Next, we compare evolutionary dynamics on the empirical hypergraphs and the hyper $dK$-series.
We use a previously proposed model of public goods game on hypergraphs, which proceeds as follows \cite{alvarez}.
Each node selects either to cooperate or defect in each round of evolutionary dynamics.
A cooperator transfers an asset $c$ to the public goods of hyperedge $e$, where $|e| \ge 2$. 
A defector does not contribute to the public goods.
The total investment in $e$ is $n_{\text{C}} c$, where $n_{\text{C}}$ is the number of cooperators in $e$.
Then, one multiplies the total investment by the synergy factor $R$, where $R > 1$, and then equally distributes the multiplied total investment among all the nodes in $e$.
The payoff that a cooperator and defector receives from hyperedge $e$ is equal to $\pi_{\text{C}} = R n_{\text{C}} c/|e| - c$ and $\pi_{\text{D}} = R n_{\text{C}} c/|e|$, respectively.
As in the previous study \cite{alvarez}, we assume $R = \alpha |e_j|^{\beta}$, where $\alpha > 0$ and $\beta \geq 0$.

We numerically simulate the evolutionary public goods game on the given hypergraph as follows.
Initially, each node is independently cooperator or defector with a probability of 0.5 each.
In each round, we first uniformly randomly select a node $v_i$, whose strategy (i.e., cooperation or defection) may be updated, with probability $1/N$ and then select a hyperedge $e_j$ to which $v_i$ belongs with probability $1/k_i$ uniformly at random. 
We continue this selection procedure until we select a hyperedge with $|e_j| \ge 2$.
We have confirmed that each node belongs to at least one hyperedge with $|e_j| \geq 2$ in all cases.
Then, all the nodes that belong to $e_j$ play the public goods game just once in each of the hyperedges to which they belong.
Each node accumulates the payoffs from all the games that the node plays. 
Then, we divide the accumulated payoff by the number of games that the node has played.
We denote by $\pi_i$ the payoff of node $v_i$.
Node $v_i$ adopts the strategy of the node that has gained the largest payoff in hyperedge $e_j$, denoted by $v_{i'}$, with probability $(\pi_{i'} - \pi_{i}) / \Delta$.
When $\beta < 1$, we set 
\begin{align}
\Delta =
\begin{cases}
\alpha \tilde{s}_{\text{min}}^{\beta - 1} (\tilde{s}_{\text{min}} - 1) - \alpha s_{\text{max}}^{\beta - 1} + 1 & \text{if }\alpha \leq \frac{2}{\tilde{s}_{\text{min}}^{\beta - 1} + s_{\text{max}}^{\beta - 1}}, \\
\alpha \tilde{s}_{\text{min}}^{\beta} - 1 & \text{otherwise},
\end{cases}
\label{eq:8}
\end{align}
where $\tilde{s}_{\text{min}} = \max \{ s_{\text{min}}, 2 \}$, and $s_{\text{max}}$ and $s_{\text{min}}$ are the largest and smallest sizes of the hyperedge, respectively.
When $\beta \geq 1$, we set
\begin{align}
\Delta =
\begin{cases}
\alpha s_{\text{max}}^{\beta - 1} (s_{\text{max}} - 1) - \alpha \tilde{s}_{\text{min}}^{\beta - 1} + 1 & \text{if }\alpha \leq \frac{2}{\tilde{s}_{\text{min}}^{\beta - 1} + s_{\text{max}}^{\beta - 1}}, \\
\alpha s_{\text{max}}^{\beta} - 1 & \text{otherwise}.
\end{cases}
\label{eq:9}
\end{align}
Equations \eqref{eq:8} and \eqref{eq:9} guarantees that the probability $(\pi_{i'} - \pi_i)/\Delta$ is normalized (see Ref. \cite{alvarez} for details).
If $\pi_{i'} \leq \pi_i$, node $v_i$ does not adopt the strategy of $v_{i'}$.
For the given $\alpha$ and $\beta$ values, we measure the fraction of cooperators as the average over the $(10^6+1)$st and $(10^6 + 10^3)$th rounds in a single run and over 100 runs.
In the case of the hyper $dK$-series, we generate an independent bipartite graph for each run.

In Figs. \ref{fig:7}(a) and \ref{fig:7}(b), we set $\beta=0$ and compare the fraction of cooperators on the primary-school hypergraph and the hyper $dK$-series.
We set $d_e = 0$ in Fig. \ref{fig:7}(a) and $d_e = 1$ in Fig. \ref{fig:7}(b).
The results for the empirical hypergraph shown in Figs. \ref{fig:7}(a) and \ref{fig:7}(b) are the same.
We make the following observations.
First, at both $d_e$ values, the node's pairwise degree correlation present in the empirical hypergraph promotes the cooperation but the node's degree distribution or the profile of the redundancy coefficient does not.
Second, the fraction of cooperators in the hyper $dK$-series with any $d_v$ and $d_e = 0$ is considerably smaller than that in the empirical hypergraph. 
In contrast, the fraction of cooperators in the hyper $dK$-series with $d_e = 1$ is generally close to that in the empirical hypergraph.
Therefore, destroying the distribution of the hyperedge's size in the original hypergraph suppresses cooperation.
In fact, the size distribution of the hyperedge is a stronger determinant of the amount of cooperation than any of the node's local properties investigated (i.e., the degree distribution, pairwise degree correlation, and redundancy coefficient).

Figures \ref{fig:7}(c) and \ref{fig:7}(d) show the results for $\beta = 1$.
We make the following observations.
First, when $d_e = 0$ (see Fig. \ref{fig:7}(c)), preserving the node's degree correlation and redundancy of the original hypergraph individually enhances cooperation. 
However, when one destroys the degree correlation (i.e., $d_v = 0$ or $1$), there is less cooperation than in the original hypergraph.
Furthermore, intriguingly, the hyper $dK$-series with $(d_v, d_e) = (2, 0)$ and $(2.5, 0)$ realize more cooperation than on the original hypergraph, suggesting that destroying the network structure that is higher-order than the degree-correlation and redundancy promotes cooperation.
Second, there is less cooperation when the distribution of the hyperedge's size is preserved (i.e., $d_e=1$; Fig. \ref{fig:7}(d)) than destroyed (i.e., $d_e$ = 0; Fig. \ref{fig:7}(c)).
This result is opposite to that for $\beta = 0$ (see Figs. \ref{fig:7}(a) and \ref{fig:7}(b)).
Third, similarly to the case of $d_e = 0$, the preservation of the node's degree correlation and redundancy (but not higher-order structure) of the original hypergraph individually increases cooperation in the case of $d_e=1$.
In particular, hyper $dK$-series with $(d_v,d_e)=(2.5,1)$ realizes more cooperation than on the original hypergraph (see the purple line in Fig. \ref{fig:7}(d)).
A comparison between Figs. \ref{fig:7}(c) and \ref{fig:7}(d) suggests that, no matter whether the distribution of the hyperedge's size is destroyed or preserved, destroying the structure that is higher-order than the node's redundancy by randomization yields more cooperation than in the original hypergraph.
All these results qualitatively remain the same for the high-school hypergraph (see Figs. \ref{fig:7}(e)--\ref{fig:7}(h)).

The critical point $\alpha = \alpha_c(\beta)$ separating the defection and cooperation phases is analytically calculated as follows \cite{alvarez}:
\begin{align}
\alpha_c(\beta) = \frac{1}{\sum_{s = \tilde{s}_{\text{min}}}^{s_{\text{max}}} \tilde{p}(s) s^{\beta - 1}},
\label{eq:10}
\end{align}
where $\tilde{p}(s) = p(s) / \sum_{s=\tilde{s}_{\text{min}}}^{s_{\text{max}}} p(s)$, and $p(s)$ represents the fraction of hyperedges of size $s$.
Note that it holds that $\sum_{s=\tilde{s}_{\text{min}}}^{s_{\text{max}}} \tilde{p}(s) = 1$.
In the infinite well-mixed population, the evolutionary dynamics converge to full defection and full cooperation when $\alpha < \alpha_c(\beta)$ and $\alpha > \alpha_c(\beta)$, respectively.

When $\beta = 0$, the primary-school hypergraph yields $\alpha_c(0) \approx 2.31$. 
Roughly consistent with this, the fraction of cooperators on the empirical hypergraph reaches $\approx$ 1.0 at $\alpha \approx 2.5$ in our simulations (see the red lines in Figs. \ref{fig:7}(a) and \ref{fig:7}(b)).
The corresponding hyper $dK$-series with any $d_v$ and $d_e = 0$ leads to $\alpha_c(0) \approx 2.77$, which underestimates the threshold obtained from the numerical simulations, i.e., $\alpha \approx 3.3$ (see Fig. \ref{fig:7}(a)).
However, Eq. \eqref{eq:10} and our numerical results are consistent in the sense that the critical point in terms of $\alpha$ for the hyper $dK$-series with $d_e = 0$ is larger than that for the empirical hypergraph.
The hyper $dK$-series with any $d_v$ and $d_e = 1$ has the same analytically determined threshold, $\alpha_c(0) \approx 2.31$, as the empirical hypergraph because these hypergraphs have the same distribution of the hyperedge's size.
This result is also consistent with our numerical result that the fraction of cooperators reaches $\approx$ 1.0 at $\alpha \approx 2.5$ in the hyper $dK$-series with any $d_v$ and $d_e = 1$ (see Fig. \ref{fig:7}(b)).
When $\beta = 1$, Eq. \eqref{eq:10} predicts that $\alpha_c(1) = 1.0$ regardless of $d_v$ and the size distribution of the hyperedge (therefore, regardless of $d_e$). 
This result is consistent with our numerical results shown in Figs. \ref{fig:7}(c) and \ref{fig:7}(d).

For the high-school hypergraph, we obtain $\alpha_c(0) \approx 2.23$ for the empirical hypergraph and the hyper $dK$-series with $d_e = 1$, $\alpha_c(0) \approx 2.75$ for the hyper $dK$-series with $d_e = 0$, and $\alpha_c(1) = 1.0$ for the empirical and synthetic hypergraphs. 
In our numerical simulations, we obtain $\alpha_c(0) \approx 2.5$ for for the empirical hypergraph (see the red lines in Figs. \ref{fig:7}(e) and \ref{fig:7}(f)) and the hyper $dK$-series with $d_e = 1$ (see Fig. \ref{fig:7}(f)), $\alpha_c(0) \approx 3.3$ for the hyper $dK$-series with $d_e = 0$ (see Fig. \ref{fig:7}(e)), and $\alpha_c(1) \approx 1.0$ for the empirical and synthetic hypergraphs (see Figs. \ref{fig:7}(g) and \ref{fig:7}(h)).
These results are qualitatively the same as those for the primary-school hypergraph. 

\begin{figure*}[t]
      \begin{minipage}{0.31\hsize}
        \begin{center}
        \subfigure[]{
          \includegraphics[height=\figsizea]{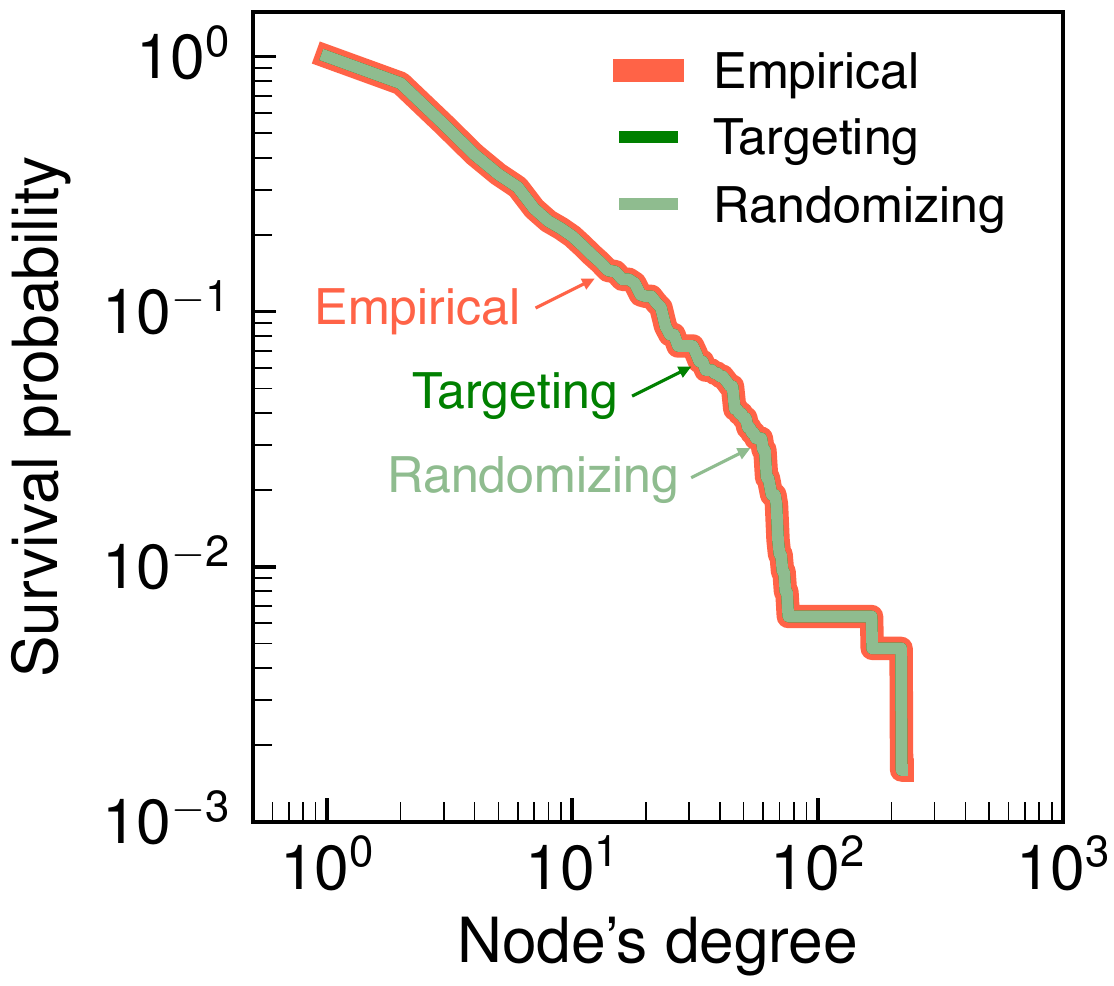}}
        \end{center}
      \end{minipage}
      \begin{minipage}{0.31\hsize}
        \begin{center}
        \subfigure[]{
        \includegraphics[height=\figsizea]{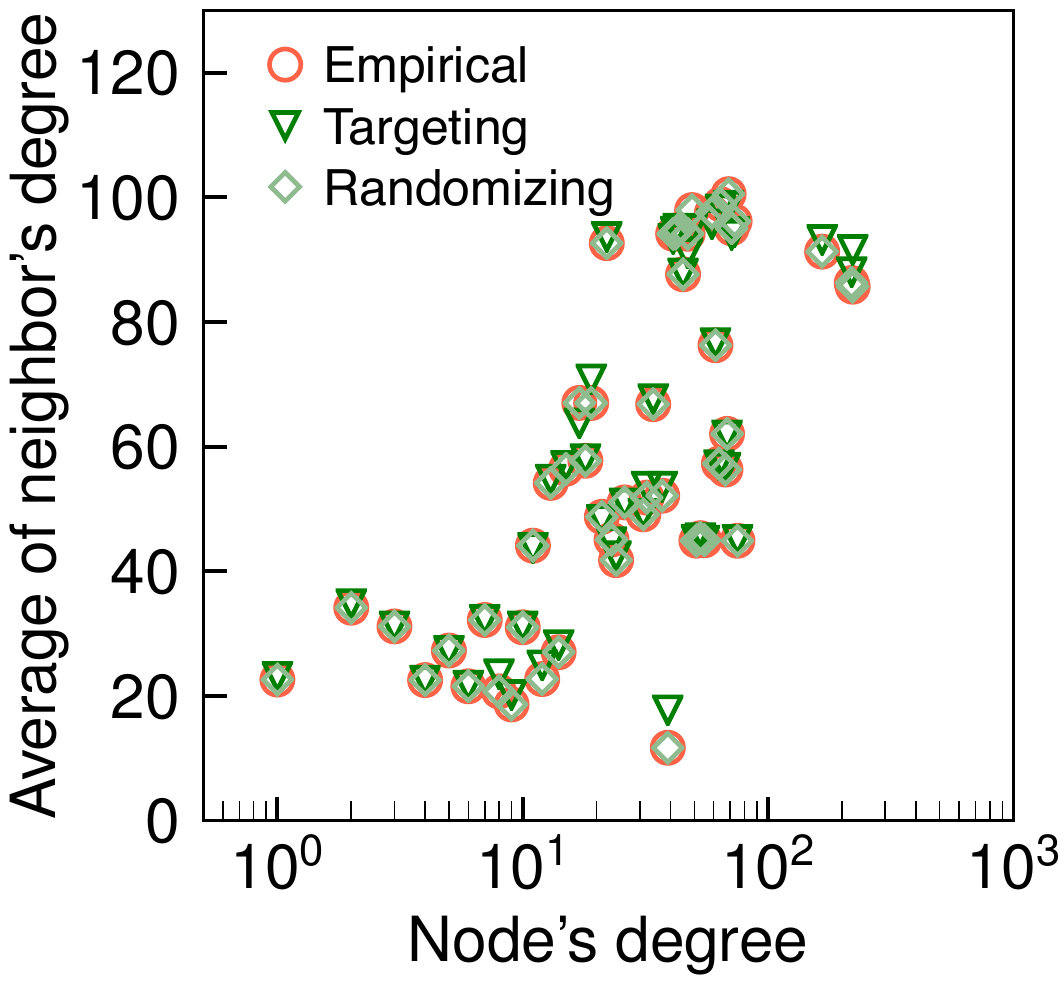}}
        \end{center}
      \end{minipage}
      \hspace{2mm}
      \begin{minipage}{0.31\hsize}
        \begin{center}
        \subfigure[]{
          \includegraphics[height=\figsizea]{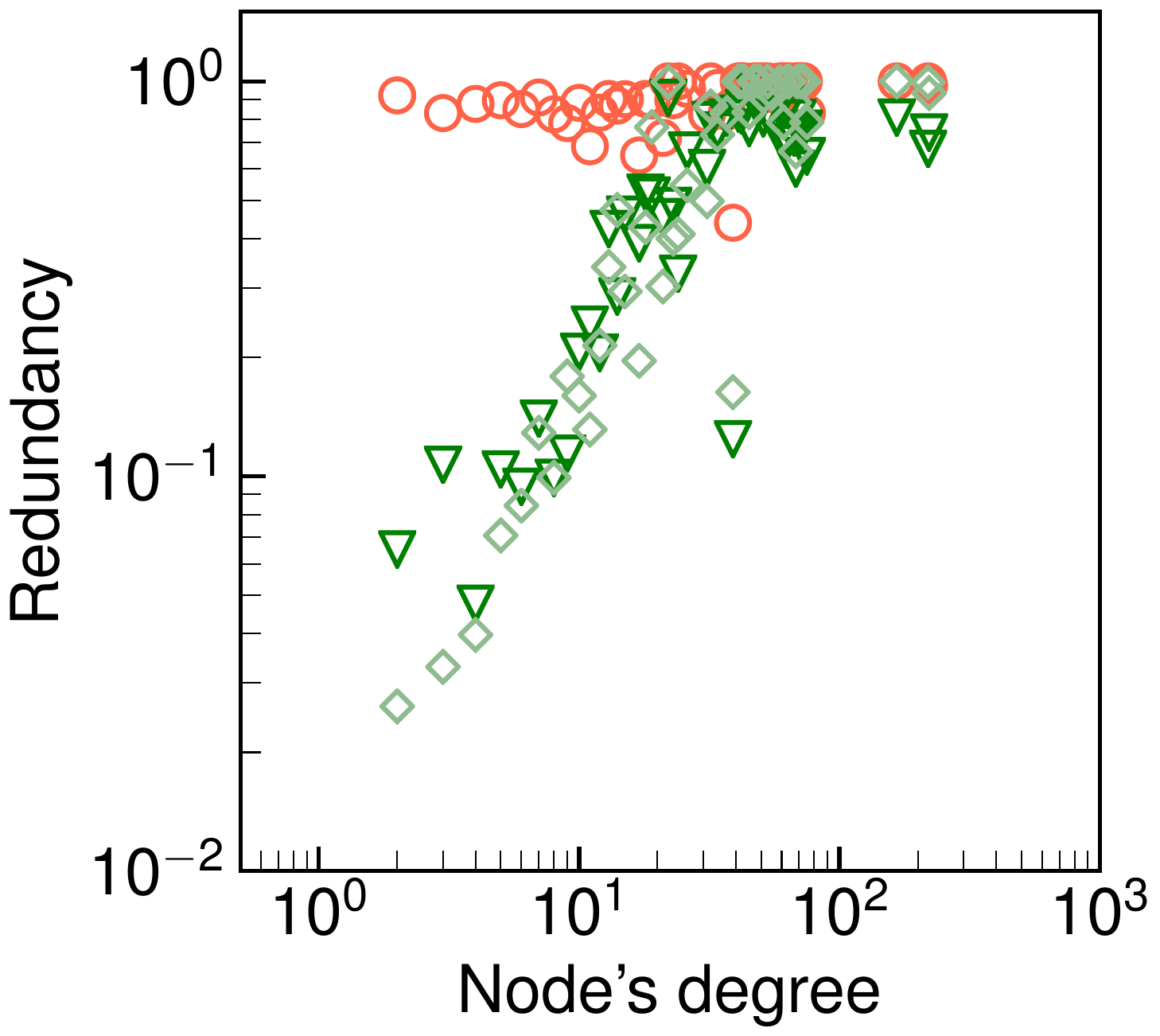}}
        \end{center}
      \end{minipage}
      
      \vspace{3mm}
      \begin{minipage}{0.31\hsize}
        \begin{center}
        \subfigure[]{
          \includegraphics[height=\figsizea]{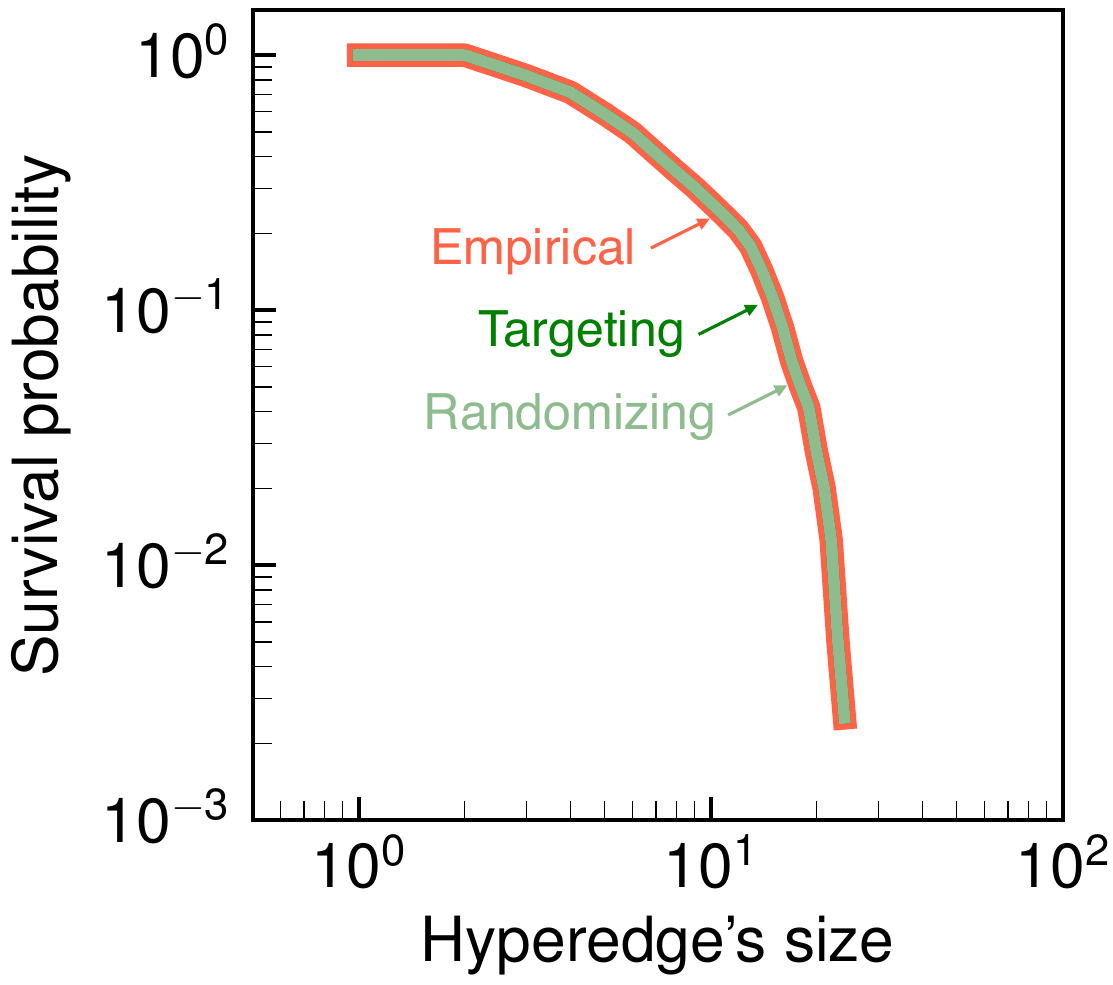}}
        \end{center}
      \end{minipage}
      \hspace{-1mm}
      \begin{minipage}{0.31\hsize}
        \begin{center}
        \subfigure[]{
          \includegraphics[height=\figsizea]{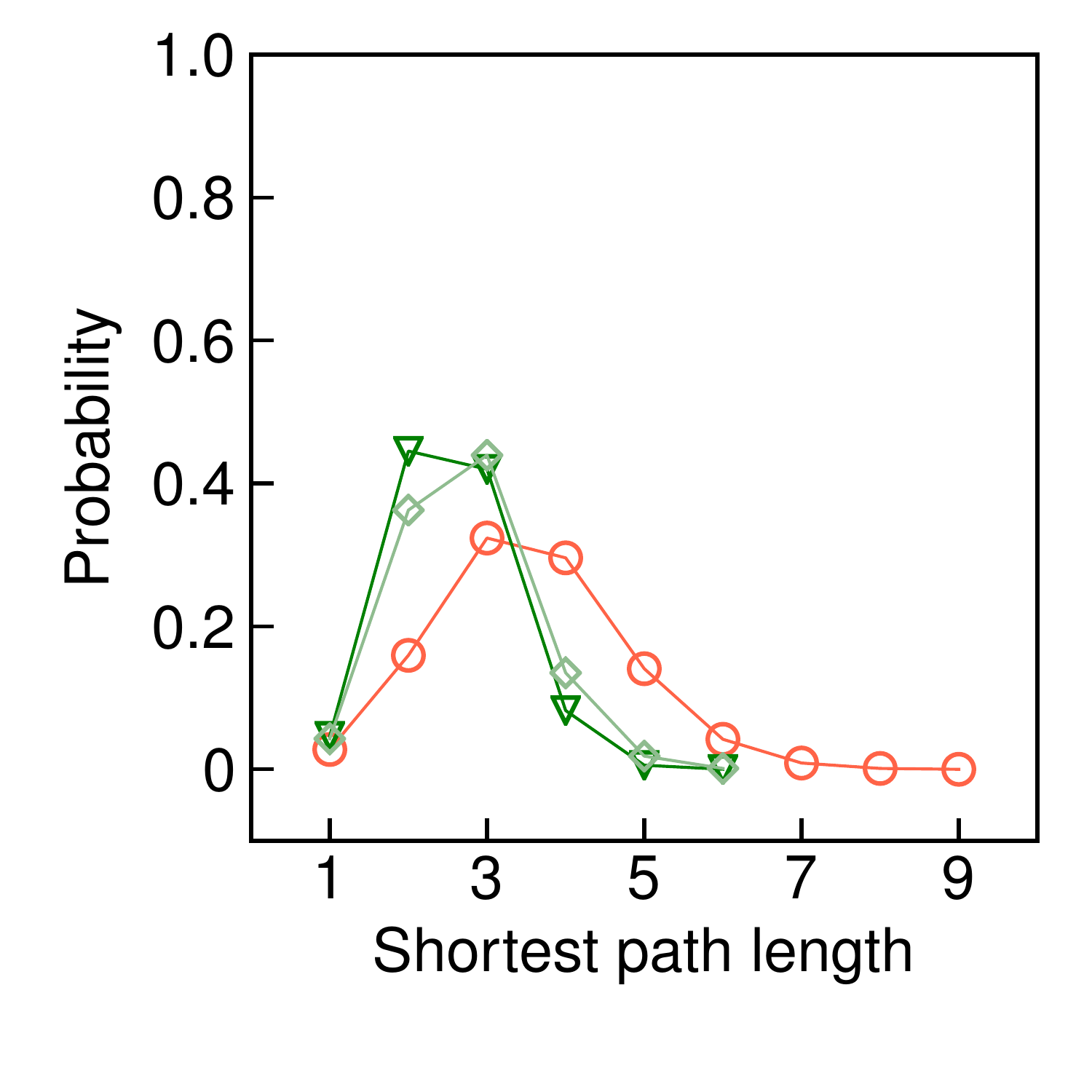}}
        \end{center}
      \end{minipage}
      \hspace{2mm}
      \begin{minipage}{0.31\hsize}
        \begin{center}
        \subfigure[]{
        \includegraphics[height=\figsizea]{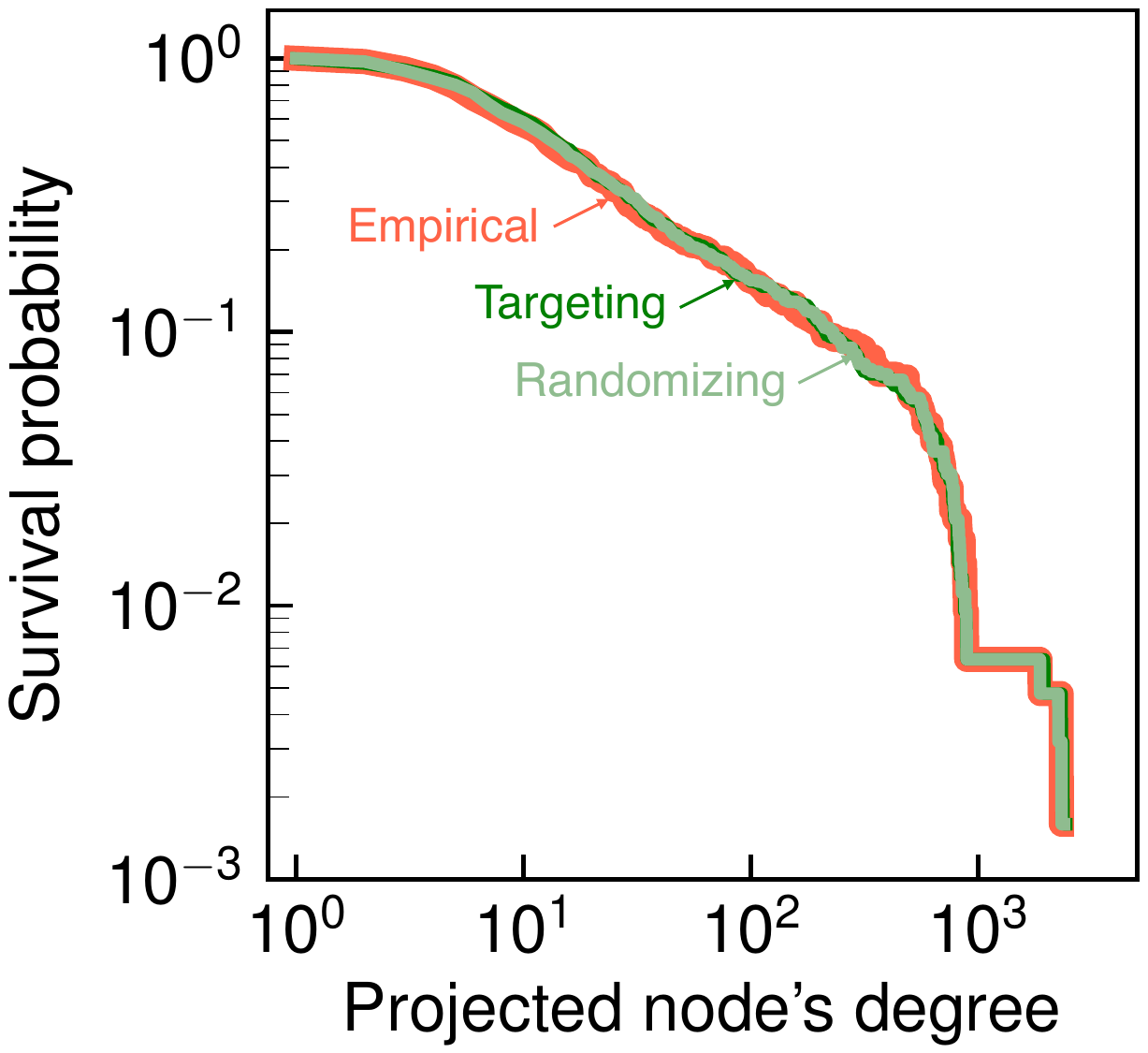}}
        \end{center}
      \end{minipage}
      \caption{Comparison between the targeting-rewiring and randomizing-rewiring processes for the drug hypergraph. We set $(d_v, d_e) = (2, 1)$. (a) Cumulative degree distribution of the node, (b) average degree of the nearest neighbors of nodes with degree $k$, (c) degree-dependent redundancy coefficient of the node, (d) cumulative size distribution of the hyperedge, (e) distribution of the shortest path length between nodes, and (f) cumulative degree distribution of the one-mode projection. We indicate the curves behind other curves by the arrow and label wherever multiple curves completely or almost overlap each other.}
      \label{fig:8}
\end{figure*}

\begin{figure*}
	\begin{minipage}{0.24\hsize}
        \begin{center}
        \subfigure[]{
          \includegraphics[height=\figsizeb]{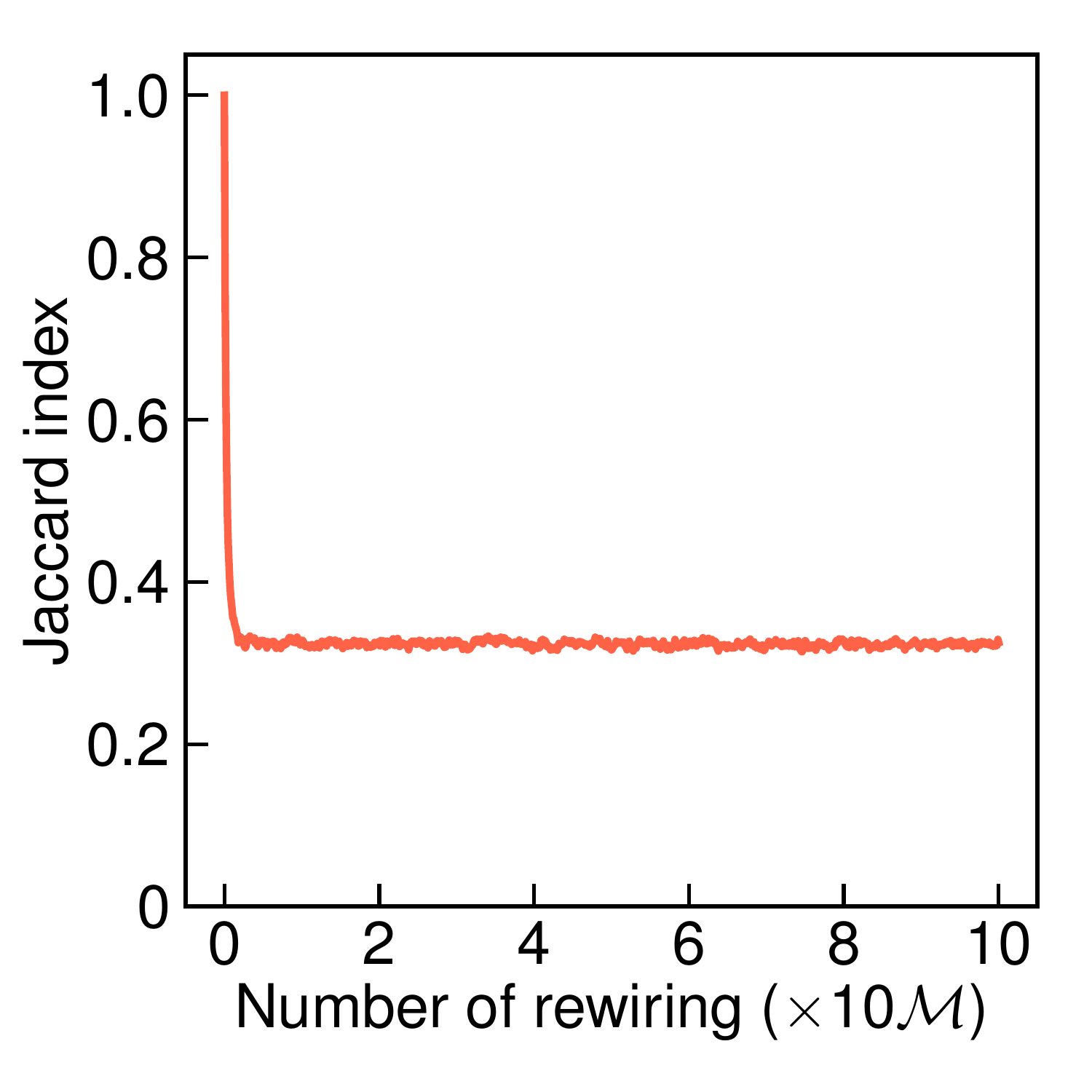}}
        \end{center}
	\end{minipage}
	\begin{minipage}{0.24\hsize}
        \begin{center}
        \subfigure[]{
          \includegraphics[height=\figsizeb]{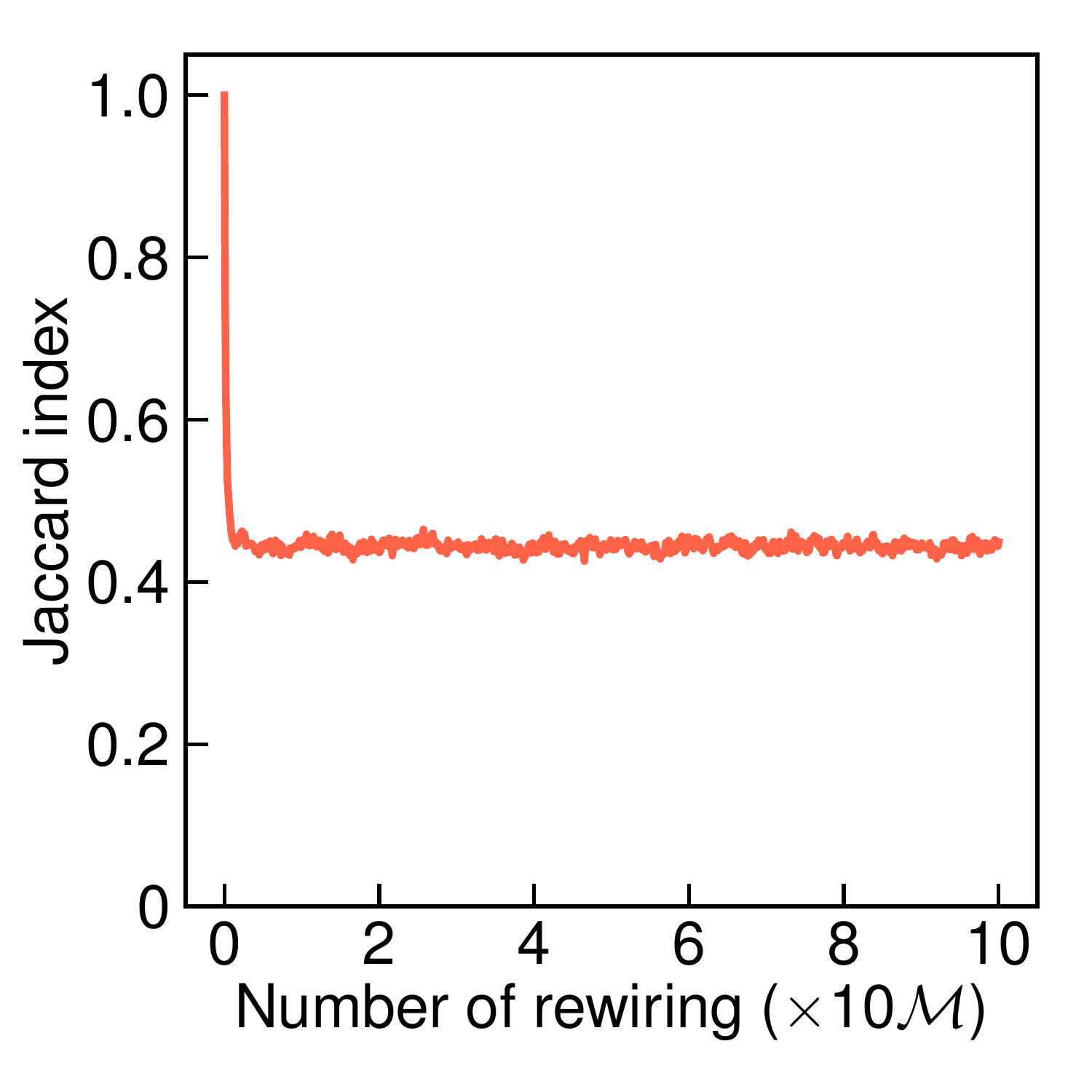}}
        \end{center}
	\end{minipage}
	\begin{minipage}{0.24\hsize}
        \begin{center}
        \subfigure[]{
          \includegraphics[height=\figsizeb]{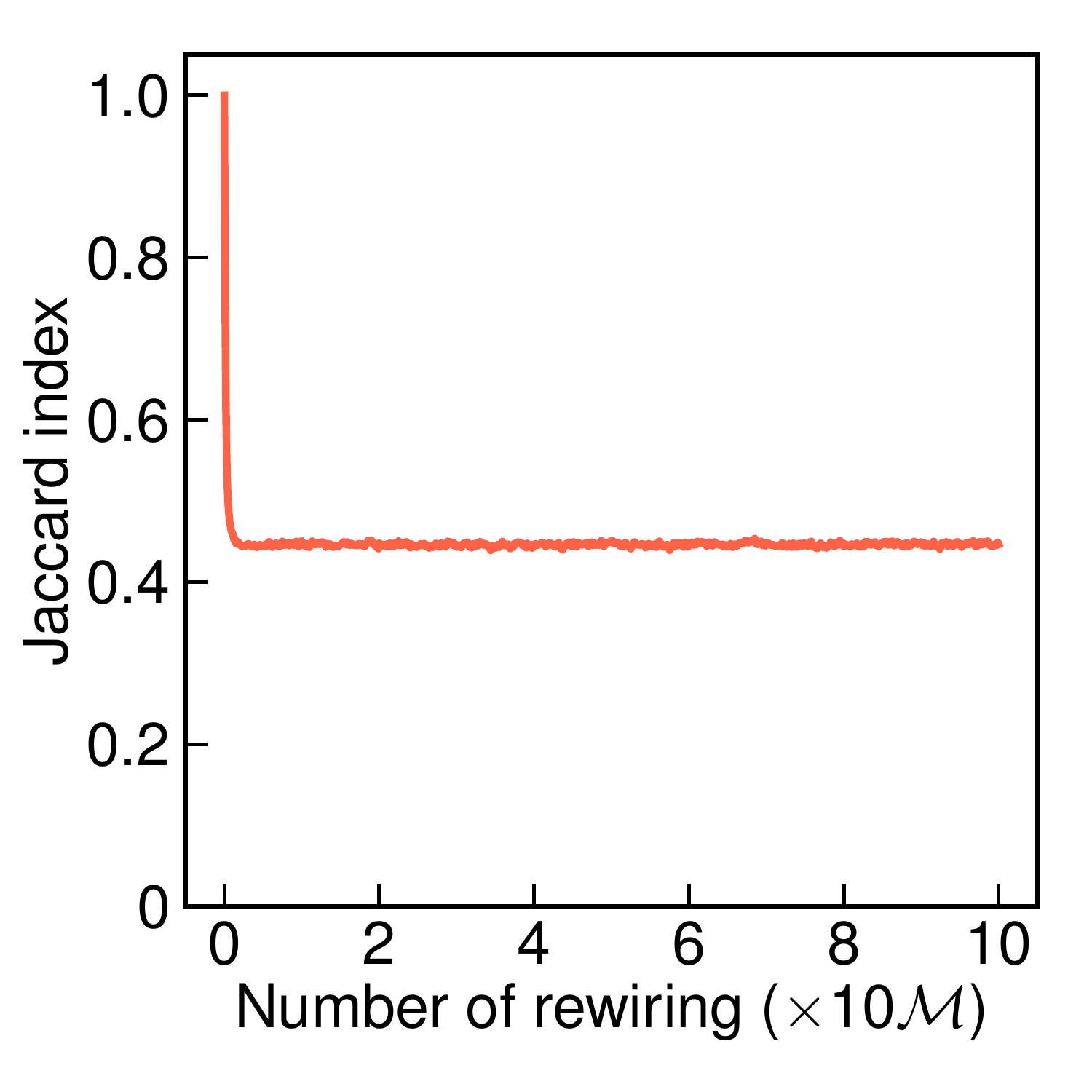}}
        \end{center}
	\end{minipage}
	\begin{minipage}{0.24\hsize}
        \begin{center}
        \subfigure[]{
          \includegraphics[height=\figsizeb]{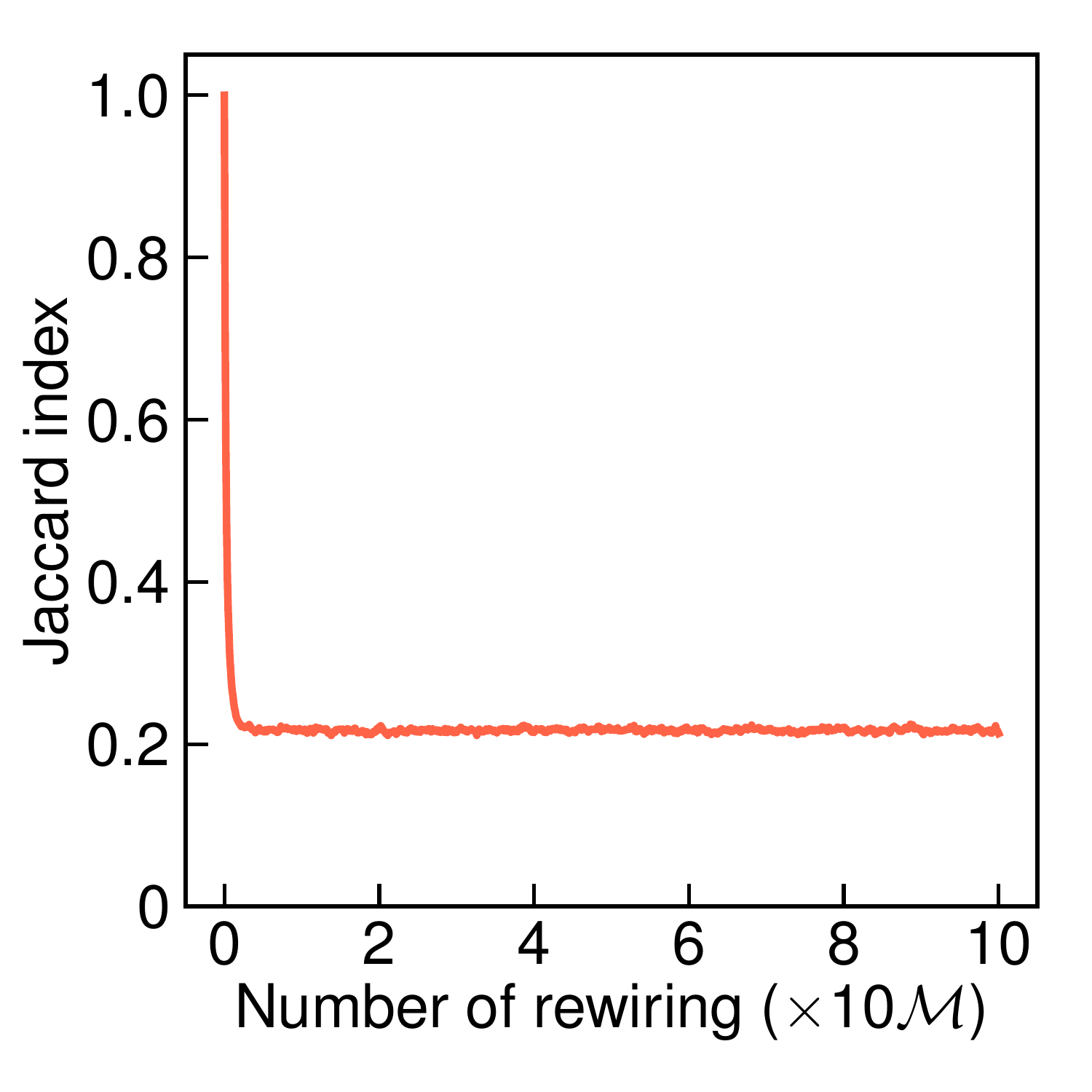}}
        \end{center}
	\end{minipage}
      \caption{The Jaccard index between a set of edges of the empirical hypergraph and that of the hypergraph generated under the randomizing rewiring. We set $(d_v, d_e) = (2, 1)$. (a) Drug, (b) Enron, (c) primary-school, and (d) high-school. The Jaccard index between the sets of edges is given by $|\mathcal{E} \cap \tilde{\mathcal{E}}|/|\mathcal{E} \cup \tilde{\mathcal{E}}|$, where $\mathcal{E}$ and $\tilde{\mathcal{E}}$ are the set of edges in the original and synthetic hypergraphs, respectively. 
In calculating the Jaccard index, we removed multiplicity of edges in $\tilde{\mathcal{E}}$.}
      \label{fig:9}
\end{figure*}

\section{Conclusion}
We proposed a family of reference models for hypergraphs called the hyper $dK$-series.
The hyper $dK$-series preserves the local properties of nodes and hyperedges in the given hypergraph to different extents.
We empirically showed that the hyper $dK$-series preserves the properties of nodes and hyperedges, as intended, across different hypergraph data sets.
We also showcased its use as reference models in investigating epidemic spreading and evolution of cooperation on hypergraphs.
Models of dynamical processes on hypergraphs, such as the epidemic spreading \cite{bodo, suo2018, jhun2019, landry2020}, evolutionary dynamics \cite{alvarez, burgio2020}, opinion dynamics \cite{neuhauser2020, hickok2021, sahasrabuddhe2021}, and synchronization \cite{lucas2020, mulas2020, de2021, salova2021}, have been proposed.
Deploying the hyper $dK$-series to studies of various models of dynamics is expected to better reveal how the dynamics depend on the specific structural properties of the given hypergraphs.

Up to our numerical efforts, we found that the hyper $dK$-series with a larger $d_v$ value better approximates the distribution of the shortest path length between nodes for the empirical hypergraphs.
However, as expected, even the hyper $dK$-series with the largest $d_v$ value (i.e., $d_v=2.5$) does not accurately approximate the distribution of the shortest path length.
In particular, we found that the average shortest path length for the hypergraphs generated by the hyper $dK$-series with $d_v = 2.5$ is smaller than that for the empirical hypergraph for all the four data sets (e.g., the drug hypergraph has the average shortest length of 3.53, whereas the hyper $dK$-series has 3.03 for $(d_v, d_e)=(2.5, 0)$ and 2.77 for $(d_v, d_e)=(2.5,1)$).
The community structure is one of network structures that is higher-order than the redundancy coefficient of the node and likely increases the shortest path length between nodes.
Extending the hyper $dK$-series to reference models that additionally preserve the community structure warrants future work.
To this end, it may be useful to employ a family of stochastic block models with the community structure for bipartite graphs \cite{doreian2004, ball2011, larremore2014, yen2020} or hypergraphs \cite{ghoshdastidar2017, ahn2018, ke2019, chodrow2021}.

\section*{Acknowledgments}
KN was supported by JSPS KAKENHI Grant Number JP21J10415.
KS was supported by JSPS KAKENHI Grant Number JP21H04872.
NM acknowledges support from AFOSR European Office (under Grant No. FA9550-19-1-7024), the Sumitomo Foundation, and the Nakatani Foundation.


\appendix
\addcontentsline{toc}{section}{Appendices}
\section*{Appendix}

\section{Comparison of the targeting rewiring and randomizing rewiring for $(d_v, d_e) = (2, 1)$}

In this section we compare the targeting-rewiring and randomizing-rewiring processes with $(d_v, d_e) = (2,1)$.
We show the distributions of the six quantities for the two rewiring processes for the drug hypergraph in Fig. \ref{fig:8}.
Both rewiring processes exactly preserve the degree distribution of the node and the size distribution of the hyperedge of the original bipartite graph (see Figs. \ref{fig:8}(a) and \ref{fig:8}(d)).
The randomizing-rewiring process exactly preserves $k_{\text{nn}}(k)$, whereas the targeting-rewiring process only approximately preserves it (see Fig. \ref{fig:8}(b)).
The two rewiring methods produce similar networks in terms of the degree-dependent redundancy coefficient, the distribution of the shortest path length between nodes, and the degree distribution of the one-mode projection, as shown in Figs. \ref{fig:8}(c), \ref{fig:8}(e), and \ref{fig:8}(f), respectively.

We also compare the two rewiring processes in terms of the overlap of the edges of the empirical hypergraph and those of the synthetic hypergraphs.
Figure \ref{fig:9}(a) shows the Jaccard index between sets of edges in the drug hypergraph and the hypergraph generated by the randomizing rewiring as a function of the number of rewiring attempts.
The figure indicates that the Jaccard index steadily decreases as the randomizing rewiring proceeds.
However, it plateaus at $\approx$ 0.32, which implies that a set of edges in the synthetic bipartite graph is not sufficiently shuffled due to the constraints that each edge rewiring step has to preserve $P(k, k')$ in addition to the degree of each node.
The Jaccard index similarly plateaus at $\approx$ 0.45, $\approx$ 0.45, and $\approx$ 0.21 for the Enron, primary-school, and high-school hypergraphs, respectively (see Figs. \ref{fig:9}(b), \ref{fig:9}(c), and \ref{fig:9}(d), respectively).
In contrast, the Jaccard index is $\approx$ 0.036, $\approx$ 0.016, $\approx$ 0.006, $\approx$ 0.005 under the targeting rewiring for the drug, Enron, primary-school, and high-school hypergraphs, respectively.
Therefore, we conclude that the randomizing rewiring does not sufficiently shuffle the edges of the input hypergraph.


\newpage

\begin{center}
\vspace*{12pt}
{\Large Supplementary Materials for:\\
\vspace{12pt}
Randomizing hypergraphs preserving degree correlation and local clustering}
\vspace{12pt} \\
\end{center}

\setcounter{figure}{0}
\setcounter{table}{0}
\setcounter{section}{0}

\renewcommand{\thesection}{S\arabic{section}}
\renewcommand{\thefigure}{S\arabic{figure}}
\renewcommand{\thetable}{S\arabic{table}}

\begin{center}
\author{Kazuki Nakajima, Kazuyuki Shudo, Naoki Masuda}
\vspace{24pt} \\
\end{center}

\section{Size of the largest connected component of hypergraphs generated by hyper $dK$-series}
We measured how the size (i.e., number of nodes) of the largest connected component of the empirical hypergraphs changes by randomization using the hyper $dK$-series.
We show in Table S1 the size of the largest connected component of hypergraphs the hyper $dK$-series generates, divided by that of the original hypergraph.
The table indicates that we barely lose nodes in the largest connected component by the randomization.

\section{Statistical test for the structural properties of hypergraphs generated by hyper $dK$-series}

In this section, we statistically test whether the hyper $dK$-series changes each structural property of a given hypergraph.
Consider a combination of any of the four empirical hypergraphs, any $(d_v, d_e)$ pair, and any of the six structural properties. 
To carry out a $t$-test, we first generate 100 pairs of independent hypergraphs using the hyper $dK$-series.
Second, we measure the distance between the two hypergraphs in each pair in terms of the distance measure for the selected structural property.
We denote by $\mu^{\text{rand}}$ and $\sigma^{\text{rand}}$ the mean and standard deviation, respectively, of the distance calculated on the basis of the 100 pairs of randomized hypergraphs.
Third, we generate another 100 hypergraphs using the hyper $dK$-series with the same $(d_v, d_e)$.
Fourth, we measure the distance between the empirical hypergraph and each of the 100 hypergraphs in terms of the selected structural property.
We denote by $\mu^{\text{emp}}$ and $\sigma^{\text{emp}}$ the mean and standard deviation, respectively, of the distance between a randomized hypergraph and the empirical hypergraph calculated on the basis of the 100 pairs.
Finally, we calculate the effect size for the $t$-test, called the Cohen's $d$ \cite{cohen1988}, as
\begin{align}
d = \frac{\mu^{\text{emp}} - \mu^{\text{rand}}}{\sqrt{\frac{(\sigma^{\text{emp}})^2 + (\sigma^{\text{rand}})^2}{2}}}. \tag{S1}
\end{align}
We define $d = 0$ if both $\mu^{\text{emp}} - \mu^{\text{rand}}$ and $(\sigma^{\text{emp}})^2 + (\sigma^{\text{rand}})^2$ are equal to zero.
We regard the effect size to be very small ($d = \pm 0.01$), small ($d = \pm 0.2$), medium ($d = \pm 0.5$), large ($d = \pm 0.8$), very large ($d = \pm 1.2$), and huge ($d = \pm 2.0$) \cite{cohen1988, sawilowsky2009}.

Table \ref{table:2} shows $\mu^{\text{rand}}$, $\sigma^{\text{rand}}$, $\mu^{\text{emp}}$, $\sigma^{\text{emp}}$, and Cohen's $d$ for the cumulative degree distribution of the node.
The effect size is huge when $(d_v, d_e) = (0, 0)$ and $(0, 1)$ for all the four empirical hypergraphs because the hyper $dK$-series with $(d_v, d_e) =$ (0, 0) and (0, 1) destroys the degree of each node.
For the other $(d_v, d_e)$ values, the effect size is zero because the hyper $dK$-series with $d_v \in \{1, 2, 2.5\}$ exactly preserves the degree of each node.

Table \ref{table:3} shows the results for the average degree of the nearest neighbors of nodes with degree $k$.
The effect size is huge when $d_v$ is 0 or 1 because the hyper $dK$-series with these $d_v$ values destroys the degree correlation.
When $d_v$ is 2 or 2.5, the hyper $dK$-series intends to preserve the degree correlation of the node.
However, Table \ref{table:3} indicates that the effect size ranges from medium to huge values, depending on the empirical network and the $d_e$ value. 
This is because the $\sigma^{\text{rand}}$ and $\sigma^{\text{emp}}$ are small.
Nevertheless, the Cohen's $d$ values in these cases are much smaller than those for $d_v = 0$ and 1.

Table \ref{table:4} shows the results for the degree-dependent redundancy coefficient of the node.
The effect size is huge when $d_v$ is 0, 1, or 2 because the hyper $dK$-series with these $d_v$ values destroys the redundancy of the node of a given hypergraph.
When $d_v$ is 2.5, the hyper $dK$-series intends to preserve the redundancy of the node.
However, the effect size is huge for all the four hypergraphs and $d_e$ values.
As in the case of the degree correlation, this is because $\sigma^{\text{rand}}$ and $\sigma^{\text{emp}}$ are small.
Nevertheless, similar to Table \ref{table:3}, $d$ is much smaller with $d_v = 2.5$ than with $d_v \le 2$.

Table \ref{table:5} shows the results for the cumulative size distribution of the hyperedge.
The effect size is huge when $d_e = 0$ for all the four empirical hypergraphs because the hyper $dK$-series with $d_e = 0$ destroys the size of each hyperedge.
When $d_e = 1$, the effect size is equal to zero because the hyper $dK$-series with $d_e = 1$ exactly preserves the size of each hyperedge.

Tables \ref{table:6} and \ref{table:7} show the results for the distribution of the shortest path length between nodes and those for the cumulative degree distribution of the one-mode projection, respectively.
For both properties, the effect size is huge in almost all cases. 
This result is consistent with the fact that the hyper $dK$-series does not intend to preserve these two properties. 
However, the $d$ value is considerably smaller when $d_v$ or $d_e$ is larger in most cases.

\renewcommand{\refname}{Supplementary References}

\newpage
\begin{table}
\caption{Relative size of the largest connected component of hypergraphs generated by the hyper $dK$-series. We show the mean $\pm$ standard deviation (SD) for each parameter set. We calculated the mean and the standard deviation on the basis of 100 randomized hypergraphs.}
\begin{center}
\label{table:1}
	\begin{tabular}{X || Y | Z}\hline
	Data & \centering $(d_v, d_e)$ & Mean $\pm$ SD \rule[0mm]{0mm}{3.5mm} \\ \hline
	\multirow{8}{*}{drug} & $(0, 0)$ & 1.000 $\pm$ 0.000 \\ 
	& $(1, 0)$ & 0.999 $\pm$ 0.001 \\ 
	& $(2, 0)$ & 0.984 $\pm$ 0.014 \\ 
	& $(2.5, 0)$ & 0.980 $\pm$ 0.015 \\ 
	& $(0, 1)$ & 0.999 $\pm$ 0.001 \\ 
	& $(1, 1)$ & 0.999 $\pm$ 0.001 \\ 
	& $(2, 1)$ & 0.974 $\pm$ 0.009 \\ 
	& $(2.5, 1)$ & 0.952 $\pm$ 0.014 \\ \hline
	\multirow{8}{*}{Enron} & $(0, 0)$ & 1.000 $\pm$ 0.000  \\ 
	& $(1, 0)$ & 0.999 $\pm$ 0.001 \\ 
	& $(2, 0)$ & 1.000 $\pm$ 0.000 \\ 
	& $(2.5, 0)$ & 0.999 $\pm$ 0.001 \\ 
	& $(0, 1)$ & 1.000 $\pm$ 0.000 \\ 
	& $(1, 1)$ & 1.000 $\pm$ 0.000 \\ 
	& $(2, 1)$ & 0.999 $\pm$ 0.001 \\ 
	& $(2.5, 1)$ & 0.999 $\pm$ 0.001 \\ \hline
	\multirow{8}{*}{primary-school} & $(0, 0)$ & 1.000 $\pm$ 0.000 \\ 
	& $(1, 0)$ & 1.000 $\pm$ 0.000 \\
	& $(2, 0)$ & 1.000 $\pm$ 0.000 \\ 
	& $(2.5, 0)$ & 1.000 $\pm$ 0.000 \\
	& $(0, 1)$ & 1.000 $\pm$ 0.000 \\
	& $(1, 1)$ & 1.000 $\pm$ 0.000 \\
	& $(2, 1)$ & 1.000 $\pm$ 0.000 \\
	& $(2.5, 1)$ & 1.000 $\pm$ 0.000 \\ \hline
	\multirow{8}{*}{high-school} & $(0, 0)$ & 1.000 $\pm$ 0.000 \\ 
	& $(1, 0)$ & 0.999 $\pm$ 0.001 \\ 
	& $(2, 0)$ & 1.000 $\pm$ 0.000 \\ 
	& $(2.5, 0)$ & 1.000 $\pm$ 0.000 \\ 
	& $(0, 1)$ & 1.000 $\pm$ 0.000 \\ 
	& $(1, 1)$ & 1.000 $\pm$ 0.000 \\ 
	& $(2, 1)$ & 1.000 $\pm$ 0.000 \\ 
	& $(2.5, 1)$ & 1.000 $\pm$ 0.000 \\ \hline
	\end{tabular}
\end{center}
\end{table}

\newpage

\newpage
\begin{table*}
\caption{Effect size for the cumulative degree distribution of the node.}
\begin{center}
\label{table:2}
	\begin{tabular}{A || B | C C C C C}\hline
	Data & \centering $(d_v, d_e)$ & $\mu^{\text{rand}}$ & $\sigma^{\text{rand}}$ & $\mu^{\text{emp}}$ & $\sigma^{\text{emp}}$ & Cohen's $d$ \rule[2mm]{0mm}{3mm} \\ \hline
	\multirow{8}{*}{drug}
& $(0, 0)$ & 0.028 & 0.010 & 0.603 & 0.008 & 62.88\\
& $(1, 0)$ & 0.000 & 0.000 & 0.000 & 0.000 & 0.000 \\
& $(2, 0)$ & 0.000 & 0.000 & 0.000 & 0.000 & 0.000 \\
& $(2.5, 0)$ & 0.000 & 0.000 & 0.000 & 0.000 & 0.000 \\
& $(0, 1)$ & 0.027 & 0.008 & 0.602 & 0.008 & 69.04\\
& $(1, 1)$ & 0.000 & 0.000 & 0.000 & 0.000 & 0.000 \\
& $(2, 1)$ & 0.000 & 0.000 & 0.000 & 0.000 & 0.000 \\
& $(2.5, 1)$ & 0.000 & 0.000 & 0.000 & 0.000 & 0.000 \\
\hline
\multirow{8}{*}{Enron}
& $(0, 0)$ & 0.062 & 0.018 & 0.421 & 0.019 & 19.58\\
& $(1, 0)$ & 0.000 & 0.000 & 0.000 & 0.000 & 0.000 \\
& $(2, 0)$ & 0.000 & 0.000 & 0.000 & 0.000 & 0.000 \\
& $(2.5, 0)$ & 0.000 & 0.000 & 0.000 & 0.000 & 0.000 \\
& $(0, 1)$ & 0.062 & 0.020 & 0.421 & 0.016 & 19.98\\
& $(1, 1)$ & 0.000 & 0.000 & 0.000 & 0.000 & 0.000 \\
& $(2, 1)$ & 0.000 & 0.000 & 0.000 & 0.000 & 0.000 \\
& $(2.5, 1)$ & 0.000 & 0.000 & 0.000 & 0.000 & 0.000 \\
\hline
\multirow{8}{*}{primary-school}
& $(0, 0)$ & 0.053 & 0.014 & 0.374 & 0.009 & 27.06\\
& $(1, 0)$ & 0.000 & 0.000 & 0.000 & 0.000 & 0.000 \\
& $(2, 0)$ & 0.000 & 0.000 & 0.000 & 0.000 & 0.000 \\
& $(2.5, 0)$ & 0.000 & 0.000 & 0.000 & 0.000 & 0.000 \\
& $(0, 1)$ & 0.052 & 0.013 & 0.373 & 0.009 & 27.86\\
& $(1, 1)$ & 0.000 & 0.000 & 0.000 & 0.000 & 0.000 \\
& $(2, 1)$ & 0.000 & 0.000 & 0.000 & 0.000 & 0.000 \\
& $(2.5, 1)$ & 0.000 & 0.000 & 0.000 & 0.000 & 0.000 \\
\hline
\multirow{8}{*}{high-school}
& $(0, 0)$ & 0.045 & 0.011 & 0.312 & 0.010 & 25.51\\
& $(1, 0)$ & 0.000 & 0.000 & 0.000 & 0.000 & 0.000 \\
& $(2, 0)$ & 0.000 & 0.000 & 0.000 & 0.000 & 0.000 \\
& $(2.5, 0)$ & 0.000 & 0.000 & 0.000 & 0.000 & 0.000 \\
& $(0, 1)$ & 0.042 & 0.011 & 0.311 & 0.010 & 25.69\\
& $(1, 1)$ & 0.000 & 0.000 & 0.000 & 0.000 & 0.000 \\
& $(2, 1)$ & 0.000 & 0.000 & 0.000 & 0.000 & 0.000 \\
& $(2.5, 1)$ & 0.000 & 0.000 & 0.000 & 0.000 & 0.000 \\
\hline
	\end{tabular}
\end{center}
\end{table*}

\newpage
\begin{table*}
\caption{Effect size for the average degree of the nearest neighbors of nodes with degree $k$.}
\begin{center}
\label{table:3}
	\begin{tabular}{A || B | C C C C C}\hline
	Data & \centering $(d_v, d_e)$ & $\mu^{\text{rand}}$ & $\sigma^{\text{rand}}$ & $\mu^{\text{emp}}$ & $\sigma^{\text{emp}}$ & Cohen's $d$ \rule[2mm]{0mm}{3mm} \\ \hline
\multirow{8}{*}{drug}
& $(0, 0)$ & 0.146 & 0.065 & 0.946 & 0.004 & 17.35\\
& $(1, 0)$ & 0.058 & 0.007 & 0.395 & 0.008 & 46.99\\
& $(2, 0)$ & 0.015 & 0.002 & 0.042 & 0.003 & 10.05\\
& $(2.5, 0)$ & 0.015 & 0.002 & 0.042 & 0.003 & 10.05\\
& $(0, 1)$ & 0.142 & 0.060 & 0.946 & 0.004 & 18.80\\
& $(1, 1)$ & 0.049 & 0.006 & 0.394 & 0.006 & 58.56\\
& $(2, 1)$ & 0.013 & 0.002 & 0.022 & 0.002 & 3.644\\
& $(2.5, 1)$ & 0.013 & 0.002 & 0.022 & 0.002 & 3.644\\
\hline
\multirow{8}{*}{Enron}
& $(0, 0)$ & 0.246 & 0.075 & 0.794 & 0.018 & 10.07\\
& $(1, 0)$ & 0.052 & 0.005 & 0.194 & 0.007 & 22.52\\
& $(2, 0)$ & 0.015 & 0.002 & 0.013 & 0.002 & -1.087\\
& $(2.5, 0)$ & 0.015 & 0.002 & 0.013 & 0.002 & -1.087\\
& $(0, 1)$ & 0.248 & 0.081 & 0.795 & 0.019 & 9.300\\
& $(1, 1)$ & 0.051 & 0.005 & 0.194 & 0.006 & 25.25\\
& $(2, 1)$ & 0.028 & 0.004 & 0.029 & 0.003 & 0.425\\
& $(2.5, 1)$ & 0.028 & 0.004 & 0.029 & 0.003 & 0.425\\
\hline
\multirow{8}{*}{primary-school}
& $(0, 0)$ & 0.257 & 0.049 & 0.838 & 0.017 & 15.99\\
& $(1, 0)$ & 0.021 & 0.001 & 0.089 & 0.002 & 41.77\\
& $(2, 0)$ & 0.006 & 0.001 & 0.007 & 0.001 & 0.844\\
& $(2.5, 0)$ & 0.006 & 0.001 & 0.007 & 0.001 & 0.844\\
& $(0, 1)$ & 0.259 & 0.058 & 0.840 & 0.017 & 13.56\\
& $(1, 1)$ & 0.026 & 0.002 & 0.089 & 0.002 & 32.88\\
& $(2, 1)$ & 0.010 & 0.001 & 0.014 & 0.001 & 4.733\\
& $(2.5, 1)$ & 0.010 & 0.001 & 0.014 & 0.001 & 4.733\\
\hline
\multirow{8}{*}{high-school}
& $(0, 0)$ & 0.209 & 0.051 & 0.710 & 0.014 & 13.38\\
& $(1, 0)$ & 0.034 & 0.004 & 0.114 & 0.003 & 21.34\\
& $(2, 0)$ & 0.011 & 0.002 & 0.010 & 0.002 & -0.565\\
& $(2.5, 0)$ & 0.011 & 0.002 & 0.010 & 0.002 & -0.565\\
& $(0, 1)$ & 0.211 & 0.061 & 0.708 & 0.015 & 11.23\\
& $(1, 1)$ & 0.042 & 0.005 & 0.114 & 0.004 & 16.33\\
& $(2, 1)$ & 0.020 & 0.003 & 0.025 & 0.002 & 2.007\\
& $(2.5, 1)$ & 0.020 & 0.003 & 0.025 & 0.002 & 2.007\\
\hline
	\end{tabular}
\end{center}
\end{table*}

\newpage
\begin{table*}
\caption{Effect size for the degree-dependent redundancy coefficient of the node.}
\begin{center}
\label{table:4}
	\begin{tabular}{A || B | C C C C C}\hline
	Data & \centering $(d_v, d_e)$ & $\mu^{\text{rand}}$ & $\sigma^{\text{rand}}$ & $\mu^{\text{emp}}$ & $\sigma^{\text{emp}}$ & Cohen's $d$ \rule[2mm]{0mm}{3mm} \\ \hline
\multirow{8}{*}{drug}
& $(0, 0)$ & 0.362 & 0.154 & 0.975 & 0.004 & 5.642\\
& $(1, 0)$ & 0.129 & 0.013 & 0.638 & 0.008 & 47.62\\
& $(2, 0)$ & 0.049 & 0.006 & 0.430 & 0.005 & 68.89\\
& $(2.5, 0)$ & 0.042 & 0.005 & 0.136 & 0.006 & 16.30\\
& $(0, 1)$ & 0.328 & 0.109 & 0.956 & 0.006 & 8.113\\
& $(1, 1)$ & 0.124 & 0.013 & 0.508 & 0.008 & 36.74\\
& $(2, 1)$ & 0.063 & 0.008 & 0.394 & 0.006 & 47.09\\
& $(2.5, 1)$ & 0.050 & 0.006 & 0.135 & 0.006 & 14.77\\
\hline
\multirow{8}{*}{Enron}
& $(0, 0)$ & 0.379 & 0.082 & 0.949 & 0.005 & 9.835\\
& $(1, 0)$ & 0.278 & 0.045 & 0.765 & 0.011 & 14.88\\
& $(2, 0)$ & 0.104 & 0.027 & 0.418 & 0.014 & 14.81\\
& $(2.5, 0)$ & 0.045 & 0.011 & 0.058 & 0.011 & 1.186\\
& $(0, 1)$ & 0.420 & 0.082 & 0.943 & 0.006 & 9.013\\
& $(1, 1)$ & 0.323 & 0.049 & 0.752 & 0.012 & 11.99\\
& $(2, 1)$ & 0.158 & 0.027 & 0.499 & 0.013 & 16.28\\
& $(2.5, 1)$ & 0.093 & 0.017 & 0.148 & 0.015 & 3.530\\
\hline
\multirow{8}{*}{primary-school}
& $(0, 0)$ & 0.324 & 0.047 & 0.927 & 0.008 & 17.83\\
& $(1, 0)$ & 0.147 & 0.009 & 0.547 & 0.006 & 54.16\\
& $(2, 0)$ & 0.058 & 0.005 & 0.371 & 0.008 & 46.30\\
& $(2.5, 0)$ & 0.034 & 0.004 & 0.206 & 0.007 & 29.40\\
& $(0, 1)$ & 0.326 & 0.054 & 0.970 & 0.003 & 16.91\\
& $(1, 1)$ & 0.150 & 0.012 & 0.808 & 0.002 & 78.27\\
& $(2, 1)$ & 0.106 & 0.008 & 0.564 & 0.004 & 68.08\\
& $(2.5, 1)$ & 0.058 & 0.005 & 0.115 & 0.005 & 11.72\\
\hline
\multirow{8}{*}{high-school}
& $(0, 0)$ & 0.326 & 0.048 & 0.910 & 0.005 & 17.26\\
& $(1, 0)$ & 0.304 & 0.079 & 0.741 & 0.031 & 7.296\\
& $(2, 0)$ & 0.129 & 0.019 & 0.409 & 0.015 & 16.32\\
& $(2.5, 0)$ & 0.013 & 0.005 & 0.030 & 0.005 & 3.792\\
& $(0, 1)$ & 0.359 & 0.059 & 0.965 & 0.002 & 14.57\\
& $(1, 1)$ & 0.393 & 0.228 & 0.895 & 0.019 & 3.103\\
& $(2, 1)$ & 0.225 & 0.034 & 0.761 & 0.008 & 22.00\\
& $(2.5, 1)$ & 0.044 & 0.007 & 0.064 & 0.008 & 2.697\\
\hline
	\end{tabular}
\end{center}
\end{table*}

\newpage
\begin{table*}
\caption{Effect size for the cumulative size distribution of the hyperedge.}
\begin{center}
\label{table:5}
	\begin{tabular}{A || B | C C C C C}\hline
	Data & \centering $(d_v, d_e)$ & $\mu^{\text{rand}}$ & $\sigma^{\text{rand}}$ & $\mu^{\text{emp}}$ & $\sigma^{\text{emp}}$ & Cohen's $d$ \rule[2mm]{0mm}{3mm} \\ \hline
\multirow{8}{*}{drug}
& $(0, 0)$ & 0.023 & 0.007 & 0.249 & 0.011 & 24.41\\
& $(1, 0)$ & 0.022 & 0.007 & 0.252 & 0.010 & 26.86\\
& $(2, 0)$ & 0.022 & 0.007 & 0.252 & 0.010 & 26.86\\
& $(2.5, 0)$ & 0.022 & 0.007 & 0.252 & 0.010 & 26.86\\
& $(0, 1)$ & 0.000 & 0.000 & 0.000 & 0.000 & 0.000 \\
& $(1, 1)$ & 0.000 & 0.000 & 0.000 & 0.000 & 0.000 \\
& $(2, 1)$ & 0.000 & 0.000 & 0.000 & 0.000 & 0.000 \\
& $(2.5, 1)$ & 0.000 & 0.000 & 0.000 & 0.000 & 0.000 \\
\hline
\multirow{8}{*}{Enron}
& $(0, 0)$ & 0.016 & 0.006 & 0.163 & 0.007 & 23.91\\
& $(1, 0)$ & 0.016 & 0.006 & 0.164 & 0.006 & 24.36\\
& $(2, 0)$ & 0.016 & 0.006 & 0.164 & 0.006 & 24.36\\
& $(2.5, 0)$ & 0.016 & 0.006 & 0.164 & 0.006 & 24.36\\
& $(0, 1)$ & 0.000 & 0.000 & 0.000 & 0.000 & 0.000 \\
& $(1, 1)$ & 0.000 & 0.000 & 0.000 & 0.000 & 0.000 \\
& $(2, 1)$ & 0.000 & 0.000 & 0.000 & 0.000 & 0.000 \\
& $(2.5, 1)$ & 0.000 & 0.000 & 0.000 & 0.000 & 0.000 \\
\hline
\multirow{8}{*}{primary-school}
& $(0, 0)$ & 0.005 & 0.002 & 0.304 & 0.003 & 124.64\\
& $(1, 0)$ & 0.005 & 0.002 & 0.304 & 0.003 & 125.55\\
& $(2, 0)$ & 0.005 & 0.002 & 0.304 & 0.003 & 125.55\\
& $(2.5, 0)$ & 0.005 & 0.002 & 0.304 & 0.003 & 125.55\\
& $(0, 1)$ & 0.000 & 0.000 & 0.000 & 0.000 & 0.000 \\
& $(1, 1)$ & 0.000 & 0.000 & 0.000 & 0.000 & 0.000 \\
& $(2, 1)$ & 0.000 & 0.000 & 0.000 & 0.000 & 0.000 \\
& $(2.5, 1)$ & 0.000 & 0.000 & 0.000 & 0.000 & 0.000 \\
\hline
\multirow{8}{*}{high-school}
& $(0, 0)$ & 0.006 & 0.003 & 0.325 & 0.004 & 96.33\\
& $(1, 0)$ & 0.007 & 0.003 & 0.324 & 0.003 & 105.77\\
& $(2, 0)$ & 0.007 & 0.003 & 0.324 & 0.003 & 105.77\\
& $(2.5, 0)$ & 0.007 & 0.003 & 0.324 & 0.003 & 105.77\\
& $(0, 1)$ & 0.000 & 0.000 & 0.000 & 0.000 & 0.000 \\
& $(1, 1)$ & 0.000 & 0.000 & 0.000 & 0.000 & 0.000 \\
& $(2, 1)$ & 0.000 & 0.000 & 0.000 & 0.000 & 0.000 \\
& $(2.5, 1)$ & 0.000 & 0.000 & 0.000 & 0.000 & 0.000 \\
\hline
	\end{tabular}
\end{center}
\end{table*}

\newpage
\begin{table*}
\caption{Effect size for the distribution of the shortest path length between nodes.}
\begin{center}
\label{table:6}
	\begin{tabular}{A || B | C C C C C}\hline
	Data & \centering $(d_v, d_e)$ & $\mu^{\text{rand}}$ & $\sigma^{\text{rand}}$ & $\mu^{\text{emp}}$ & $\sigma^{\text{emp}}$ & Cohen's $d$ \rule[2mm]{0mm}{3mm} \\ \hline
\multirow{8}{*}{drug}
& $(0, 0)$ & 0.006 & 0.005 & 1.577 & 0.005 & 334.86\\
& $(1, 0)$ & 0.019 & 0.012 & 1.331 & 0.016 & 92.60\\
& $(2, 0)$ & 0.043 & 0.024 & 0.762 & 0.032 & 25.39\\
& $(2.5, 0)$ & 0.074 & 0.041 & 0.437 & 0.052 & 7.711\\
& $(0, 1)$ & 0.005 & 0.003 & 1.609 & 0.005 & 435.53\\
& $(1, 1)$ & 0.018 & 0.013 & 1.416 & 0.019 & 85.59\\
& $(2, 1)$ & 0.043 & 0.023 & 0.788 & 0.029 & 28.82\\
& $(2.5, 1)$ & 0.077 & 0.036 & 0.590 & 0.045 & 12.61\\
\hline
\multirow{8}{*}{Enron}
& $(0, 0)$ & 0.012 & 0.009 & 0.607 & 0.010 & 61.19\\
& $(1, 0)$ & 0.013 & 0.009 & 0.489 & 0.004 & 70.38\\
& $(2, 0)$ & 0.012 & 0.007 & 0.434 & 0.007 & 60.01\\
& $(2.5, 0)$ & 0.016 & 0.009 & 0.352 & 0.014 & 28.21\\
& $(0, 1)$ & 0.009 & 0.006 & 0.655 & 0.009 & 82.50\\
& $(1, 1)$ & 0.015 & 0.009 & 0.486 & 0.005 & 66.53\\
& $(2, 1)$ & 0.013 & 0.008 & 0.438 & 0.008 & 51.94\\
& $(2.5, 1)$ & 0.020 & 0.012 & 0.370 & 0.012 & 28.35\\
\hline
\multirow{8}{*}{primary-school}
& $(0, 0)$ & 0.005 & 0.004 & 0.860 & 0.005 & 191.98\\
& $(1, 0)$ & 0.005 & 0.004 & 0.706 & 0.005 & 166.63\\
& $(2, 0)$ & 0.004 & 0.003 & 0.372 & 0.003 & 124.09\\
& $(2.5, 0)$ & 0.004 & 0.003 & 0.346 & 0.004 & 93.71\\
& $(0, 1)$ & 0.004 & 0.003 & 0.538 & 0.003 & 163.88\\
& $(1, 1)$ & 0.003 & 0.003 & 0.435 & 0.003 & 154.58\\
& $(2, 1)$ & 0.003 & 0.002 & 0.244 & 0.003 & 108.59\\
& $(2.5, 1)$ & 0.004 & 0.003 & 0.035 & 0.000 & 16.02\\
\hline
\multirow{8}{*}{high-school}
& $(0, 0)$ & 0.004 & 0.003 & 0.534 & 0.000 & 218.95\\
& $(1, 0)$ & 0.006 & 0.003 & 0.529 & 0.002 & 184.54\\
& $(2, 0)$ & 0.005 & 0.003 & 0.510 & 0.003 & 185.10\\
& $(2.5, 0)$ & 0.006 & 0.003 & 0.504 & 0.003 & 159.05\\
& $(0, 1)$ & 0.002 & 0.001 & 0.534 & 0.000 & 677.59\\
& $(1, 1)$ & 0.004 & 0.002 & 0.514 & 0.003 & 228.79\\
& $(2, 1)$ & 0.003 & 0.002 & 0.494 & 0.002 & 253.58\\
& $(2.5, 1)$ & 0.004 & 0.003 & 0.440 & 0.003 & 156.22\\
\hline
	\end{tabular}
\end{center}
\end{table*}

\newpage
\begin{table*}
\caption{Effect size for the cumulative degree distribution of the one-mode projection.}
\begin{center}
\label{table:7}
	\begin{tabular}{A || B | C C C C C}\hline
	Data & \centering $(d_v, d_e)$ & $\mu^{\text{rand}}$ & $\sigma^{\text{rand}}$ & $\mu^{\text{emp}}$ & $\sigma^{\text{emp}}$ & Cohen's $d$ \rule[2mm]{0mm}{3mm} \\ \hline
\multirow{8}{*}{drug}
& $(0, 0)$ & 0.036 & 0.009 & 0.663 & 0.008 & 73.88\\
& $(1, 0)$ & 0.026 & 0.006 & 0.237 & 0.009 & 27.67\\
& $(2, 0)$ & 0.027 & 0.008 & 0.086 & 0.010 & 6.477\\
& $(2.5, 0)$ & 0.029 & 0.009 & 0.080 & 0.011 & 5.106\\
& $(0, 1)$ & 0.035 & 0.010 & 0.700 & 0.008 & 75.60\\
& $(1, 1)$ & 0.031 & 0.009 & 0.313 & 0.012 & 27.07\\
& $(2, 1)$ & 0.030 & 0.008 & 0.052 & 0.009 & 2.609\\
& $(2.5, 1)$ & 0.032 & 0.009 & 0.045 & 0.009 & 1.532\\
\hline
\multirow{8}{*}{Enron}
& $(0, 0)$ & 0.087 & 0.025 & 0.379 & 0.011 & 14.88\\
& $(1, 0)$ & 0.049 & 0.011 & 0.074 & 0.008 & 2.553\\
& $(2, 0)$ & 0.041 & 0.011 & 0.085 & 0.012 & 3.792\\
& $(2.5, 0)$ & 0.046 & 0.011 & 0.071 & 0.011 & 2.274\\
& $(0, 1)$ & 0.072 & 0.017 & 0.390 & 0.017 & 18.47\\
& $(1, 1)$ & 0.058 & 0.013 & 0.081 & 0.012 & 1.880\\
& $(2, 1)$ & 0.044 & 0.010 & 0.057 & 0.010 & 1.297\\
& $(2.5, 1)$ & 0.050 & 0.012 & 0.050 & 0.010 & -0.070\\
\hline
\multirow{8}{*}{primary-school}
& $(0, 0)$ & 0.069 & 0.018 & 0.697 & 0.015 & 38.60\\
& $(1, 0)$ & 0.038 & 0.006 & 0.370 & 0.008 & 44.64\\
& $(2, 0)$ & 0.027 & 0.006 & 0.359 & 0.005 & 60.54\\
& $(2.5, 0)$ & 0.028 & 0.005 & 0.359 & 0.006 & 58.58\\
& $(0, 1)$ & 0.054 & 0.015 & 0.385 & 0.011 & 24.59\\
& $(1, 1)$ & 0.029 & 0.005 & 0.044 & 0.004 & 3.198\\
& $(2, 1)$ & 0.024 & 0.004 & 0.032 & 0.003 & 2.326\\
& $(2.5, 1)$ & 0.025 & 0.004 & 0.032 & 0.004 & 1.738\\
\hline
\multirow{8}{*}{high-school}
& $(0, 0)$ & 0.056 & 0.018 & 0.628 & 0.014 & 35.84\\
& $(1, 0)$ & 0.035 & 0.008 & 0.361 & 0.013 & 31.38\\
& $(2, 0)$ & 0.027 & 0.006 & 0.321 & 0.008 & 41.20\\
& $(2.5, 0)$ & 0.030 & 0.006 & 0.320 & 0.008 & 42.14\\
& $(0, 1)$ & 0.047 & 0.012 & 0.347 & 0.009 & 28.38\\
& $(1, 1)$ & 0.027 & 0.005 & 0.072 & 0.005 & 9.151\\
& $(2, 1)$ & 0.024 & 0.004 & 0.050 & 0.004 & 6.595\\
& $(2.5, 1)$ & 0.024 & 0.004 & 0.049 & 0.004 & 6.194\\
\hline
	\end{tabular}
\end{center}
\end{table*}


\begin{thebibliography}{10}
\providecommand{\url}[1]{#1}
\csname url@samestyle\endcsname
\providecommand{\newblock}{\relax}
\providecommand{\bibinfo}[2]{#2}
\providecommand{\BIBentrySTDinterwordspacing}{\spaceskip=0pt\relax}
\providecommand{\BIBentryALTinterwordstretchfactor}{4}
\providecommand{\BIBentryALTinterwordspacing}{\spaceskip=\fontdimen2\font plus
\BIBentryALTinterwordstretchfactor\fontdimen3\font minus
  \fontdimen4\font\relax}
\providecommand{\BIBforeignlanguage}[2]{{%
\expandafter\ifx\csname l@#1\endcsname\relax
\typeout{** WARNING: IEEEtran.bst: No hyphenation pattern has been}%
\typeout{** loaded for the language `#1'. Using the pattern for}%
\typeout{** the default language instead.}%
\else
\language=\csname l@#1\endcsname
\fi
#2}}
\providecommand{\BIBdecl}{\relax}
\BIBdecl

\bibitem{boccaletti}
S.~Boccaletti, V.~Latora, Y.~Moreno, M.~Chavez, and D.-U. Hwang, ``Complex
  networks: Structure and dynamics,'' \emph{Phys. Rep.}, vol. 424, pp.
  175--308, 2006.

\bibitem{barrat2008}
A.~Barrat, M.~Barth{\'e}lemy, and A.~Vespignani, \emph{Dynamical Processes on
  Complex Networks}.\hskip 1em plus 0.5em minus 0.4em\relax Cambridge, UK:
  Cambridge University Press, 2008.

\bibitem{latora2017}
V.~Latora, V.~Nicosia, and G.~Russo, \emph{Complex Networks: {P}rinciples,
  Methods and Applications}.\hskip 1em plus 0.5em minus 0.4em\relax Cambridge,
  UK: Cambridge University Press, 2017.

\bibitem{newman_networks}
M.~E.~J. Newman, \emph{Networks. Second Edition}.\hskip 1em plus 0.5em minus
  0.4em\relax Oxford, UK: Oxford University Press, 2018.

\bibitem{stehle}
J.~Stehlé, N.~Voirin, A.~Barrat, C.~Cattuto, L.~Isella, J.-F. Pinton,
  M.~Quaggiotto, W.~Van~den Broeck, C.~Régis, B.~Lina, and P.~Vanhems,
  ``High-resolution measurements of face-to-face contact patterns in a primary
  school,'' \emph{PLOS ONE}, vol.~6, 2011, article No. e23176.

\bibitem{mastrandrea}
R.~Mastrandrea, J.~Fournet, and A.~Barrat, ``Contact patterns in a high school:
  {A} comparison between data collected using wearable sensors, contact diaries
  and friendship surveys,'' \emph{PLOS ONE}, vol.~10, 2015, article No.
  e0136497.

\bibitem{klimt}
B.~Klimt and Y.~Yang, ``The {E}nron corpus: A new dataset for email
  classification research,'' in \emph{European Conference on Machine Learning},
  2004, pp. 217--226.

\bibitem{newman2001_2}
M.~E.~J. Newman, ``The structure of scientific collaboration networks,''
  \emph{Proc. Natl. Acad. Sci. USA}, vol.~98, pp. 404--409, 2001, article No.
  e23176.

\bibitem{patania}
A.~Patania, G.~Petri, and F.~Vaccarino, ``The shape of collaborations,''
  \emph{EPJ Data Sci.}, vol.~6, 2017, article No. 18.

\bibitem{vasilyeva2021}
E.~Vasilyeva, A.~Kozlov, K.~Alfaro-Bittner, D.~Musatov, A.~M. Raigorodskii,
  M.~Perc, and S.~Boccaletti, ``Multilayer representation of collaboration
  networks with higher-order interactions,'' \emph{Sci. Rep.}, vol.~11, 2021,
  article No. 5666.

\bibitem{klamt}
S.~Klamt, U.-U. Haus, and F.~Theis, ``Hypergraphs and cellular networks,''
  \emph{PLOS Comput. Biol.}, vol.~5, 2009, article No. e1000385.

\bibitem{estrada}
E.~Estrada and G.~J. Ross, ``Centralities in simplicial complexes.
  {A}pplications to protein interaction networks,'' \emph{J. Theor. Biol.},
  vol. 438, pp. 46--60, 2018.

\bibitem{benson}
A.~R. Benson, R.~Abebe, M.~T. Schaub, A.~Jadbabaie, and J.~Kleinberg,
  ``Simplicial closure and higher-order link prediction,'' \emph{Proc. Natl.
  Acad. Sci. USA}, vol. 115, pp. E11\,221--E11\,230, 2018.

\bibitem{battiston}
F.~Battiston, G.~Cencetti, I.~Iacopini, V.~Latora, M.~Lucas, A.~Patania, J.-G.
  Young, and G.~Petri, ``Networks beyond pairwise interactions: {S}tructure and
  dynamics,'' \emph{Phys. Rep.}, vol. 874, pp. 1--92, 2020.

\bibitem{newman2001_3}
M.~E.~J. Newman, ``Scientific collaboration networks. {I}. {N}etwork
  construction and fundamental results,'' \emph{Phys. Rev. E}, vol.~64, 2001,
  article No. 016131.

\bibitem{barabasi2002}
A.~L. Barabási, H.~Jeong, Z.~Néda, E.~Ravasz, A.~Schubert, and T.~Vicsek,
  ``Evolution of the social network of scientific collaborations,''
  \emph{Physica A}, vol. 311, pp. 590--614, 2002.

\bibitem{ramasco}
J.~J. Ramasco, S.~N. Dorogovtsev, and R.~Pastor-Satorras, ``Self-organization
  of collaboration networks,'' \emph{Phys. Rev. E}, vol.~70, 2004, article No.
  036106.

\bibitem{gomez}
J.~Gómez-Gardeñes, M.~Romance, R.~Criado, D.~Vilone, and A.~Sánchez,
  ``Evolutionary games defined at the network mesoscale: The public goods
  game,'' \emph{Chaos}, vol.~21, 2011, article No. 016113.

\bibitem{benson2016}
A.~R. Benson, D.~F. Gleich, and J.~Leskovec, ``Higher-order organization of
  complex networks,'' \emph{Science}, vol. 353, pp. 163--166, 2016.

\bibitem{giusti}
C.~Giusti, R.~Ghrist, and D.~S. Bassett, ``Two's company, three (or more) is a
  simplex,'' \emph{J. Comput. Neurosci.}, vol.~41, pp. 1--14, 2016.

\bibitem{grilli}
J.~Grilli, G.~Barab{\'a}s, M.~J. Michalska-Smith, and S.~Allesina,
  ``Higher-order interactions stabilize dynamics in competitive network
  models,'' \emph{Nature}, vol. 548, pp. 210--213, 2017.

\bibitem{lambiotte2019}
R.~Lambiotte, M.~Rosvall, and I.~Scholtes, ``From networks to optimal
  higher-order models of complex systems,'' \emph{Nat. Phys.}, vol.~15, pp.
  313--320, 2019.

\bibitem{chodrow}
P.~S. Chodrow, ``Configuration models of random hypergraphs,'' \emph{J. Complex
  Netw.}, vol.~8, 2020, article No. cnaa018.

\bibitem{schaub}
M.~T. Schaub, A.~R. Benson, P.~Horn, G.~Lippner, and A.~Jadbabaie, ``Random
  walks on simplicial complexes and the normalized {H}odge 1-{L}aplacian,''
  \emph{SIAM Rev.}, vol.~62, pp. 353--391, 2020.

\bibitem{yoon}
S.~Yoon, H.~Song, K.~Shin, and Y.~Yi, ``How much and when do we need
  higher-order information in hypergraphs? {A} case study on hyperedge
  prediction,'' in \emph{Proceedings of The Web Conference 2020}, 2020, pp.
  2627--2633.

\bibitem{lee2021}
G.~Lee, M.~Choe, and K.~Shin, ``How do hyperedges overlap in real-world
  hypergraphs? - {P}atterns, measures, and generators,'' in \emph{Proceedings
  of the Web Conference 2021}, 2021, pp. 3396--3407.

\bibitem{cimini2019}
G.~Cimini, T.~Squartini, F.~Saracco, D.~Garlaschelli, A.~Gabrielli, and
  G.~Caldarelli, ``The statistical physics of real-world networks,'' \emph{Nat.
  Rev. Phys.}, vol.~1, pp. 58--71, 2019.

\bibitem{molloy}
M.~Molloy and B.~Reed, ``A critical point for random graphs with a given degree
  sequence,'' \emph{Random Struct. Algorithms}, vol.~6, pp. 161--180, 1995.

\bibitem{newman2001}
M.~E.~J. Newman, S.~H. Strogatz, and D.~J. Watts, ``Random graphs with
  arbitrary degree distributions and their applications,'' \emph{Phys. Rev. E},
  vol.~64, 2001, article No. 026118.

\bibitem{fosdick}
B.~K. Fosdick, D.~B. Larremore, J.~Nishimura, and J.~Ugander, ``Configuring
  random graph models with fixed degree sequences,'' \emph{SIAM Rev.}, vol.~60,
  pp. 315--355, 2018.

\bibitem{watts}
D.~J. Watts and S.~H. Strogatz, ``Collective dynamics of ‘small-world’
  networks,'' \emph{Nature}, vol. 393, pp. 440--442, 1998.

\bibitem{milo}
R.~Milo, S.~Shen-Orr, S.~Itzkovitz, N.~Kashtan, D.~Chklovskii, and U.~Alon,
  ``Network motifs: {S}imple building blocks of complex networks,''
  \emph{Science}, vol. 298, pp. 824--827, 2002.

\bibitem{newman2002}
M.~E.~J. Newman, ``Assortative mixing in networks,'' \emph{Phys. Rev. Lett.},
  vol.~89, 2002, article No. 208701.

\bibitem{newman2006}
------, ``Modularity and community structure in networks,'' \emph{Proc. Natl.
  Acad. Sci. USA}, vol. 103, pp. 8577--8582, 2006.

\bibitem{maslov}
S.~Maslov, K.~Sneppen, and A.~Zaliznyak, ``Detection of topological patterns in
  complex networks: correlation profile of the internet,'' \emph{Physica A},
  vol. 333, pp. 529--540, 2004.

\bibitem{serrano}
M.~{\'A}ngeles~Serrano and M.~Bogu\~n\'a, ``Tuning clustering in random
  networks with arbitrary degree distributions,'' \emph{Phys. Rev. E}, vol.~72,
  2005, article No. 036133.

\bibitem{mahadevan}
P.~Mahadevan, D.~Krioukov, K.~Fall, and A.~Vahdat, ``Systematic topology
  analysis and generation using degree correlations,'' \emph{SIGCOMM Comput.
  Commun. Rev.}, vol.~36, pp. 135--146, 2006.

\bibitem{newman2009}
M.~E.~J. Newman, ``Random graphs with clustering,'' \emph{Phys. Rev. Lett.},
  vol. 103, 2009, article No. 058701.

\bibitem{karrer}
B.~Karrer and M.~E.~J. Newman, ``Random graphs containing arbitrary
  distributions of subgraphs,'' \emph{Phys. Rev. E}, vol.~82, 2010, article No.
  066118.

\bibitem{stanton}
I.~Stanton and A.~Pinar, ``Constructing and sampling graphs with a prescribed
  joint degree distribution,'' \emph{ACM J. Exp. Algorithmics}, vol.~17, 2012,
  article No. 3.5.

\bibitem{gjoka_2_5_k}
M.~Gjoka, M.~Kurant, and A.~Markopoulou, ``2.5{K}-graphs: {F}rom sampling to
  generation,'' in \emph{2013 Proceedings of IEEE INFOCOM}, 2013, pp.
  1968--1976.

\bibitem{bassler}
K.~E. Bassler, C.~I. Del~Genio, P.~L. Erd{\H{o}}s, I.~Mikl{\'{o}}s, and
  Z.~Toroczkai, ``Exact sampling of graphs with prescribed degree
  correlations,'' \emph{New J. Phys.}, vol.~17, 2015, article No. 083052.

\bibitem{orsini}
C.~Orsini, M.~M. Dankulov, P.~Colomer{-}de{-}Simón, A.~Jamakovic,
  P.~Mahadevan, A.~Vahdat, K.~E. Bassler, Z.~Toroczkai, M.~Boguñá,
  G.~Caldarelli, S.~Fortunato, and D.~Krioukov, ``Quantifying randomness in
  real networks,'' \emph{Nat. Commun.}, vol.~6, 2015, article No. 8627.

\bibitem{saracco2015}
F.~Saracco, R.~Di~Clemente, A.~Gabrielli, and T.~Squartini, ``Randomizing
  bipartite networks: the case of the {W}orld {T}rade {W}eb,'' \emph{Sci.
  Rep.}, vol.~5, 2015, article No. 10595.

\bibitem{boroojeni}
A.~A. Boroojeni, J.~Dewar, T.~Wu, and J.~M. Hyman, ``Generating bipartite
  networks with a prescribed joint degree distribution,'' \emph{J. Complex
  Netw.}, vol.~5, pp. 839--857, 2017.

\bibitem{saracco2017}
F.~Saracco, M.~J. Straka, R.~Di~Clemente, A.~Gabrielli, G.~Caldarelli, and
  T.~Squartini, ``Inferring monopartite projections of bipartite networks: {A}n
  entropy-based approach,'' \emph{New J. Phys.}, vol.~19, 2017, article No.
  053022.

\bibitem{de}
G.~F. de~Arruda, G.~Petri, and Y.~Moreno, ``Social contagion models on
  hypergraphs,'' \emph{Phys. Rev. Research}, vol.~2, 2020, article No. 023032.

\bibitem{alvarez}
U.~Alvarez-Rodriguez, F.~Battiston, G.~F. de~Arruda, Y.~Moreno, M.~Perc, and
  V.~Latora, ``Evolutionary dynamics of higher-order interactions in social
  networks,'' \emph{Nat. Human Behav.}, vol.~5, pp. 586--595, 2021.

\bibitem{pastor}
R.~Pastor-Satorras, A.~V\'azquez, and A.~Vespignani, ``Dynamical and
  correlation properties of the {I}nternet,'' \emph{Phys. Rev. Lett.}, vol.~87,
  2001, article No. 258701.

\bibitem{latapy}
M.~Latapy, C.~Magnien, and N.~D. Vecchio, ``Basic notions for the analysis of
  large two-mode networks,'' \emph{Soc. Netw.}, vol.~30, pp. 31--48, 2008.

\bibitem{guillaume}
J.-L. Guillaume and M.~Latapy, ``Bipartite graphs as models of complex
  networks,'' \emph{Physica A}, vol. 371, pp. 795--813, 2006.

\bibitem{pena}
J.~Peña and Y.~Rochat, ``Bipartite graphs as models of population structures
  in evolutionary multiplayer games,'' \emph{PLOS ONE}, vol.~7, 2012, article
  No. e44514.

\bibitem{tarissan}
F.~Tarissan, B.~Quoitin, P.~M{\'e}rindol, B.~Donnet, J.~J. Pansiot, and
  M.~Latapy, ``Towards a bipartite graph modeling of the internet topology,''
  \emph{Comput. Netw.}, vol.~57, pp. 2331--2347, 2013.

\bibitem{payrato2019}
C.~Payrat\'o-Borr\`as, L.~Hern\'andez, and Y.~Moreno, ``Breaking the spell of
  nestedness: {T}he entropic origin of nestedness in mutualistic systems,''
  \emph{Phys. Rev. X}, vol.~9, 2019, article No. 031024.

\bibitem{er}
P.~Erd{\H{o}}s and A.~R{\'e}nyi, ``On random graphs {I},'' \emph{Publ. Math.
  Debrecen}, vol.~6, pp. 290--297, 1959.

\bibitem{zhou2011}
W.~Zhou and L.~Nakhleh, ``Properties of metabolic graphs: {B}iological
  organization or representation artifacts?'' \emph{BMC Bioinformatics},
  vol.~12, 2011, article No. 132.

\bibitem{ramasco2006}
J.~J. Ramasco and S.~A. Morris, ``Social inertia in collaboration networks,''
  \emph{Phys. Rev. E}, vol.~73, 2006, article No. 016122.

\bibitem{li2007}
M.~Li, J.~Wu, D.~Wang, T.~Zhou, Z.~Di, and Y.~Fan, ``Evolving model of weighted
  networks inspired by scientific collaboration networks,'' \emph{Physica A},
  vol. 375, pp. 355--364, 2007.

\bibitem{bodo}
{\'A}.~Bod{\'o}, G.~Y. Katona, and P.~L. Simon, ``{SIS} epidemic propagation on
  hypergraphs,'' \emph{Bull. Math. Biol.}, vol.~78, pp. 713--735, 2016.

\bibitem{suo2018}
Q.~Suo, J.-L. Guo, and A.-Z. Shen, ``Information spreading dynamics in
  hypernetworks,'' \emph{Physica A}, vol. 495, pp. 475--487, 2018.

\bibitem{jhun2019}
B.~Jhun, M.~Jo, and B.~Kahng, ``{Simplicial {SIS} model in scale-free uniform
  hypergraph},'' \emph{J. Stat. Mech.}, vol. 2019, 2019, article No. 123207.

\bibitem{iacopini}
I.~Iacopini, G.~Petri, A.~Barrat, and V.~Latora, ``Simplicial models of social
  contagion,'' \emph{Nat. Commun.}, vol.~10, 2019, article No. 2485.

\bibitem{landry2020}
N.~W. Landry and J.~G. Restrepo, ``The effect of heterogeneity on hypergraph
  contagion models,'' \emph{Chaos}, vol.~30, 2020, article No. 103117.

\bibitem{burgio2020}
G.~Burgio, J.~T. Matamalas, S.~Gómez, and A.~Arenas, ``Evolution of
  cooperation in the presence of higher-order interactions: {F}rom networks to
  hypergraphs,'' \emph{Entropy}, vol.~22, 2020, article No. 744.

\bibitem{neuhauser2020}
L.~Neuh\"auser, A.~Mellor, and R.~Lambiotte, ``Multibody interactions and
  nonlinear consensus dynamics on networked systems,'' \emph{Phys. Rev. E},
  vol. 101, 2020, article No. 032310.

\bibitem{hickok2021}
A.~Hickok, Y.~Kureh, H.~Z. Brooks, M.~Feng, and M.~A. Porter, ``A
  bounded-confidence model of opinion dynamics on hypergraphs,'' \emph{arXiv
  preprint arXiv:2102.06825}, 2021.

\bibitem{sahasrabuddhe2021}
R.~Sahasrabuddhe, L.~Neuhäuser, and R.~Lambiotte, ``Modelling non-linear
  consensus dynamics on hypergraphs,'' \emph{J. Phys. Complexity}, vol.~2,
  2021, article No. 025006.

\bibitem{lucas2020}
M.~Lucas, G.~Cencetti, and F.~Battiston, ``Multiorder laplacian for
  synchronization in higher-order networks,'' \emph{Phys. Rev. Research},
  vol.~2, 2020, article No. 033410.

\bibitem{mulas2020}
R.~Mulas, C.~Kuehn, and J.~Jost, ``Coupled dynamics on hypergraphs: {M}aster
  stability of steady states and synchronization,'' \emph{Phys. Rev. E}, vol.
  101, 2020, article No. 062313.

\bibitem{de2021}
G.~F. de~Arruda, M.~Tizzani, and Y.~Moreno, ``Phase transitions and stability
  of dynamical processes on hypergraphs,'' \emph{Commun. Phys.}, vol.~4, 2021,
  article No. 24.

\bibitem{salova2021}
A.~Salova and R.~M. D'Souza, ``Cluster synchronization on hypergraphs,''
  \emph{arXiv preprint arXiv:2101.05464}, 2021.

\bibitem{doreian2004}
P.~Doreian, V.~Batagelj, and A.~Ferligoj, ``Generalized blockmodeling of
  two-mode network data,'' \emph{Soc. Netw.}, vol.~26, pp. 29--53, 2004.

\bibitem{ball2011}
B.~Ball, B.~Karrer, and M.~E.~J. Newman, ``Efficient and principled method for
  detecting communities in networks,'' \emph{Phys. Rev. E}, vol.~84, 2011,
  article No. 036103.

\bibitem{larremore2014}
D.~B. Larremore, A.~Clauset, and A.~Z. Jacobs, ``Efficiently inferring
  community structure in bipartite networks,'' \emph{Phys. Rev. E}, vol.~90,
  2014, article No. 012805.

\bibitem{yen2020}
T.-C. Yen and D.~B. Larremore, ``Community detection in bipartite networks with
  stochastic block models,'' \emph{Phys. Rev. E}, vol. 102, 2020, article No.
  032309.

\bibitem{ghoshdastidar2017}
D.~Ghoshdastidar and A.~Dukkipati, ``Consistency of spectral hypergraph
  partitioning under planted partition model,'' \emph{Ann. Stat.}, vol.~45, pp.
  289--315, 2017.

\bibitem{ahn2018}
K.~Ahn, K.~Lee, and C.~Suh, ``Hypergraph spectral clustering in the weighted
  stochastic block model,'' \emph{IEEE J. Selected Topics in Signal
  Processing}, vol.~12, pp. 959--974, 2018.

\bibitem{ke2019}
Z.~T. Ke, F.~Shi, and D.~Xia, ``Community detection for hypergraph networks via
  regularized tensor power iteration,'' \emph{arXiv preprint arXiv:1909.06503},
  2019.

\bibitem{chodrow2021}
P.~S. Chodrow, N.~Veldt, and A.~R. Benson, ``Generative hypergraph clustering:
  {F}rom blockmodels to modularity,'' \emph{Science Advances}, vol.~7, 2021,
  article No. eabh1303.

\end{thebibliography}

\begin{thebibliography}{1}

\bibitem{cohen1988}
Jacob Cohen.
\newblock {\em Statistical Power Analysis for the Behavioral Sciences}.
\newblock Routledge Academic, New York, NY, 1988.

\bibitem{sawilowsky2009}
Shlomo~S Sawilowsky.
\newblock New effect size rules of thumb.
\newblock {\em Journal of Modern Applied Statistical Methods}, 8:26, 2009.

\end{thebibliography}
\end{document}